\documentclass[sigconf, nonacm]{acmart}

\usepackage{soul}
\usepackage{xspace}
\usepackage{listings}

\usepackage{tikz}
\usetikzlibrary{positioning}
\usepackage{xcolor}
\usepackage{booktabs}
\usepackage{cleveref}
\usepackage{csquotes}

\crefname{section}{Sec.}{Secs.}
\crefname{subsection}{Sec.}{Secs.}
\crefname{subsubsection}{Sec.}{Secs.}
\crefname{figure}{Fig.}{Figs.}
\crefname{table}{Table}{Tables}
\crefname{equation}{Eq.}{Eqs.}
\crefname{appendix}{App.}{Apps.}
\crefname{algorithm}{Alg.}{Algs.}
\crefname{theorem}{Thm.}{Thms.}
\crefname{lemma}{Lem.}{Lems.}
\crefname{definition}{Def.}{Defs.}
\crefname{example}{Ex.}{Exs.}

\usepackage[ruled,linesnumbered,vlined]{algorithm2e}
\usepackage{algorithmic}
\usepackage{enumitem}
\usepackage{pifont}
\usepackage{makecell} 
\usepackage{booktabs}
\usepackage{array}
\usepackage{multicol}
\usepackage{multirow}
\usepackage{graphicx}
\usepackage{subcaption}
\usepackage{colortbl}
\usepackage[most]{tcolorbox}

\newcommand{\techreport}[1]{#1}
\newcommand{\papertext}[1]{}

\newcolumntype{C}[1]{>{\centering\arraybackslash}p{#1}}
\newcolumntype{L}[1]{>{\raggedright\arraybackslash}p{#1}}

\newcommand{\topic}[1]{\vspace{-3.5pt}\smallskip \smallskip \noindent{\bf #1.}}

\newcommand{\ttt}[1]{{\small \texttt{#1}}\xspace}
\newcommand{\tinytt}[1]{{\scriptsize \texttt{#1}}\xspace}

\newcommand{\mtt}[1]{{\small \texttt{#1}}\xspace}

\newcommand{\revision}[1]{{#1}}

\definecolor{darkred}{RGB}{140,15,70}
 \newcommand{\com}[1]{
 \smallskip \noindent \textcolor{darkred}{{\em #1}}}

\newif\ifrebuttal
\rebuttalfalse %

\definecolor{ronecolor}{RGB}{0,80,170}        %
\definecolor{rtwocolor}{RGB}{50,120,50}           %
\definecolor{rfourcolor}{RGB}{190,70,0}       %
\ifrebuttal
  \renewcommand{\revision}[1]{\textcolor{darkred}{#1}}
  \newcommand{\rone}[1]{\textcolor{ronecolor}{#1}}
  \newcommand{\rtwo}[1]{\textcolor{rtwocolor}{#1}}
  \newcommand{\rfour}[1]{\textcolor{rfourcolor}{#1}}
\else
  \renewcommand{\revision}[1]{#1}
  \newcommand{\rone}[1]{#1}
  \newcommand{\rtwo}[1]{#1}
  \newcommand{\rfour}[1]{#1}
\fi

\newcommand{\comone}[1]{
\smallskip \noindent \textcolor{ronecolor}{{\em #1}}}
\newcommand{\comtwo}[1]{
\smallskip \noindent \textcolor{rtwocolor}{{\em #1}}}
\newcommand{\comfour}[1]{
\smallskip \noindent \textcolor{rfourcolor}{{\em #1}}}

\newcommand{\cmark}{\ding{51}} %
\newcommand{\xmark}{\ding{55}} %

\definecolor{opcolor}{RGB}{0,100,80} %
\newcommand{\op}[1]{\textcolor{opcolor}{\small \texttt{#1}}\xspace}

\definecolor{rewrite}{HTML}{0C815A}
\definecolor{rewritered}{HTML}{EF4444}

\newcommand{\dnum}[1]{\tikz[baseline=-0.7ex]{\node[circle,fill=violet,text=white,inner sep=0pt,font=\fontsize{5}{5}\selectfont\bfseries,minimum size=1em] {#1};}\,}

\newcommand{\newcat}{\textsuperscript{\dag}}  %
\newcommand{\paramsens}{\textsuperscript{\ddag}}  %

\usepackage{pifont}
\newcommand{\circled}[1]{\ding{\numexpr171+#1\relax}}

\usepackage{hyperref}

\SetAlgoSkip{-15pt}
\usepackage{titlesec}
\titleformat{\section}
  {\normalfont\Large\bfseries}{\thesection}{1em}{\MakeUppercase}
\titlespacing*{\section}{0pt}{1ex}{0.75ex}
\titlespacing*{\subsection}{0pt}{0.85ex}{0.75ex}
\titlespacing*{\subsubsection}{0pt}{0.75ex}{0.75ex}
\newcommand\vldbpagestyle{plain}

\begin{document}

\ifrebuttal
\twocolumn

\begin{center}
{\LARGE \bf Summary for Revised Paper
``Multi-Objective Agentic Rewrites for Unstructured Data Processing''}
\end{center}

\noindent We thank the reviewers and meta-reviewer for their detailed and insightful feedback. We are happy to hear that reviewers appreciated the timeliness of the problem and the substantial system implementation.
We appreciate the opportunity to address their feedback in a revision. Over the last couple of months, we have prepared a new version of the paper based on their suggestions, with changes throughout the paper and {\em significant updates to the evaluation, including {\bf four new experiments} and {\bf three new analyses}}. The prominent changes include:

\smallskip
\noindent {\em {\bf A. New experiments and baselines.}} We ran four new experiments, each with new implementations as baselines: (1) we compared MOAR's Pareto frontier to that achievable by model substitution alone across all workloads; (2) we evaluated open-source LLMs as the optimization agent instead of gpt-5; (3) we ran MOAR with only the DocETL-V1 directives; and (4) we introduced MOAR (no search), an ablation that gives the simple LLM agent access to all MOAR directives without MOAR's new search algorithm.

\smallskip
\noindent {\em {\bf B. New analyses of experimental results.}} We performed three new analyses: (1) to understand which rewrites matter most, we analyzed which rewrite directives appear in the top Pareto-optimal pipelines per workload; (2) to understand optimization efficiency, we measured the fraction of explored pipelines that never contribute to the Pareto frontier per workload; and (3) we performed an error analysis for the Biodex workload to find that low accuracy reflects low ground-truth label quality, suggesting very limited further accuracy improvements (we publish our analysis in \href{https://ucbepic--biodex-labeling-web.modal.run/}{\color{blue!60!black}this interactive dashboard}).

\smallskip
\noindent {\em {\bf C. Clarified rewrite directive design space and correctness (Section 3).}} We added a discussion motivating the design space of rewrite directives, formalized the typing and composability constraints that directives must satisfy, and expanded the descriptions of new directive categories.

\smallskip
\noindent {\em {\bf D. Clarified contributions relative to prior work.}} To better articulate what is new relative to DocETL-V1 and other systems, we revised the introduction, the search algorithm presentation, and the related work.

\smallskip
\noindent {\em {\bf E. Reorganized evaluation and added more discussion (Section 5).}} We moved the optimization overhead and scalability analysis into the main paper, added design principles for effective semantic operator pipelines, and linked to agent prompt source files for reproducibility.

\smallskip
\noindent {\em {\bf F. Revised Conclusion and Future Work section.}} We revised the Conclusion and Future Work section in the technical report, with concrete strategies for reducing optimization cost.

\smallskip
\noindent Overall, these changes have resulted in substantial updates to Sections 3, 4, and 5, a revised Conclusion and Future Work section in the technical report~\cite{moartechreport}, and expanded appendices. To accommodate the new content while adhering to the 12-page limit, we tightened the introduction and condensed workload descriptions into a summary table, moving full descriptions to the technical repor~\cite{moartechreport}t.

The changes in the paper are color-coded by reviewer: \textcolor{ronecolor}{\textbf{blue}} for Reviewer 1, \textcolor{rtwocolor}{\textbf{green}} for Reviewer 2, \textcolor{rfourcolor}{\textbf{orange}} for Reviewer 4, and \textcolor{darkred}{\textbf{dark red}} for changes addressing multiple reviewers. We start by addressing each point raised in the reviews, referenced in the corresponding reviewer color.

\section*{Meta-Reviewer}

\com{1. Clarify the paper’s contributions in relation to the authors’ prior work. (R4D1)}

We have revised the introduction, the search algorithm presentation (Section 4), and the related work to better articulate what is new relative to DocETL-V1 and other systems. See R4D1 for details.

\com{2. Strengthen the current results and clarify the missing details. (R4D3, R1Ds)}

We have strengthened our results with new experiments (e.g., restricting rewrites to model substitution, MOAR but only with DocETL-V1's directives), new metrics (e.g., raw recall for BlackVault, error analysis for Biodex), and deeper analysis of optimization efficiency and directive usage across workloads. We have also clarified details throughout Sections 2--5, including ground truth requirements, budget definition, and input formats. See R4D3 and R1D1--D8 for details.

\com{3. Discuss the correctness guarantees. (R2O2)}

We have added a discussion of the rewrite design space in Section 3, formalized the typing and composability constraints that directives must satisfy in the appendix, and reported accuracy generalization from the optimization sample to the held-out test set in the appendix. See R2O2 for details.

\com{4. Conduct additional experiments: i) for quantitative evaluation on overhead and scalability; ii) for sensitivity analysis on the impact of backbone LLMs, including open-source ones. (R2O3, R2O4)}

We have moved the optimization overhead and scalability analysis into Table 7 in Section 5, including estimated execution costs at full dataset scale and million-document scale. For sensitivity analysis, we evaluated MOAR with four open-source LLMs as the agent on the CUAD workload; Kimi K2.5 and Llama-4-Maverick produce Pareto frontiers comparable to gpt-5. See R2O3 and R2O4 for details.

\section*{Reviewer 1}

\comone{(W1) The paper silently assumes pipelines that are linear (at least based on the definition of pipelines in Section 2)}

The reviewer is correct that our formal definition describes linear pipelines. This is intentional, done to simplify the presentation. All our techniques generalize to tree-structured pipelines (e.g., with ops such as joins or union). While DocETL supports a binary \ttt{equijoin}, we have observed that our users do not use \ttt{equijoin}, and so we opted to focus on linear pipelines in our text to simplify exposition. That being said, we have added a note in Section 2.1 to clarify why we use the phrase ``sequence'' or imply linear pipelines---we thank the reviewer for catching this ambiguity.

\ul{In our revision, Section 2.1 now includes a note clarifying that linearity is for simplicity and our techniques generalize to trees.}

\comone{(W2) There is also an evaluation budget that is assumed it's given as the pipelines need to be executed on sampled data.}

We do define the evaluation budget $B$ in Section 2.3 and discuss it further in Section 4. However, we agree it could be more prominent. \ul{In our revision, Figure 2 now visually depicts the evaluation budget as an input to the optimizer}; thank you for the suggestion.

\comone{(W3) There is also a silent assumption that the documents to be analyzed can only be in JSON format.}

Users can input data in JSON or CSV formats; internally, DocETL represents each document as a set of key-value pairs. Users can serialize unstructured data (e.g., PDFs or text files) by storing the content as the value for a given key. Additionally, CSVs or relational data can easily be transformed into JSON as well, with one key per column.

\ul{In our revision, Section 2.1 now clarifies the input formats and internal representation.}

\comone{(D1) It's not clear where the ground truth is found to measure the accuracy of executed pipelines which are evaluated on sample of data.}

Thank you for raising this. In Section 2.3 we mention that the user provides the accuracy function $a(\cdot)$. We now clarify that if this function requires ground truth labels, the user provides them for the evaluation sample $D_o$; in practice, we have found this is reasonable for users to do on a small sample (e.g., 40 documents) when the optimized pipeline will process thousands. However, in many cases ground truth is not required---e.g., precision can be computed by checking for hallucinations compared to the source text (as in our BlackVault experiment, where we simply check if the extracted output is not present in the original text), or an LLM judge can assess output quality (as in our prior work \cite{shankar2025docetl}). 

\ul{In our revision, Section 2.3 now clarifies that ground truth labels are optional; users provide the accuracy function $a(\cdot)$, which may or may not require them.}

\comone{(D2) What is the cost for optimization compared to the initial user's pipeline. I can imagine a scenario which could have been cheaper to just execute the initial pipeline without any optimization.}

Thank you for the suggestion. In our revision, we have added the comparison to the user-specified plan in the optimization cost table (\Cref{tab:overhead}), reporting MOAR's optimization cost as a multiple of the initial plan cost on the full dataset. Though at smaller scales it may indeed be cheaper to skip optimization, the more fundamental issue is that the user-specified plan is often not accurate enough to be useful---users in our deployments prioritize accuracy, and the user-specified plan is {\em always} dominated by MOAR's Pareto frontier (see Figure 3), meaning MOAR finds plans that are both cheaper and more accurate. At million-document scale, the optimization cost becomes negligible relative to execution cost---across our workloads, optimization averages less than 0.8\% of the expected execution cost at 1M documents (\Cref{tab:overhead}). Moreover, the increase in accuracy relative to the original plan can be as high as 800\% (in BlackVault).

\ul{In our revision, Table 7 now includes optimization costs with a comparison to user-specified plan costs on the full dataset.}

\comone{(D3) The introduction felt very technical when reading it for the first time. I would suggest to have some more high level description and then dive into the details.}

Thank you for this feedback; we have revised the introduction according to the suggestion. Specifically, we (1) revised the opening of ``Limitations of Existing Systems'' to first explain what query optimizers do at a high level, (2) revised the ``MOAR optimizer'' paragraph to lead with a high-level summary of what MOAR does before describing its components, and (3) moved some technical details (e.g., progressive disclosure) to later sections.

\ul{In our revision, the introduction now leads with high-level descriptions before technical details.}

\comone{(D4) How can an inexperienced user define the evaluation budget?}

This is a great question. We have found that users (e.g., the public defender data team mentioned in Example 1.1) struggle to define a monetary budget for optimization---they don't know what to expect, since the cost depends on unfamiliar factors like LLM pricing and data characteristics. However, it is easier for them to conceptualize ``how many different pipelines will the optimizer explore,'' i.e., logical rewrites or reformulations of the task. Hence we define the budget in terms of the number of pipelines to explore. We have added a note in Section 2.3 explaining this choice. In follow-up work, we are exploring better ways to set the budget, including varying optimization time/latency and interactive or human-in-the-loop optimization. Or, users could inspect optimized pipelines and the Pareto frontier to determine where to prioritize exploration, or to stop (à la online aggregation). 

\ul{In our revision, Section 2.3 now explains how and why the budget is defined as number of pipelines.}

\comone{(D5) The pipeline definition implies a linear pipeline but then an equijoin is mentioned as an operator type. Which is the case? Are joins allowed? In that case, the pipelines should not be defined as a sequence of operators.}

See response to R1W1.

\comone{(D6) In Section 4.3.1, it is mentioned that the documentation includes an explanation of when to apply the directives and what scenarios benefit from them. How are these guidelines extracted? Are they manually curated?}

Yes, the use case guidance for each directive is manually curated based on our experience with users and pipelines from a deployment of DocWrangler, our interactive IDE for creating DocETL pipelines~\cite{shankar2025steering}. Importantly, they do not overlap with the pipelines in our experiments. We have revised Section 4.3.1 to clarify this. Due to space constraints, we do not include the full directive documentation in the paper, but it is available in the \href{https://github.com/ucbepic/docetl/tree/2bf97c66/docetl/reasoning_optimizer/directives}{codebase}.

\ul{In our revision, Section 4 now clarifies that the use case guidance is manually curated based on prior deployment experience, and provides links to the directive documentation in the codebase.}

\comone{(D7) In the evaluation, there is a claim that MOAR discovers the highest-accuracy pipeline on every workload. This claim is incorrect as not all possible pipelines are evaluated. I think the authors wanted to say in comparison with the baselines.}

The reviewer is correct; thank you for catching this. \ul{In our revision, Section 5 now clarifies that the claim is relative to baselines.}

\comone{(D8) It is much better to have the optimization overheads in the main paper and not in a technical report. This is very important information for the paper. In contrast, the long description of the workloads could be in the technical report.}

Thank you for this suggestion. \ul{In our revision, we have moved the optimization overhead results into Section 5. We have also condensed the workload descriptions into a summary table in Section 5, with full descriptions in an appendix in the technical report.}

\section*{Reviewer 2}

\comtwo{(O1) The paper has not systematically investigated the design space. It is unclear how optimal MOAR is. In addition, it is unclear if the extended rewrite directives are expressive enough for general workload. A systematic analysis of the design space and the dimensions of requirements in the problem of rewriting is crucial.}

We appreciate this feedback. We agree it would be good to have a systematic analysis of the design space for rewriting in semantic operator pipelines; we have added to Section 3 and summarize our new text here. Overall, the design space is truly infinite, so no optimizer can guarantee finding the optimal pipeline. This is analogous to traditional top-down query optimizers like Cascades~\cite{Graefe1995TheCF}: there is no guarantee that Cascades' transformation rules are complete or sufficient for optimality, yet the framework is effective because it is extensible (i.e., new rules can be added as needed). MOAR follows the same extensible library approach, by contributing an extensible optimization framework for unstructured data pipelines (e.g., search algorithm, multi-objective reward tracking, and agent-based instantiation), not just the specific set of rewrite rules. That said, our directive library covers a broad range of rewrite categories---cost reduction (e.g., operator fusion, code substitution, model selection), accuracy improvement (e.g., task decomposition, data decomposition, prompt rewriting), and combinations of both (e.g., code-based document compression reduces cost while focusing LLM attention on relevant content). To push toward completeness, we include an ``arbitrary rewrite'' directive that allows the agent to propose any change without directive scaffolding; this directive was selected only once across all experiments, suggesting the existing categories may already cover the rewrites the agent finds useful.

Another important aspect of the design space is formalizing the requirements for valid rewrites. Each rewrite directive must satisfy typing and composability constraints---for example, if a directive decomposes $\ttt{map}_x$ with output schema $s_x$ into two sub-maps $\ttt{map}_y$ and $\ttt{map}_z$, we can verify that $s_y \cup s_z = s_x$. In our revision, we formalize these constraints in Appendix~\ref{sec:detailed-rewrite} of our technical report~\cite{moartechreport}.

Moreover, in our revision, we add a new analysis of which directives appear in the top-5 accuracy Pareto-optimal pipelines across workloads in Section 5. Across 28 pipelines, 11 distinct directives are selected; model substitution appears in 25 of 28 pipelines but never alone, with clarify instructions (13 pipelines), document chunking (11), and code-based document compression (9) being the most common non-model-substitution directives.

\ul{We have made the following changes in our revision: we added a discussion of the rewrite design space and the extensible library approach in Section 3, formalized typing and composability constraints for rewrite directives in Appendix B of our technical report, and added an analysis of directive usage across Pareto-optimal pipelines in Section 5.}

\comtwo{(O2) The correctness guarantees are unclear. Rewrites are validated only through sample-based accuracy checks. There is no formal or systematic guarantee that synthesized code, fused operators, or prompt rewrites preserve semantics. This is indeed difficult and remains an open problem for LLMs, but the authors should discuss how to mitigate the risks.}

The reviewer raises an important point. We note that no existing semantic operator optimizer provides formal correctness guarantees---any rewrite, even a simple one like changing the model, can change accuracy, and all systems (e.g., LOTUS, ABACUS) validate rewrites empirically via sample-based accuracy checks, as we do. Specifically for MOAR, some aspects of rewrite correctness can be formally verified and some cannot. As discussed in R2.O1, each directive must satisfy typing and composability constraints; we formalize these for all directives in Appendix~\ref{sec:detailed-rewrite} of our technical report, with a summary in \Cref{tab:rewrite_directives} in the main paper. However, we cannot formally verify that when a directive decomposes an operator with prompt $p_x$ into two operators with prompts $p_y$ and $p_z$, these new prompts faithfully capture the intent of the original; this remains an open challenge for semantic operators, and we rely on sample-based validation, like existing systems.

Moreover, there is a risk that accuracies on the optimization sample ($D_o$) do not generalize to the held-out test set ($D_T$). In the revised paper, we measured the difference in accuracies between $D_o$ and $D_T$: across all Pareto-optimal pipelines and workloads, the average accuracy difference is less than 5 percentage points. We report the full per-workload breakdown in Appendix D of our technical report. Generalization from small samples is a known challenge shared by other semantic operator optimizers,
(e.g., ABACUS) that also rely on sample-based validation, and, of course, in any approach involving supervised learning, and we see improving this as promising future work.

\ul{In our revision, we have discussed what can and cannot be formally verified about rewrites in Section 3, formalized typing and composability constraints in Appendix B, and reported accuracy generalization from $D_o$ to $D_T$ in Appendix D in our technical report.}

\comtwo{(O3) The overhead and scalability are unclear. In MOAR, each iteration involves LLM calls, pipeline execution on samples, and bandit state. The paper did not quantify wall-clock (or monetary cost of OpenAI API calls) overhead for processing pipelines. Each workload uses 40 documents for optimization and 100 for testing. It is unclear how MOAR performs on larger settings, for instance, million-document scale and complex pipelines.}

Thank you for raising this. The optimization costs, including monetary/API costs and wall-clock time, were detailed in our technical report. However, we agree these details are important, so we have moved them into the main paper (Table~\ref{tab:overhead} in Section 5). We note that the pipelines in our evaluation are indeed complex and representative of real-world scale: for example, CUAD requires extracting 40+ distinct clause types per document, and Biodex involves multi-label classification over 20,000+ possible adverse reaction labels.

We have also added estimated costs for running the highest-accuracy pipeline discovered by MOAR at full dataset scale and at million-document scale. At million-document scale, expected execution costs range from \$2,987 (Game Reviews) to \$221,868 (Biodex), demonstrating that execution cost---not optimization cost---dominates at scale. Importantly, MOAR's search cost does not increase with dataset size, because optimization is performed on a fixed-size sample ($D_o$). As LLMs become cheaper, future work can explore searching over more pipelines and using larger optimization samples.

\ul{In our revision, we have moved our optimization overhead results into Section 5, added estimated execution costs at full dataset scale and million-document scale, and clarified that MOAR's search cost does not depend on the total number of documents in a dataset.}

\comtwo{(O4) The author should also evaluate MOAR on open-source LLMs. The LLM agentic approach may be sensitive to model drift and the quality of rewrites may also depend on the power of LLM itself. How MOAR performs when intercepting with weaker LLMs. I would suggest the author add a sensitiveness analysis of different LLMs.}

\begin{figure}[t]
\vspace{-10pt}
\begin{tcolorbox}[colframe=rtwocolor, colback=white, boxrule=0.5pt, left=2pt, right=2pt, top=2pt, bottom=2pt]
\begin{center}
\includegraphics[width=0.55\columnwidth]{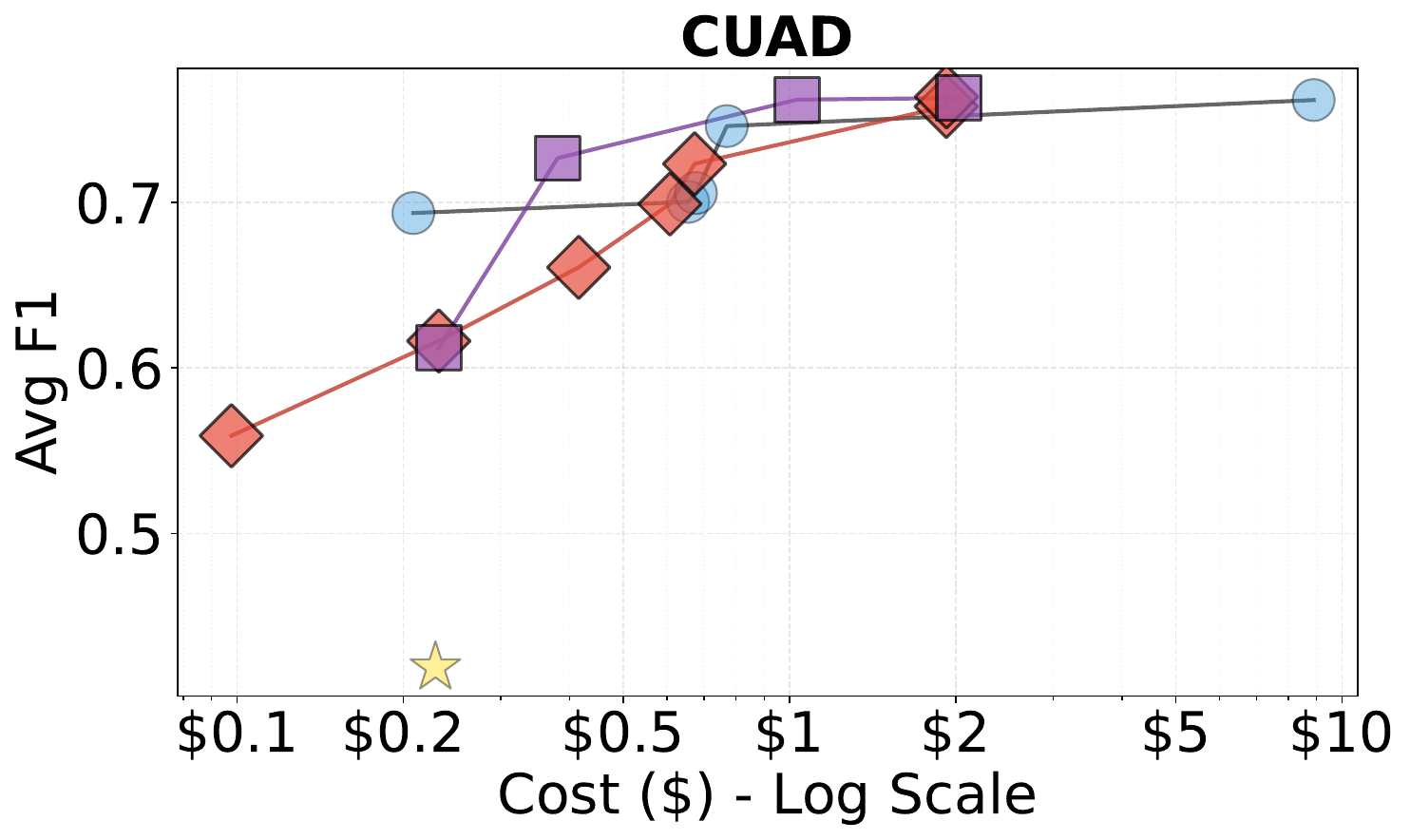}\\[1pt]
\includegraphics[width=0.7\columnwidth]{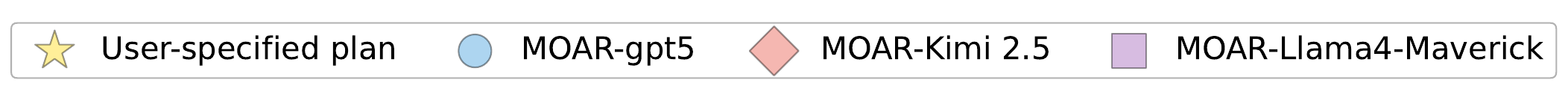}
\end{center}
\vspace{-4pt}
{\footnotesize \rtwo{\textbf{Revision Figure 5:} Pareto frontiers on CUAD when varying the LLM used to select and instantiate rewrite directives.}}

\vspace{4pt}
{\footnotesize \rtwo{\textbf{Revision Table 9:} LLMs evaluated to select and instantiate rewrite directives on CUAD.}}
\begin{center}
\scriptsize
\setlength{\tabcolsep}{3pt}
\begin{tabular}{lcccc}
\toprule
\textbf{LLM} & \textbf{Reasoning?} & \textbf{Context} & \textbf{Max Acc.} & \textbf{Cost} \\
\midrule
gpt-5 & Yes & 128k & 0.7618 & 8.8865 \\
Kimi K2.5 & Yes & 128k & 0.7637 & 1.9242 \\
Llama-4-Maverick-17B & No & 1M & 0.7630 & 2.0264 \\
Qwen3-8B & No & 32k & --- & --- \\
Llama-3.1-8B & No & 128k & --- & --- \\
\bottomrule
\end{tabular}
\end{center}

\vspace{2pt}
{\footnotesize \rtwo{\textbf{Revision Table 10:} Representative errors with Llama-3.1-8B to select and instantiate rewrite directives on CUAD.}}
\begin{center}
\scriptsize
\setlength{\tabcolsep}{3pt}
\begin{tabular}{l l p{3.5cm}}
\toprule
\textbf{Directive} & \textbf{Target} & \textbf{Error} \\
\midrule
(parse failure) & --- & Invalid control character at line 2 col 23 \\
\texttt{reduce\_gleaning} & \texttt{map} & Can only be applied to reduce ops, got map \\
\texttt{swap\_with\_code} & \texttt{map} & Can only be applied to reduce ops, got map \\
\texttt{operator\_fusion} & \texttt{map, map} & List index out of range \\
\bottomrule
\end{tabular}
\end{center}
\end{tcolorbox}
\vspace{-10pt}
\end{figure}

This is a great suggestion. In our revision, we run a new experiment evaluating MOAR with four open-source LLMs replacing the gpt-5 agent on the CUAD workload, using TogetherAI's inference platform: Kimi K2.5~\cite{team2026kimi} (a reasoning model), Llama-4-Maverick-17B-128E-Instruct-FP8~\cite{llama4} (a large mixture-of-experts model), Qwen3-8B~\cite{qwen3}, and Llama-3.1-8B-Instruct-Turbo~\cite{llama3}.

Kimi K2.5 and Llama-4-Maverick both produce Pareto frontiers comparable to gpt-5, achieving maximum accuracies of 0.7637 and 0.7630 respectively (vs.\ 0.7618 for gpt-5); we report full results in the new \Cref{fig:cuad-os,tab:agent-models} in our technical report~\cite{moartechreport} (shown here as Revision Figure~5 and Revision Table~9 for ease of reference). Two LLMs fail: Qwen3-8B's context window, 32k tokens, is too small; and Llama-3.1-8B cannot follow the agent prompt's instructions; we provide representative failure logs in \Cref{tab:failure-logs} in our technical report~\cite{moartechreport} (shown here as Revision Table~10 for ease of reference). Overall, MOAR's agent requires long context ($>$32k tokens) and strong instruction-following, but not reasoning capabilities---Llama-4-Maverick succeeds without being a reasoning model.

\ul{In our revision, in Section 5, we present a new experiment evaluating MOAR with four open-source LLMs as the agent, with more details in our technical report.}

\section*{Reviewer 4}

\comfour{(D1.1) DocETL already introduced agentic rewrite directives and an optimizer that evaluates rewritten pipelines. MOAR mainly adds a second objective (cost), expands the directive library, and replaces the optimizer with a different search heuristic. It would help readers better appreciate the contribution if the paper more clearly isolated which of these changes is primarily responsible for the reported improvements.}

Thank you for the feedback. We agree that DocETL-V1 introduced rewrite directives and a search algorithm as part of a high-level optimization framework, analogous to how the relational setting has rewrite rules and a query optimizer. But as in the relational setting, there is substantial room for novel contributions within that framework, and MOAR makes two independent ones, as discussed next.

Our paper is the first to show that query plans for semantic operator pipelines do not exhibit optimal substructure. All prior optimizers---DocETL-V1 and ABACUS, as well as traditional database optimizers---use memorization or dynamic programming, which relies on optimal substructure. MOAR's search algorithm had to be designed from scratch to account for this. In designing it, we had to adapt and extend ideas from MCTS (Monte Carlo Tree Search) for this setting; these extensions are also nontrivial, as we discuss in our response to D2.1.

In addition to the new search algorithm, this paper introduces over 17 new rewrite directives (expanding the library from 13 to over 30). These new directives emerged from over a year of DocETL deployments across journalism, law, medicine, and other domains, by analyzing how real applications failed with DocETL-V1 and distilling those failure patterns into reusable, composable directives.

Neither contribution is sufficient on its own (new optimizer, expanded directives). To demonstrate this, we ran a new experiment (summarized in Section 5 with detailed Pareto frontiers in a new \Cref{fig:directive-library} in our technical report~\cite{moartechreport}, shown here as Revision Figure~6 for ease of reference): we evaluate MOAR-V1, which uses MOAR's new search algorithm but only DocETL-V1's directives. On three workloads (BlackVault, CUAD, Sustainability), the new directives are essential---MOAR's frontier dominates MOAR-V1's, which in turn dominates DocETL-V1's, with best accuracy up to 5.3$\times$ higher on BlackVault. On two workloads (Biodex, Medec), the new search algorithm itself provides substantial improvements; e.g., MOAR-V1 already improves best accuracy by 1.3--1.4$\times$ over DocETL-V1, with the full directive library providing additional gains. Overall, both the search algorithm and the expanded directive library are necessary for strong performance across all workloads.

\begin{figure}
\vspace{-10pt}
\begin{tcolorbox}[colframe=rfourcolor, colback=white, boxrule=0.5pt, left=2pt, right=2pt, top=2pt, bottom=2pt]
    \centering
    \captionsetup[subfigure]{skip=0pt}
    \setlength{\abovecaptionskip}{2pt}
    \setlength{\belowcaptionskip}{0pt}
    \begin{subfigure}{0.39\columnwidth}
        \centering
        \includegraphics[width=\linewidth]{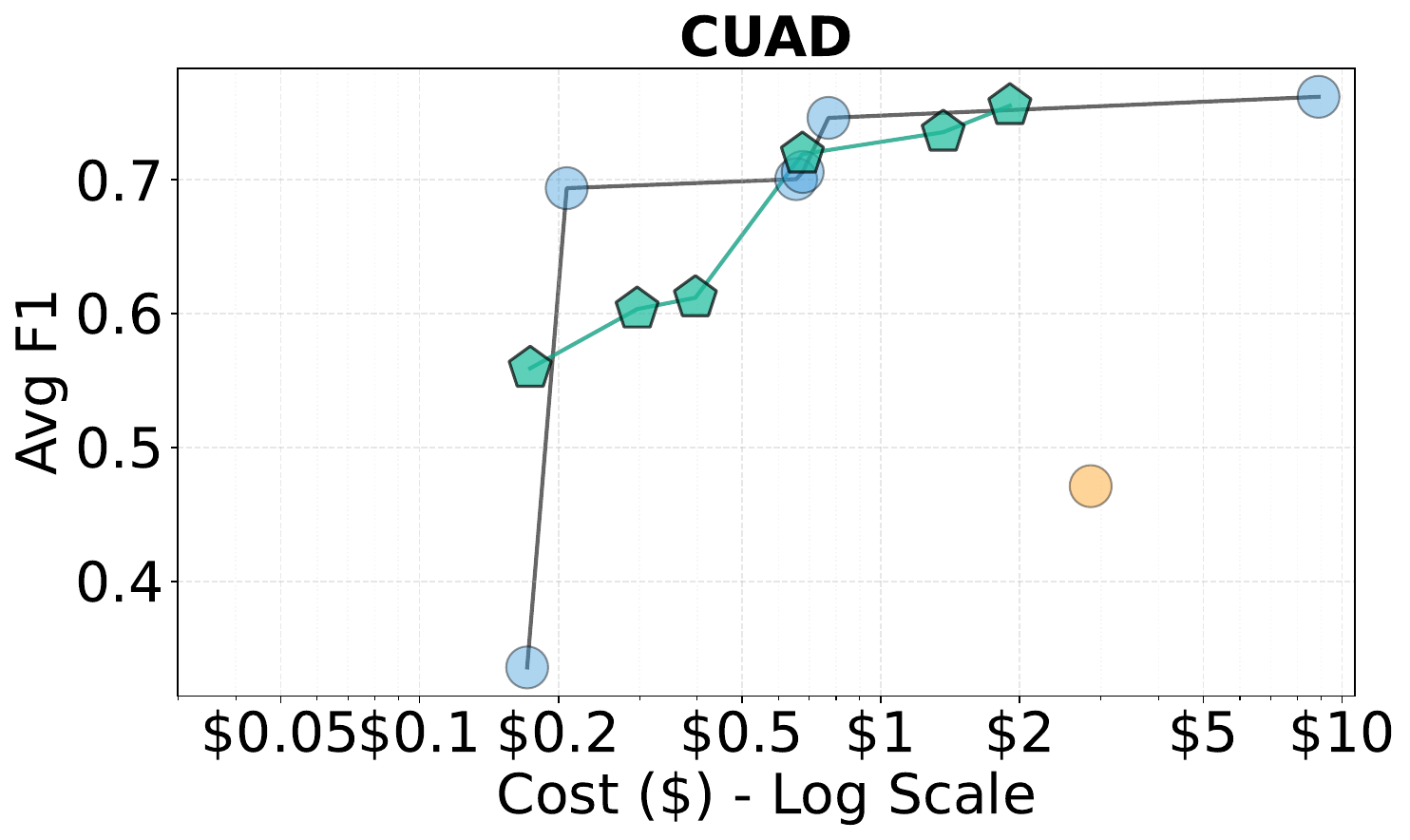}
    \end{subfigure}%
    \begin{subfigure}{0.39\columnwidth}
        \centering
        \includegraphics[width=\linewidth]{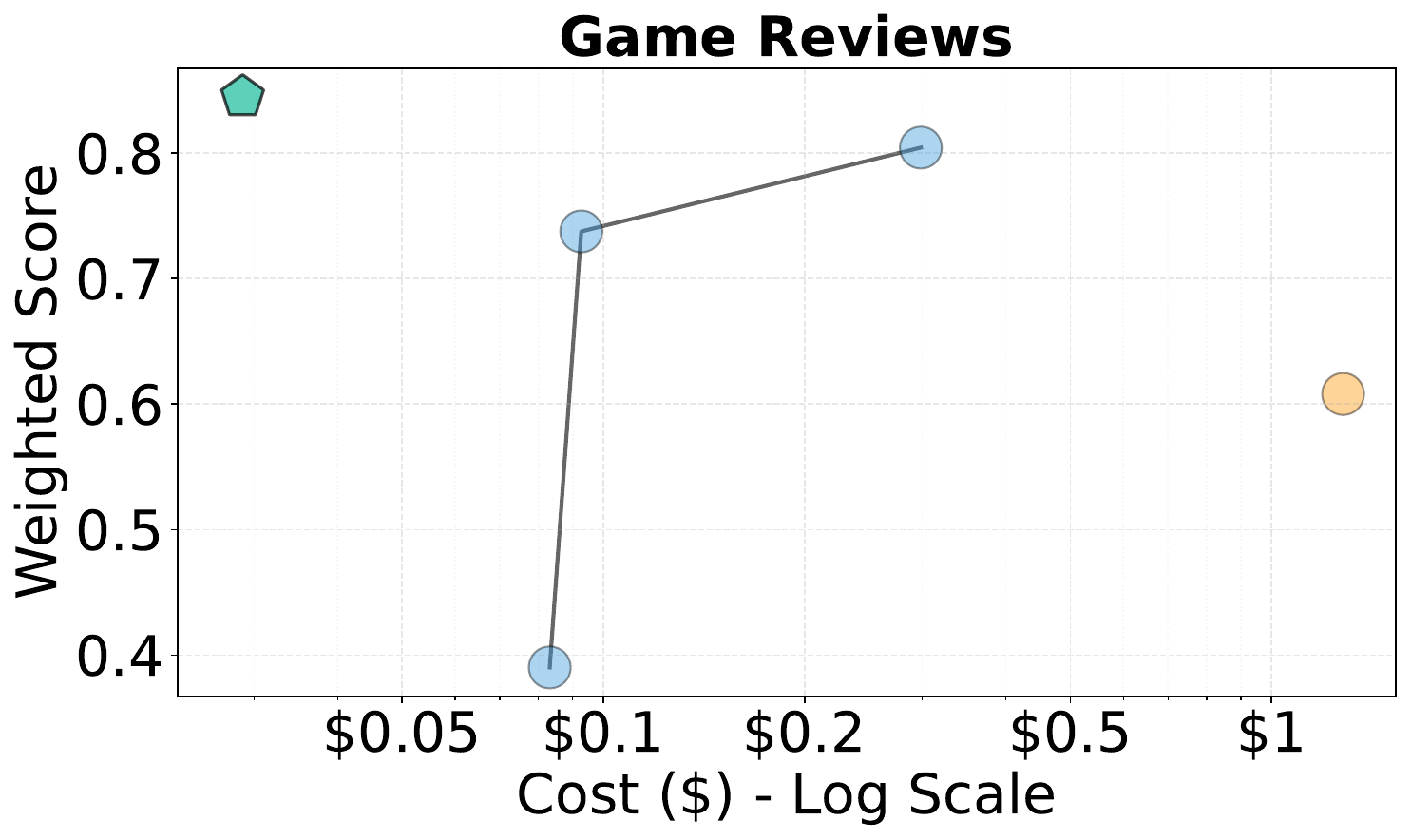}
    \end{subfigure}\\[-1mm]
    \begin{subfigure}{0.39\columnwidth}
        \centering
        \includegraphics[width=\linewidth]{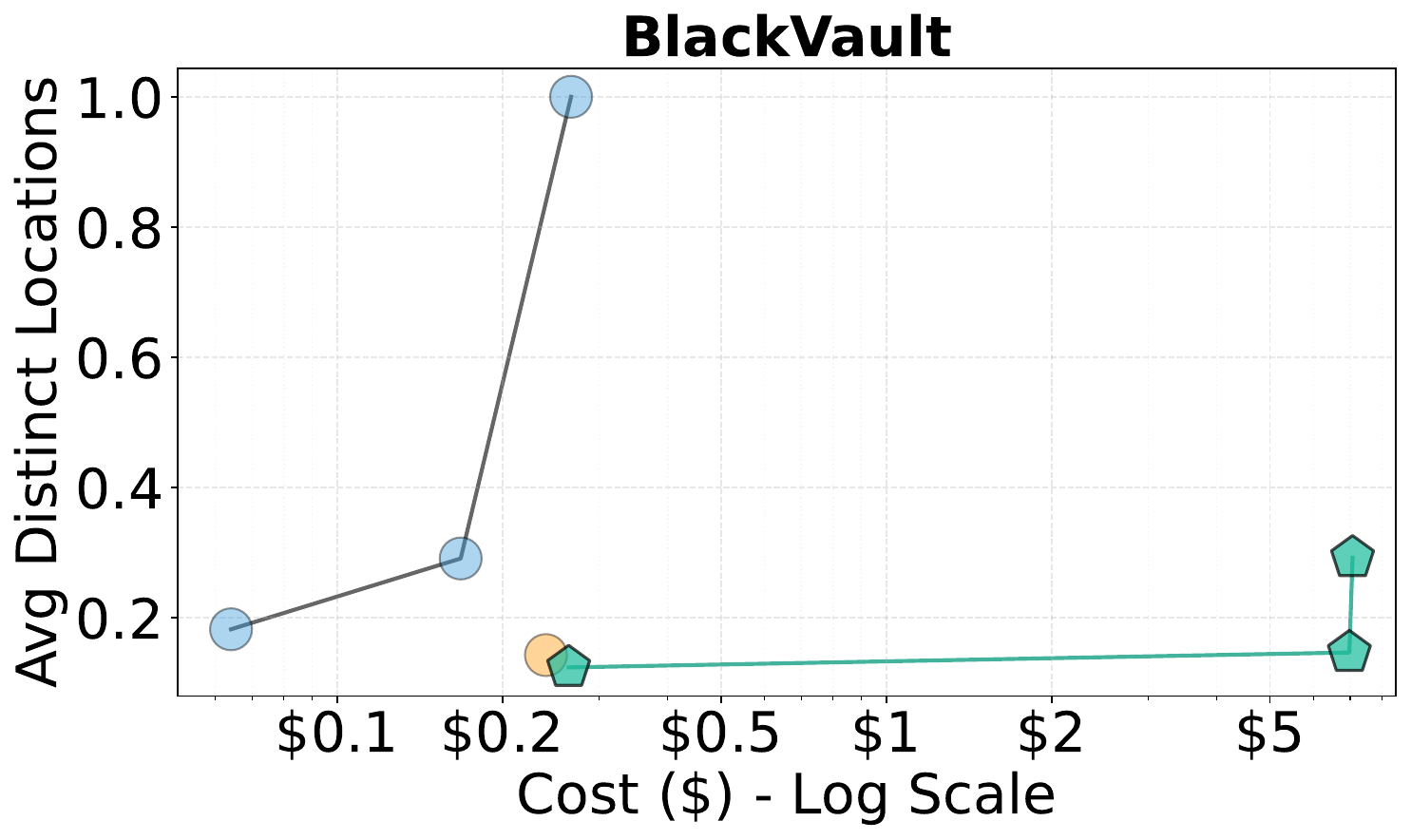}
    \end{subfigure}%
    \begin{subfigure}{0.39\columnwidth}
        \centering
        \includegraphics[width=\linewidth]{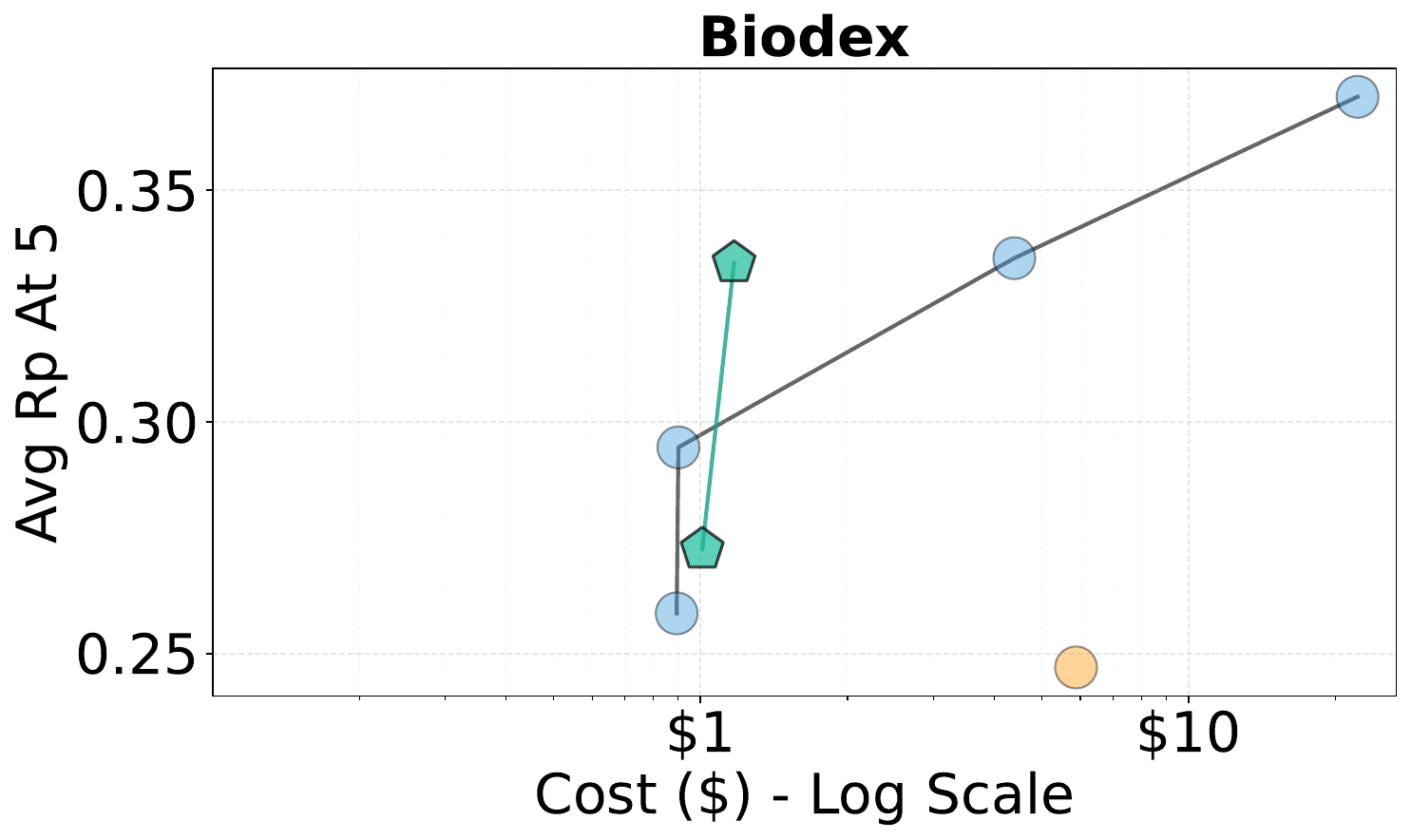}
    \end{subfigure}\\
    \begin{subfigure}{0.39\columnwidth}
        \centering
        \includegraphics[width=\linewidth]{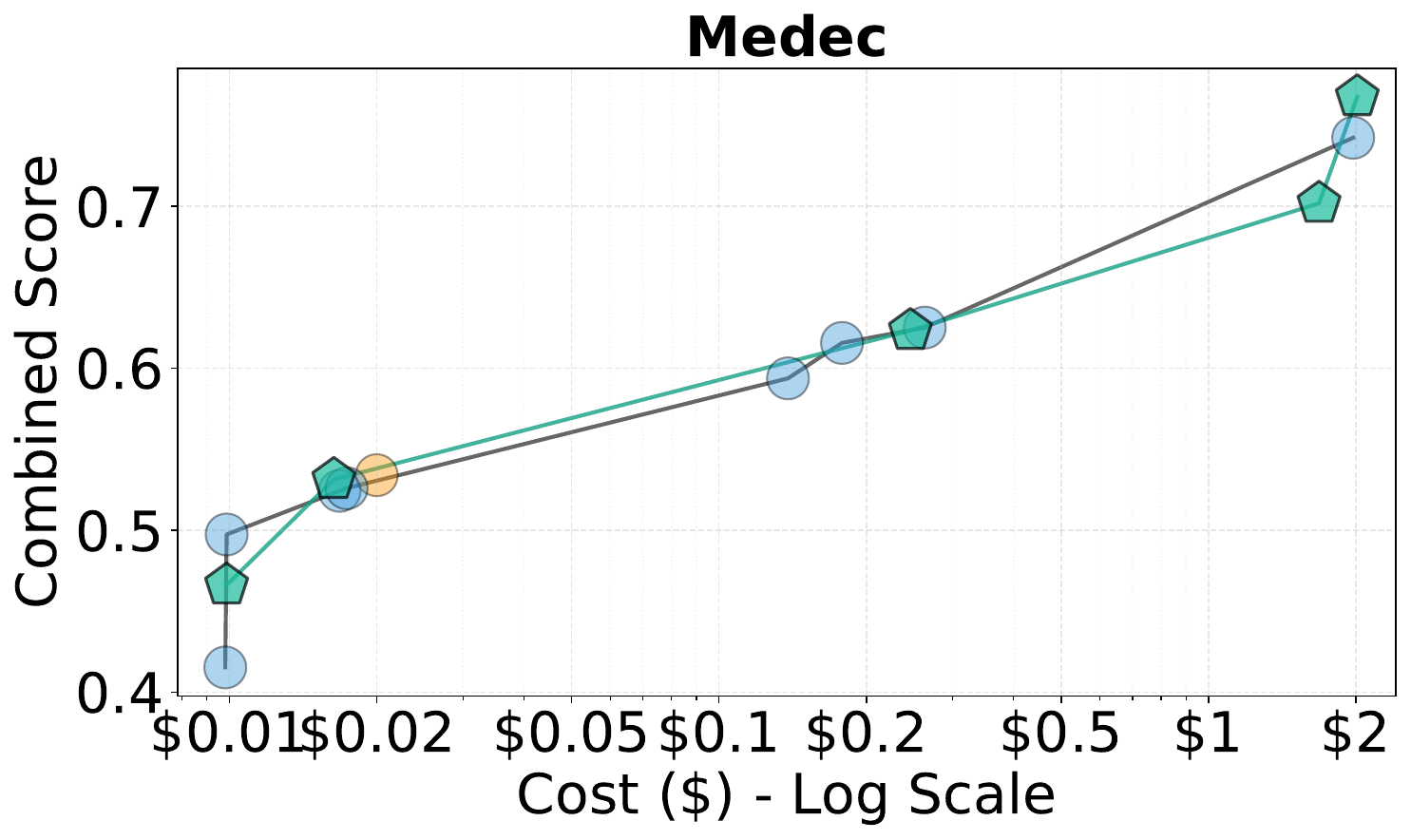}
    \end{subfigure}%
    \begin{subfigure}{0.39\columnwidth}
        \centering
        \includegraphics[width=\linewidth]{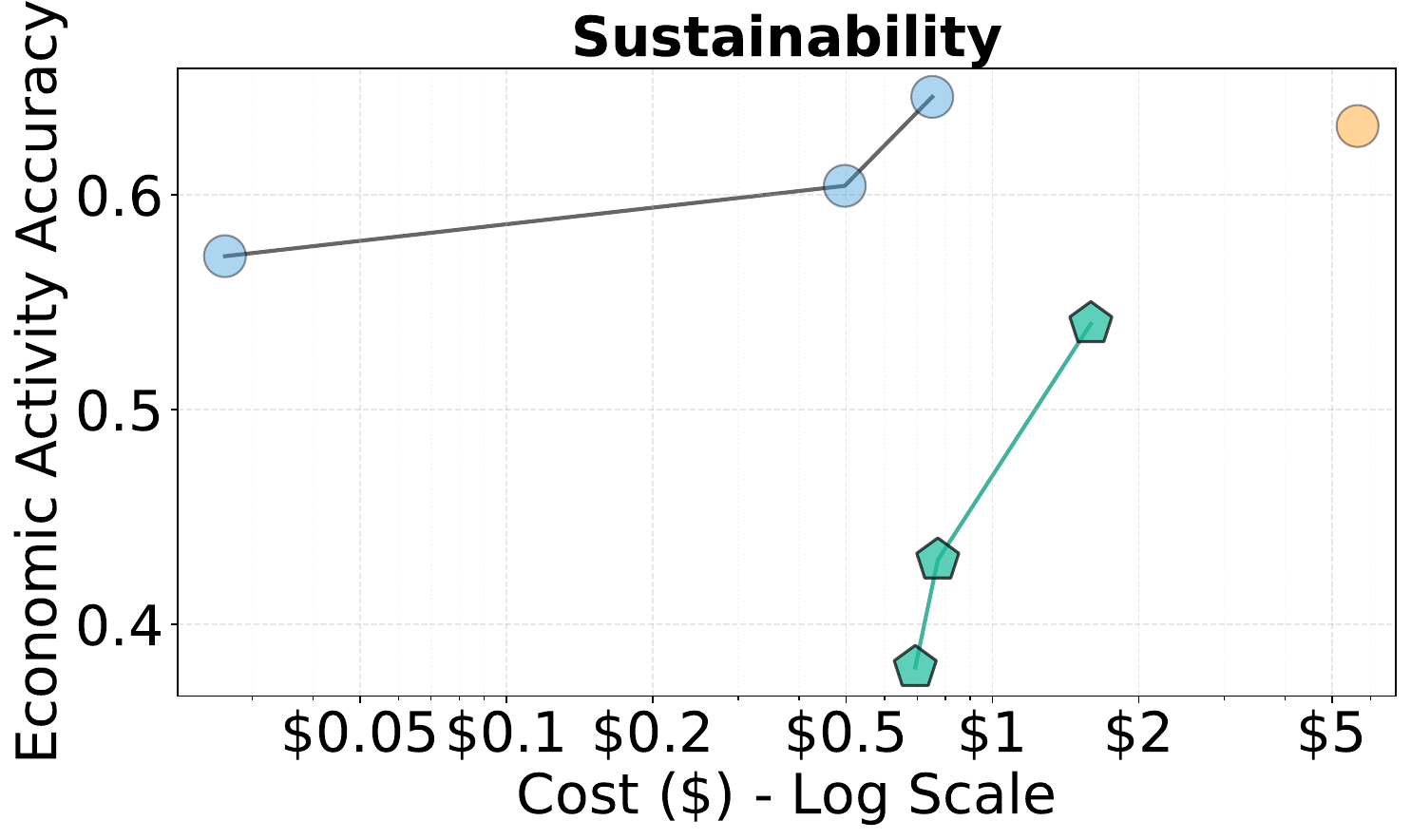}
    \end{subfigure}
    \includegraphics[width=0.5\columnwidth]{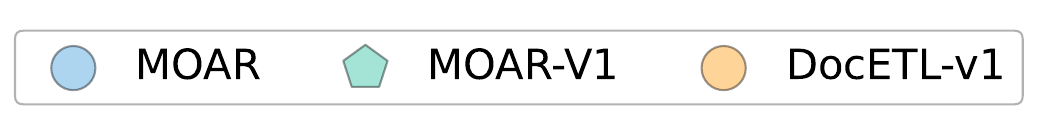}
    \vspace{-3pt}
    \par\vspace{2pt}\raggedright
    {\footnotesize \rfour{\textbf{Revision Figure 6:} Pareto frontiers for MOAR, MOAR-V1 (MOAR's search algorithm restricted to DocETL-V1's directives), and DocETL-V1.}}
\label{fig:directive-library-rev}
\end{tcolorbox}
\vspace{-10pt}
\end{figure}

\ul{In our revision, we have added a new experiment to Section 5, where we run MOAR-V1 (MOAR's search algorithm with only DocETL-V1's directives) on all six workloads. The search algorithm and the expanded directive library each independently contribute to improvements, with different workloads benefiting to different degrees from each.}

\comfour{(D1.2) More importantly, although I personally like the idea of semantic operators and find it promising, I felt that some deeper insights are still missing. In particular, beyond the substantial engineering effort, what are the key conceptual insights for designing effective semantic-operator pipelines? Section 5.3 is very helpful in understanding system-building choices and experimental results from an empirical perspective, but a more fundamental methodological discussion could further strengthen the paper.}

We agree that a deeper methodological discussion would strengthen the paper. First, we reflect on the properties of our rewrite directives. Since rewrite directives for semantic operator pipelines are analogous to rewrite rules in relational query optimizers (e.g., Calcite rules), we can ask whether they satisfy similar properties. We discuss soundness and completeness:

\topic{Soundness} In relational query optimizers, rewrite rules are, simply, sound. In our setting, soundness splits into two notions, which we will call ``type soundness'' and ``semantic soundness.'' ``Type soundness'' requires that a rewritten pipeline preserves the input-output schema of the original. Each rewrite directive has preconditions that must hold before it can apply and postconditions that are guaranteed afterward; in our revision, we formalize these for all directives in Appendix~\ref{sec:detailed-rewrite} of our technical report~\cite{moartechreport}. ``Semantic soundness'' requires that the rewritten pipeline {\em preserves the intent} of the original prompts, e.g., that splitting a map into two maps via a projection synthesis directive produces prompts that together accomplish the same task as the original. Unlike type soundness, semantic soundness cannot be formally verified. This is what makes rewrite correctness harder in the semantic operator setting than in the relational setting, and is why {\em all} semantic operator optimizers, including MOAR, ABACUS, and LOTUS, end up validating rewritten pipelines empirically, using samples.

\topic{Completeness} Like relational rewrite rules, our directive library is not complete. As discussed in our response to R2O1, however, we take steps toward high coverage: the library spans three categories (cost reduction, accuracy improvement, and combinations of both), directives are composable (applying one does not prevent applying another), and we include an ``arbitrary rewrite'' directive that allows the agent to propose {\em any} change without directive scaffolding. The ``arbitrary rewrite'' directive was selected only once across all experiments, suggesting our predefined categories may already cover the rewrites the agent finds useful, or the agent simply cannot come up with good rewrite ideas by itself.

Then, we can also reflect on what is different 
about rewrite directives compared to relational rewrites. 
One observation is that, unlike in the relational setting,
relational operators (i.e., what we call code-powered operators)
are essentially ``free'' compared to LLM-powered operators.
Offloading computation to such operators can reduce
the amount of data processed by the LLM.
In some special cases, we can entirely replace an LLM-powered operator with
a code-powered operator (directive \dnum{6} in Table~2).
More generally, we can fuse an LLM-powered map and LLM-powered filter
(in either order) into one LLM-powered map followed by a (free) code-powered filter, where the new map performs the logic in the old map {\em and} prepares new information in a way that the code-based filter can use it, e.g., by
creating a new 0/1 attribute that the code-powered filter can operate on, 
reducing the cost by half (directives \dnum{3}, \dnum{4} in Table~2). 
Or, we can prefix a code-powered operator to strip irrelevant tokens from, or reduce the size of,
each document before it reaches an LLM operator
(directives \dnum{8}, \dnum{9}, and \dnum{11} in Table~2, where \ttt{sample} is a special
code-powered operator that performs sampling).
Note, however, that in all of these cases, unlike
the relational setting, equivalence is {\em not guaranteed};
rewriting could lead to a plan that
has worse accuracy than the original.
Thus, we still need to evaluate each rewritten plan for accuracy (and cost).

We have revised the end of Section 3 to include a short summary of the above discussion (with the longer text in the tech report~\cite{moartechreport}). Finally, we have also added a ``Takeaways'' paragraph at the end of Section 5.3 that distills design principles from our empirical findings for pipeline authors. We recommend investing in specification quality: the logical plan will likely change during optimization, but a well-specified prompt helps the optimizer produce rewrites that preserve user intent (as illustrated by Medec, where a carefully crafted prompt allowed near-optimal plans to be found early); more broadly, user studies have found that specification is one of the three biggest challenges users face when authoring semantic operator pipelines, alongside understanding document contents and validating operator outputs~\cite{shankar2025steering}. We also recommend examining the Pareto frontier, since small accuracy sacrifices can yield dramatic cost savings (e.g., 91\% cost reduction for a 2\% accuracy drop on CUAD).

\ul{In our revision, we reflect on the soundness and completeness properties of rewrite directives in Section 3 (with extended discussion in the technical report), contrasting with the relational setting, and add empirical takeaways for pipeline authors in Section 5.3.}

\comfour{(D1.3) In addition, some improvements appear modest (see also D3). This raises the question of whether a non-trivial portion of the gains may come from model substitution rather than from new optimization ideas. Clarifying this would help better position the contribution.}

Yes, model substitution contributes to MOAR's gains, and this is expected: models in our pool span orders of magnitude in cost (e.g., \ttt{gpt-5-nano} vs. \ttt{gpt-5}), so selecting the right model is a key part of navigating the accuracy-cost trade-off. Moreover, searching over models is itself a hard problem that prior systems have not fully explored, because each model substitution eats into the optimization budget: MOAR uses 11 models across multiple families, vs.\ 2 for LOTUS and 1 for ABACUS in their respective experiments. That being said, we agree it would be useful to explicitly measure MOAR's gains compared to model substitution alone. So, in our revision, we construct a model-substitution-only Pareto frontier for each workload by running the user-specified pipeline with every model in our pool. Our new \Cref{sec:evaluation:sb:additional} shows that MOAR achieves higher accuracy on every workload: for instance, on Game Reviews, BlackVault, and Sustainability, MOAR's best accuracy is 1.3--8.6$\times$ higher; on CUAD, Biodex, and Medec, MOAR's best accuracy is up to 5\% higher. Regarding cost, MOAR's Pareto frontier fully dominates the model-substitution frontier on all workloads. MOAR matches the model-substitution frontier's best accuracy at 0.009--0.435$\times$ the cost on four workloads, and up to 5\% lower cost on Biodex and Medec.

\ul{In our revision, we add a model-substitution-only comparison in Section 5, showing that MOAR's gains go well beyond model selection.}

\comfour{(D1.4) The optimization overhead is also non-trivial. MOAR incurs more than twice the optimization cost of ABACUS, and these costs are discussed mainly in an appendix. Is it possible to further reduce the optimization overhead?}

We now report optimization overhead in Section 5 (previously deferred to the appendix). We also note that, unlike prior work, MOAR is not just a cost optimizer---it achieves the highest accuracy on all six workloads, with 27\% higher accuracy than ABACUS and 209\% higher than LOTUS on average. From our experience with real users, they prioritize accuracy and are often willing to incur substantial optimization costs to achieve the best plans~\cite{khattab2024dspy}.

That said, we agree that reducing optimization overhead is an important direction. In our technical report~\cite{moartechreport}, we have added a future work discussion to the conclusion with concrete strategies for reducing optimization cost. We include the paragraph here for ease of reference:

\begin{quote}
\em Looking forward, several directions could improve the efficiency of MOAR's search process. First, {\em early pruning of unpromising pipelines}: across our six workloads, only 22--44\% of explored pipelines (varying by workload) are either on the Pareto frontier or are ancestors of a frontier point in the search tree---the remaining 56--78\% never contribute to the final frontier. Running fewer samples on these pipelines or terminating their evaluation early could substantially reduce wasted optimization budget. Second, {\em estimating accuracy and cost without executing pipelines on samples}: modeling how the accuracy of individual operators composes into end-to-end accuracy could allow pruning rewrites without running the full pipeline, though modeling this composition remains an open challenge given unpredictable operator interactions. Third, {\em reducing reliance on frontier (e.g., gpt-5) LLM agents} for rewrite instantiation, e.g., by using open-source models or heuristics. Fourth, {\em user-specified accuracy lower bounds}: MOAR's search is currently biased toward high-accuracy regions of the Pareto frontier, which may under-explore cheap, moderate-accuracy plans. Allowing users to specify a minimum accuracy threshold would let the optimizer skip low-accuracy regions entirely, spending its budget on plans users actually care about.
\end{quote}

\ul{In our revision, we have moved the optimization overhead analysis into Section 5 and added a ``Future Work'' discussion to the conclusion of our revised technical report with concrete strategies for reducing optimization cost.}

\comfour{(D1.5) Finally, the analysis in Section 5.2.2 of the two cases where MOAR does not achieve cost savings for top-accuracy baseline plans is interesting. Could these cases be further analyzed and potentially leveraged to improve the current approach?}

The two cases where MOAR does not achieve cost savings are Biodex (vs.\ PZ-r\&r) and Medec (vs.\ LOTUS). For Biodex, RP@5 is low across all methods. As part of our revision, we performed a manual error analysis (see our response to R4 D3) and found that most of the apparent accuracy gap is attributable to errors in the ground truth rather than model performance; correcting the ground truth raises MOAR's RP@5 from 37.9\% to 97.5\%. Given this label quality issue, the absolute accuracy differences on Biodex are not meaningful enough to warrant further optimization effort. 

Then, for Medec, LOTUS's best accuracy is only 53.8\%, nearly 20 percentage points below MOAR's best (74.2\%); MOAR is more expensive when only matching this low accuracy, not at accuracy levels users would actually want. 
MOAR's search is biased toward high-accuracy regions of the Pareto frontier, which aligns with user priorities in our deployments but means it may under-explore cheap, moderate-accuracy plans. Incorporating user-specified accuracy lower bounds, so that the optimizer does not waste budget exploring plans below a threshold, could help focus the search. In our revised technical report~\cite{moartechreport}, we add this as a direction for future work in the conclusion.

\ul{We analyze two pipelines where MOAR does not achieve cost savings. In one case, we found labeling errors. In the other case, the baseline accuracy that MOAR is more expensive to match is far below MOAR's best. In the conclusion of our revised technical report, we propose letting users specify a minimum accuracy threshold so that the optimizer skips plans below it.}

\comfour{(D2.1) The novelty is clearly outlined in the introduction, but I found it somewhat difficult to fully connect these claims to the technical content in the main body...Many techniques used in the paper are well known. The search algorithm combines UCT- or UCB-style bandits with progressive widening, which are standard techniques, and multi-armed bandits are also used in ABACUS. The interaction with the LLM agent is motivated by progressive disclosure, a known HCI concept. While the integration of these ideas is reasonable and well engineered, it would help to more clearly articulate what is fundamentally new from an optimization perspective.}

We structured Section 4 to present each component of the search algorithm in the order it is invoked. That said, we agree it may be helpful to clarify the novelty of the search algorithm, which we describe below.

The reviewer correctly identifies that we draw on UCT and progressive widening. However, these techniques, as-is, cannot be applied to our setting; they need significant modifications. UCT-based search methods, like MCTS, traditionally assume (a) a single-objective win/loss reward and (b) independent rewards per branch of the tree (a terminal state is won or lost regardless of other branches). These assumptions don't hold in our setting, so we needed to adapt UCT. In particular, for (a), we need to optimize over a multi-objective Pareto frontier. For (b), a node's ``reward'' {\em depends on all other nodes} because the frontier is a global function of every pipeline explored so far; adding a new pipeline can change the utility of existing ones by shifting the frontier. Thus, the standard UCT reward model does not apply, so we design a custom utility metric based on marginal accuracy contribution (the vertical distance between a pipeline's accuracy and the best accuracy at comparable cost) that handles this interdependence and biases search toward high-accuracy regions of the frontier.

Regarding progressive widening: our action space is effectively infinite, and while progressive widening has been used in reinforcement learning to handle such action spaces, those systems typically run millions of iterations. Our exploration budget is multiple orders of magnitude smaller (40 iterations), so we cannot afford to try actions at random and hope to find good ones. We therefore use progressive widening to only limit the branching factor and delegate action selection to an LLM agent that reasons about pipeline semantics to prioritize which directives to try (e.g., which operator is the bottleneck, which directive is likely to help). Our system carefully manages the agent's context to include the full search history (prior rewrites and their performance across all branches of the search tree), as discussed in Section 4.3.

Overall, the resulting algorithm is not standard bandits or MCTS, or even a known extension of bandit-based approaches. It is an entirely new search algorithm. We have revised three sections to make this novelty more explicit: in Section 4, we revised the ``Algorithm Overview'' to foreground the technical insights above; in the introduction, we removed the discussion of progressive disclosure to make space; and in the related work, we clarify the distinction from ABACUS. In particular, ABACUS uses multi-armed bandits for a different problem, namely getting tighter accuracy estimates of individual semantic operator implementations in an offline profiling phase, before invoking a Cascades-based search algorithm.

\ul{In our revision, we have revised Sections 1, 4, and 6 to clarify that MOAR's search algorithm is not a standard application of UCT or progressive widening, but a new algorithm that addresses reward interdependence and action prioritization in ways these techniques do not.}

\comfour{(D2.2) In particular, directives 1 and 2 are only described at a very high level in Table 2 and are not explained elsewhere in the main paper (unless I missed them). Since rewrite directives are a central contribution, could the authors clarify whether these are intended as new rewrite directives and, if so, provide a brief explanation or example in the main text? If they are not part of the main contribution, please also kindly point it out.}

Yes, all directives in \Cref{tab:rewrite_directives} are new to MOAR, including directives 1 and 2. With over 30 directives, it is not possible to describe each one in detail in the main text; we chose to highlight the ones that are more technically interesting---e.g., harder to implement or less obvious to think of---and provide detailed descriptions of all directives in the technical report~\cite{moartechreport}. That said, we agree with the reviewer that the main text should describe the new categories more clearly. In our revision, we have clarified in Section 3 that all directives in the table are new to MOAR, and we describe each of the new directive categories (fusion and reordering, code synthesis) in more detail, including concrete examples of same-type fusion (e.g., merging two adjacent maps into a single map whose prompt performs both tasks) and map--reduce fusion (absorbing a map into a downstream reduce when the map's output does not generate the grouping keys).

\ul{In the revised Section 3, we have clarified that all directives in Table 2 are new to MOAR and expanded the descriptions of the fusion and reordering and code synthesis categories.}

\comfour{(D2.3) The paper also does not analyze the directive library systematically. While it reports that many top pipelines use modified logical plans or agent-authored code, it would be very informative to see which specific directives are most impactful and under what workload characteristics they help or hurt. This could potentially be addressed with additional experiments or a more detailed analysis.}

Thank you for this suggestion. In our revision, we analyze which specific directives appear in the top-5 accuracy Pareto-optimal pipelines per workload (28 pipelines total) and report the results in Section 5.3. We find that 11 distinct directives are selected, but a few dominate. Model substitution is the most frequent (25 pipelines) but never appears alone---every top pipeline combines it with at least one other directive. The most common non-model-substitution directives are clarify instructions (13 pipelines), document chunking (11 pipelines), code-based document compression (9 pipelines), and head/tail truncation (8 pipelines), spanning LLM-centric improvements, data decomposition, and code synthesis categories. Fusion and reordering directives appear only 4 times total. The arbitrary rewrite directive is never selected, suggesting the structured library may already cover the rewrites the agent finds useful.

\ul{In our revision, we analyze which specific directives appear in the top Pareto-optimal pipelines across workloads and report the results in Section 5.3.}

\comfour{(D3) It would help readers better appreciate the empirical results if the interpretation of Table~\ref{tab:results} were clarified.
For BlackVault, if I understand correctly, the metric is normalized by the maximum recall across methods, meaning that the top method will always appear as 1.0 by construction. Could the authors report raw (unnormalized) recall values and clarify exactly which methods are included in the normalization?...For Biodex, absolute accuracy is low across all methods, including MOAR. Is it possible to design a stronger “best-effort” logical plan (for example, retrieval or candidate generation followed by LLM reranking or verification), or other possible new solutions, and to include an error analysis to show whether there is substantial headroom and how RP@5 might be further improved?...For Medec, the gap between MOAR and the simple agent is very small, and the paper notes that the simple agent mainly benefits from switching to gpt-5. Would it be possible to add a “model-only” baseline (the original pipeline with gpt-5) and a model-fixed comparison, to more clearly demonstrate gains beyond model substitution?}

Thank you for these suggestions. In our revision, we report raw recall values for BlackVault, add a manual error analysis for Biodex, and add two new experiments for Medec. We address each workload in turn below.

\topic{BlackVault} The reviewer asked whether raw (unnormalized) recall values could be reported. We now report raw recall values in Table~\ref{tab:results} and clarify which methods are included in the normalization. Additionally, MOAR's Pareto frontier significantly exceeds the model-substitution-only frontier on this workload; see our response to D1.3 for details.

\topic{Biodex} The reviewer asked whether a stronger baseline could be designed and whether an error analysis could show whether there is room for improvement. We note that PZ-r\&r already implements the reviewer's suggestion (retrieval followed by LLM reranking), and its RP@5 of 0.296 is lower than MOAR's highest-accuracy plan (RP@5 of 0.379). DocETL does not expose retrieval as a user-facing operator, so users cannot author retrieval-augmented pipeline---retrieval is only available internally, to be considered during optimization.

Since RP@5 is low across all methods, we performed an error analysis to understand whether further improvement is possible. We sampled 50 documents and used an LLM judge (\ttt{gpt-5}) to classify each of MOAR's 324 mismatches (192 false positives, 132 false negatives) against the source biomedical papers. Of the 192 false positives, only 8 (4.2\%) are true model errors---the remaining 184 are clinically relevant reactions described in the paper but absent from the ground truth. Of the 132 false negatives, 95 are true misses (75 explicitly mentioned, 20 only implicitly), and 37 (28\%) are ground truth labels not found in the source paper at all. We manually verified a random subset of 50 judge outputs and found 100\% agreement. Correcting the ground truth raises RP@5 from 37.9\% to 97.5\%, suggesting that most of the gap is attributable to errors in the ground truth rather than model performance. We report this analysis in Section 5.2.1 of the technical report~\cite{moartechreport}; all judge outputs are available at \href{https://ucbepic--biodex-labeling-web.modal.run/}{\color{blue!60!black}this interactive web app}.

\topic{Medec} The reviewer asked whether a ``model-only'' baseline (the original pipeline run with \ttt{gpt-5}, no other changes) and a ``model-substitution-only'' comparison (the original pipeline run with each available model) could isolate gains beyond model substitution. We first clarify that the simple agent's highest-accuracy pipeline does not just substitute \ttt{gpt-5}---it also rewrites the prompt to be shorter (summarizing the few-shot examples instead of providing them verbatim). We have elaborated on this in Section 5 accordingly.

Regarding the model-only baseline: we have run the user-specified pipeline with \ttt{gpt-5} (no prompt changes or rewrite directives). This baseline achieves 8\% lower accuracy at 1.25$\times$ the cost of MOAR's highest-accuracy pipeline, and does not appear on the Pareto frontier. MOAR's highest-accuracy pipeline goes further by adding a code map to filter irrelevant text from each document before the LLM call, reducing cost while improving accuracy.

To more clearly demonstrate gains beyond model substitution, we also construct the Pareto frontier achievable by model substitution alone and compare it to MOAR's full frontier across all workloads; see our response to D1.3 for details.

\ul{Overall, in our revision, we expand on our empirical results by: (1) reporting raw recall values for BlackVault, (2) adding an error analysis of ground truth label quality for Biodex (Section 5.2.1 of the technical report}), and (3) for Medec, elaborating on the simple agent description, adding a model-only baseline, and comparing the model-substitution-only frontier against MOAR's full frontier.}

\comfour{(D4) In Section 4.3, it states that the entire rewriting decision is delegated to the LLM agent, including which directive to apply, which operators to target, and how to instantiate the rewrite. Since this is a major driver of system behavior, additional detail would help readers assess robustness and reproducibility.
Also, could the authors clarify the exact instruction format, whether multiple candidate directives or targets are sampled per step, and how frequently the agent produces invalid or low-value rewrites that consume the evaluation budget? This clarification would make the optimization process easier to reason about.}

Thank you for raising this. We address the three parts of the question below.

Regarding the instruction format and candidate sampling: at each step, the agent selects exactly one directive and one set of target operators. For most directives, this produces a single candidate pipeline. However, 6 of the 18 new directives are parameter-sensitive (marked with $\ddag$ in Table 2): these generate multiple candidate rewrites with different parameter values (e.g., different chunk sizes); MOAR evaluates all candidates on $D_o$ and keeps the highest-accuracy one in the search tree. Appendix B in our technical report~\cite{moartechreport} details the exact number of candidates for each parameter-sensitive directive and how the agent is instructed to generate them.

Regarding reproducibility: in our revision we link to the \href{https://github.com/ucbepic/docetl/blob/2bf97c66/docetl/reasoning_optimizer/agent.py}{\color{blue!60!black}exact source files} containing all agent prompts in Section 5, and we release the full search logs for all experiments at \href{https://drive.google.com/drive/folders/1Meebo7J8bhHg9b1MOP14MjQorx0OgIcF?usp=drive_link}{\color{blue!60!black}this Google Drive link}, so readers can inspect every directive selection, instantiation, and evaluation.

Regarding low-value rewrites: in our revision, we report the fraction of explored pipelines that never contribute to the Pareto frontier (i.e., are neither frontier points nor ancestors of frontier points) in Section 5.3. This fraction varies by workload: 58\% on CUAD, 79\% on Game Reviews, 76\% on BlackVault, 64\% on Biodex, 56\% on Medec, and 59\% on Sustainability. We discuss concrete strategies for reducing this wasted optimization budget in our response to D1.4 and in the conclusion of our technical report~\cite{moartechreport}.

\ul{In our revision, we clarify the rewriting and optimization process in three ways: (1) we clarify the instruction format and candidate sampling procedure in Section 4.3, (2) we provide links to full agent prompts and release full search logs in Section 5, and (3) we report the per-workload fraction of explored pipelines that never contribute to the Pareto frontier in Section 5.3.}

\comfour{(D5) Baselines: Would it be possible to design and include a stronger baseline than the simple agent, for example by equipping it with a limited set of the same rewrite directives but without the full MOAR search framework? Such a baseline could help better isolate the benefits of MOAR’s structured search.}

\begin{figure}[t]
\vspace{-10pt}
\begin{tcolorbox}[colframe=rfourcolor, colback=white, boxrule=0.5pt, left=2pt, right=2pt, top=2pt, bottom=2pt]
\raggedright
{\footnotesize \rfour{\textbf{Revision Table 5:} Best accuracy by method.}}
\vspace{1pt}
\centering
\scriptsize
\setlength{\tabcolsep}{2pt}
\begin{tabular}{lcccccc}
\toprule
\textbf{Workload} & \textbf{DocETL-V1} & \textbf{LOTUS} & \textbf{PZ} & \textbf{SA} & \rfour{\makecell{\textbf{MOAR} \\ \textbf{(no search)}}} & \textbf{MOAR}  \\
\midrule
CUAD            & 0.471 & 0.402 & \underline{0.694} & 0.521 & \rfour{0.495} & \textbf{0.762} \\
Game Reviews    & 0.608 & 0.522 & \underline{0.683} & 0.467 & \rfour{0.703} & \textbf{0.804} \\
BlackVault      & 5.339 & 3.020 & -- & 7.230 & \rfour{4.055} & \textbf{37.333} \\
(normalized)    & (0.143) & (0.081) & -- & \underline{(0.194)} & \rfour{(0.109)} & \textbf{(1.000)} \\
Biodex & 0.247 &
\makecell[l]{0.260 {\scriptsize(d)}\\ 0.202 {\scriptsize(r\&r)}} &
\makecell[l]{0.260 {\scriptsize(d)}\\ 0.296 {\scriptsize(r\&r)}} &
\underline{0.333} & \rfour{0.259} & \textbf{0.370} \\
Medec           & 0.534 & 0.538 & 0.536 & 0.726 & \rfour{\textbf{0.755}} & \rfour{\underline{0.742}} \\
Sustainability  & \underline{0.632} & 0.516 & -- & 0.543 & \rfour{0.560} & \textbf{0.646} \\
\midrule
\textbf{Avg Gain (\%)}
                & +130.71\% & +209.53\% & +26.65\% & +94.36\% & \rfour{+157.04\%} & -- \\
\bottomrule
\end{tabular}

\vspace{4pt}
\raggedright
{\footnotesize \rfour{\textbf{Revision Table 6:} MOAR cost to match baseline accuracy.}}
\vspace{1pt}
\centering
\scriptsize
\setlength{\tabcolsep}{2pt}
\begin{tabular}{lccccc}
\toprule
\textbf{Workload} & \textbf{DocETL-V1} & \textbf{LOTUS} & \textbf{PZ} & \textbf{SA} & \rfour{\makecell{\textbf{MOAR} \\ \textbf{(no search)}}}\\
\midrule
CUAD & \cellcolor{green!50}0.073$\times$ & — & \cellcolor{green!40}0.290$\times$ & \cellcolor{green!35}0.377$\times$ & \cellcolor{green!55} \rfour{0.029x} \\
Game Reviews & \cellcolor{green!50}0.072$\times$ & \cellcolor{green!50}0.071$\times$ & \cellcolor{green!65}0.003$\times$ & — & \cellcolor{green!55} \rfour{0.014x} \\
BlackVault & \cellcolor{green!40}0.267$\times$ & — & n/a & \cellcolor{green!40}0.497$\times$ & \rfour{—} \\
Biodex & \cellcolor{green!45}0.152$\times$ & \cellcolor{green!45}0.145$\times$ & \cellcolor{red!20}1.840$\times$ & \cellcolor{green!40}0.196$\times$ & \cellcolor{green!35}\rfour{0.451x} \\
Medec & \cellcolor{green!20}0.840$\times$ & \cellcolor{red!20}1.245$\times$ & \cellcolor{green!50}0.046$\times$ & \cellcolor{green!15}0.966$\times$ & \cellcolor{green!10}\rfour{0.999x} \\
Sustainability & \cellcolor{green!40}0.133$\times$ & — & n/a & \cellcolor{green!40}0.143$\times$ & \cellcolor{green!50} \rfour{0.078x} \\
\bottomrule
\end{tabular}
\end{tcolorbox}
\vspace{-10pt}
\end{figure}

Thank you for this suggestion. Our simple agent (SA) baseline already has access to DocETL-V1's rewrite directives, which is a limited set as the reviewer suggests. In our revision, we updated the SA description in Section 5.1.1 to clarify this.

However, inspired by the reviewer's comment, we introduce a stronger ablation: MOAR (no search), which gives SA access to all 18 new MOAR rewrite directives in addition to DocETL-V1's. The agent can apply any directive but must select and instantiate rewrites on its own, without MOAR's search algorithm. This directly measures how much of MOAR's improvement comes from the search algorithm vs.\ simply having access to a richer directive library. We add MOAR (no search) to our main results figure (\Cref{fig:frontier}) and to \Cref{tab:results,tab:cost-savings} (included here as Revision Tables 5 and 6 for ease of reviewing), and discuss the results in \Cref{sec:evaluation:sb:acc,sec:evaluation:sb:cost}.

On average, MOAR's highest accuracy is 2.571$\times$ that of MOAR (no search), and MOAR achieves MOAR (no search)'s best accuracy at 0.314$\times$ the cost. On five of six workloads, MOAR's Pareto frontier completely dominates MOAR (no search). On Medec, MOAR (no search) achieves the highest accuracy (0.755 vs.\ MOAR's 0.742, a difference that may be noise) by applying {\em five rewrites}: four of our predefined directives (model substitution, code-based document compression, gleaning, and prompt rewriting) and one the agent wrote from scratch (a code map to programmatically clean the output). Overall, access to a rich directive library alone is not sufficient: MOAR's search algorithm is necessary to fully exploit the directives on essentially all workloads.

\ul{In our revision, we add MOAR (no search) as an ablation in Section 5, discuss the results in Sections 5.2.1 and 5.2.2, and have updated Figure 3 and Tables 5 and 6 to include the new results.}

\comfour{(D6.1) Minor issues: The progressive disclosure technique from HCI [9] sounds interesting, but I only noticed it mentioned in the introduction. I did not find a corresponding explanation in the solution sections. Could the authors kindly point to where this is discussed in more detail, or briefly explain how the agent's interaction with directives is structured across different stages, if this is a contribution?}

\ul{Progressive disclosure is described in Section 4.3.2}: the agent first sees only directive names, descriptions, and use case guidance to choose which directive to apply, and only then loads the full instantiation schema and example application to generate concrete parameters. We have removed it from the introduction's contribution list to make space for other revisions.

\comfour{(D6.2) Minor issues: I believe the paper would be in an even stronger shape if a few additional rigorous checks could be performed. For example, in Section 5.2.2, the last sentence ends with two periods. Thanks.}

Fixed; thank you for catching this. \ul{We have corrected the double period in Section 5.2.2.}

\comfour{(W1) The contribution appears incremental relative to DocETL, and the evidence for significant gains beyond prior work could be strengthened.}

See our responses to R4.D1.1 and R4.D1.3, where we detail the novel aspects of our search algorithm and how our paper is the first query optimization paper that does not assume optimal substructure. We present a new experiment showing that both the search algorithm and the expanded directive library independently contribute to improvements, and clarify that gains go well beyond model substitution.

\comfour{(W2) Many techniques are well known, and the paper is not fully self-contained, with important details deferred to a technical report.}

Regarding novelty, see our response to R4.D2.1. Regarding self-containment, we have made several changes to make the main paper more complete: we moved the optimization overhead analysis into Section 5, replaced the detailed workload descriptions with a summary table (moving full descriptions to the technical report~\cite{moartechreport}), and expanded the description of the fusion, reordering, and code synthesis categories in Section 3 (see R4.D2.2).

\ul{We have moved optimization overhead into Section 5, condensed workload descriptions into a summary table, and expanded rewrite directive descriptions in Section 3.}

\comfour{(W3) Several critical optimization decisions are delegated to an LLM agent, but additional methodological detail would help readers better assess robustness and reproducibility.}

See our responses to R4.D4 and R2.O4.

\setcounter{page}{0}
\setcounter{figure}{0}
\setcounter{table}{0}
\fi

\papertext{\renewcommand{\baselinestretch}{0.99}}

\title{Multi-Objective Agentic Rewrites for\\Unstructured Data Processing\vspace{-1em}}
\author{Lindsey Linxi Wei$^{1\dagger}$, Shreya Shankar$^{2\dagger}$, Sepanta Zeighami$^2$,\\Yeounoh Chung$^3$, Fatma Ozcan$^3$, Aditya G. Parameswaran$^2$}
\affiliation{%
$^1$University of Washington, $^2$UC Berkeley, $^3$Systems Research @ Google\\
\url{linxiwei @ cs.washington.edu}, \{\url{shreyashankar, zeighami, adityagp}\} \url{@ berkeley.edu}, \{\url{yeounoh, fozcan}\} \url{@ google.com}}
\thanks{$^\dagger$Co-first authors. Lindsey was a visiting research intern at UC Berkeley.}

\begin{abstract}
One year ago, we open-sourced \href{https://github.com/ucbepic/docetl}{\color{blue!60!black}DocETL}, a declarative system for LLM-powered data processing that, as of \revision{March 2026}, has \revision{3.7K} GitHub stars and users across domains (e.g., journalism, law, medicine, policy, finance, and urban planning).
In DocETL, users build pipelines by composing operators described in natural language, also known as semantic operators, with an LLM executing each operator's logic.
However, due to complexity in the operator or the data it operates on, LLMs often give inaccurate results.
To this end, DocETL introduced {\em rewrite directives}, or abstract rules that guide LLM agents in rewriting pipelines by decomposing operators or data.
For example, decomposing a single \op{filter("is this email sent from an executive and discussing fraud?")} into the conjunction of two separate semantic filters may improve accuracy.
However, DocETL only optimizes for accuracy, not cost. 
{\em How do we optimize for both?} 

We present \href{https://ucbepic.github.io/docetl/optimization/moar/}{\color{blue!60!black}MOAR} (\textbf{M}ulti-\textbf{O}bjective \textbf{A}gentic \textbf{R}ewrites), a new optimizer for DocETL.
To target cost optimization, we introduce two new categories of directives and extend all three existing categories with new ones, bringing the total to over 30 directives---more than doubling what DocETL originally had.
Moreover, since operators can interact with each other unpredictably due to LLM behavior, optimizing operators or sub-pipelines individually can yield suboptimal overall plans.
Recognizing this, we design a new global search algorithm that explores rewrites in the context of entire pipelines.
Since the space of rewrites is infinite---pipelines can be rewritten in many ways, and each rewritten pipeline can itself be rewritten---our algorithm adapts a multi-armed bandit framework to prioritize which pipelines to rewrite.
Across six workloads, MOAR achieves {\bf 27\% higher accuracy} than ABACUS, the next-best optimizer, while matching its best accuracy at {\bf 55\% of its cost}.

\end{abstract}

\maketitle

\pagestyle{\vldbpagestyle}

\section{Introduction}
\label{sec:intro}

LLMs are now integrated into systems that support queries
over unstructured data, from both industry and academia~\cite{databricks-llm, snowflake-llm, HeSethi2025BigQueryAIFunctions, shankar2025docetl, patel2025semantic, liu2024declarative, jo2024thalamusdb}.  
In these LLM-powered data processing systems, a query is expressed 
as a \rone{combination} of {\em semantic operators}.
Semantic operators are data processing operators such as \ttt{map}, \ttt{reduce}, and \ttt{filter}, each described in natural language for an LLM to carry out.
Users define an initial pipeline, and the system's optimizer then determines how to execute it.
DocETL~\cite{shankar2025docetl} is one such system.
Consider the following example of a DocETL workload  
from the public defender's office  
in a major California city:  

\begin{example}[Enhancement factors in public defender workloads]
\label{ex:enhancement}
\small
Public defenders that we work with represent defendants whose sentences were increased due to {\em enhancement factors}---circumstances like threatening with a firearm, causing severe injury, or kidnapping.
To investigate whether enhancement factors are applied equitably across racial groups, defenders want to extract evidence of factors from tens of thousands of pages of police reports and trial transcripts, then compare which factors were actually charged by the court.
The pipeline is typically a single operation: {\footnotesize\textcolor{opcolor}{\texttt{map("given a description of these [eight] types of enhancement factors...list each factor present, along with supporting evidence")}}}.
\end{example}

Query optimization in settings like the one above is challenging. 
Unlike traditional data processing, 
where the optimizer only minimizes cost, 
the query plan must also be {\em accurate}. 
If the accuracy is too low (e.g., below 95\% precision as per guidelines from public defenders), the plan's output is not useful. 
The optimizer should surface high-accuracy plans that span a range of costs,  
so users can select the one best aligned with their budget.

\begin{table}[t]
\centering
\footnotesize
\begin{tabular}{lccc}
\toprule
\textbf{System} & \textbf{Multi-Objective} & \textbf{Rewrite Coverage} & \textbf{Global Search} \\
\midrule
LOTUS~\cite{patel2025semantic}   & \xmark & \xmark & No search \\
ABACUS~\cite{russo2025abacus}    & \cmark & \xmark & \xmark \\
DocETL-V1~\cite{shankar2025docetl}  & \xmark & \cmark & \xmark \\
\textbf{DocETL--MOAR}             & \cmark & \cmark\cmark & \cmark \\
\bottomrule
\end{tabular}
\caption{Comparison of semantic operator system query optimizers. 
MOAR (ours) is multi-objective, 
covers a broad space of rewrites, and searches without assuming optimal substructure.}
\label{tab:optimizer-comparison}
\vspace{-25pt}
\end{table}

\begin{figure*}[t]
  \centering
  \vspace{-20pt}
  \includegraphics[width=0.85\textwidth]{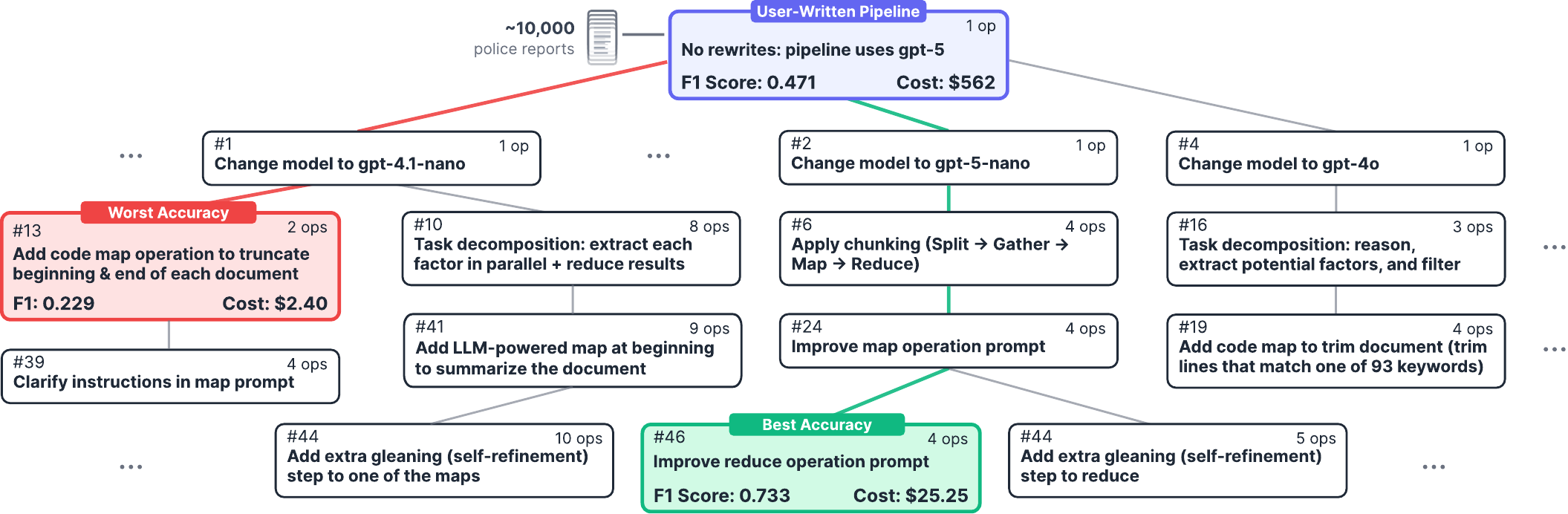}
  \caption{Sample of pipelines explored by MOAR when optimizing the pipeline in \Cref{ex:enhancement}.
  The user-written pipeline (top, purple) contains a single \ttt{map} operator that extracts all enhancement factors.  
  Each node is a pipeline variant produced by a rewrite (e.g., model change, code synthesis, task decomposition, data decomposition).  
  The ``$\hdots$'' symbols denote other explored pipelines.  
  The best pipeline (\#46, green) achieves the highest F1 score while also costing less than the user's original plan.
  }
  \label{fig:moar-search-tree}
  \vspace{-10pt}
\end{figure*}

\topic{Limitations of Existing Systems}
\rone{Query optimizers in data systems search over rewrites (i.e., transformations that restructure queries) to find efficient plans.
Search typically relies on the principle of {\em optimal substructure}:} an optimal plan can be constructed from optimal solutions to its subplans.\footnote{Even when physical properties like sort order make subplan costs context-dependent, we can restore optimal substructure, by treating each (expression, property) pair as a distinct subproblem.}
For example, the Cascades framework~\cite{Graefe1995TheCF} organizes subexpressions into equivalence groups, and the optimal plan for a subexpression is reused wherever it appears.
However, in LLM-powered data processing, even when optimizing only for accuracy, the same subplan can produce outputs of varying benefit, {\em depending on the subplans that precede or follow it.}
For instance, in \Cref{ex:enhancement}, suppose we decompose the pipeline into a \ttt{map} per factor type, followed by a \ttt{reduce} that unifies the evidence.
A \ttt{map} that extracts smaller or fewer text spans may achieve higher accuracy in isolation, but a \ttt{map} that includes more surrounding context may yield better overall accuracy if the downstream \ttt{reduce} can leverage that context to deduplicate extractions or filter out false positives.
Which \ttt{map} implementation is optimal depends on which \ttt{reduce} implementation that follows---so we cannot optimize them independently.
{\bf \em More generally, since individual subplan choices interact in unpredictable ways, composing locally optimal subplans can yield globally suboptimal plans.}

Existing optimizers for LLM-powered data processing have limitations as summarized in \Cref{tab:optimizer-comparison}.
LOTUS~\cite{patel2025semantic} only minimizes cost, requiring users to author an accurate plan first, which is difficult for users.
LOTUS ignores the pipeline search problem entirely by considering only one optimized implementation per operator.
\rone{ABACUS~\cite{russo2025abacus} performs cost-based search using rules from traditional data processing (e.g., filter pushdown) and new implementation rules (i.e., logical-to-physical) for semantic operators (e.g., model selection, prompting strategies, ensembling).}
ABACUS intelligently samples operator implementations to estimate their cost and accuracy, then uses the Cascades search framework~\cite{Graefe1995TheCF}---which relies on optimal substructure---to find Pareto-optimal query plans.
Our original DocETL optimizer~\cite{shankar2025docetl} (that we call DocETL-V1) relaxes the optimal substructure assumption slightly, but still optimizes operators from upstream to downstream, 
\rone{thus missing rewrites where the best upstream choice depends on downstream operators.}
Moreover, DocETL-V1 optimizes only for accuracy, not cost.

\topic{The MOAR optimizer} \rone{In this paper, we present MOAR, an optimizer that automatically discovers pipeline variants that improve accuracy over the user's original while offering a range of costs. MOAR explores rewrites and returns a Pareto frontier of options spanning different accuracy-cost trade-offs, all under a limited evaluation budget.}
MOAR consists of two key components: {\em (1)} a library of rewrite directives that define transformations over pipelines, and {\em (2)} a novel search algorithm that does not assume optimal substructure.
For {\em (1)}, following DocETL, we use {\em rewrite directives}---abstract rules that are instantiated by LLM agents into concrete pipeline rewrites based on task semantics and sample data.
We substantially extend DocETL's directive library to target both accuracy and cost.
For {\em (2)}, we model the space of pipelines (i.e., query plans) as a graph where each node is a complete pipeline, and each edge applies a rewrite to produce a new one. MOAR iteratively {\em (i)} selects a promising pipeline,  
{\em (ii)} applies a rewrite to produce a variant, and  
{\em (iii)} evaluates it on samples to estimate accuracy and cost.

\topic{New Rewrite Directives} The first challenge in designing MOAR was creating rewrite directives that reduce cost while maintaining or improving accuracy.
We introduce two new categories of directives: {\em code synthesis}, which replaces LLM-powered operators with synthesized Python code, and {\em operator fusion and reordering}, which combines multiple operators into fewer ones or reorders them---with 4 and 5 directives in each.
For operator fusion, same-type operators can obviously be fused (e.g., two \ttt{map\text{s}} combined by unioning output schemas and merging prompts). But different-type operators can be fused too: a \ttt{map} followed by a \ttt{filter} can become a single \ttt{map} whose prompt incorporates the filter logic and generates an additional Boolean attribute, followed by a non-LLM based predicate that drops documents based on that attribute.
\rfour{We also extend DocETL's three existing categories with new cost-reducing directives.
For example, we extend projection synthesis (a category of directives that insert \ttt{map} operations prior to more complex tasks) with directives that insert \ttt{map\text{s}} to {\em compress} documents, reducing the amount of text that downstream semantic operators handle.
MOAR adds 18 new directives to DocETL, bringing the total to over 30---capturing rewrites that are informed by our real deployments but have not been systematically explored in prior work.}

\techreport{Then, during the rewrite process, we observed that when the LLM agent receives the full specifications of all 30+ directives simultaneously in its prompt, it struggles to select appropriate rewrites.
Inspired by {\em progressive disclosure} from HCI~\cite{CarrollRosson1987ProgressiveDisclosure}---a technique that reduces cognitive load in user interfaces by revealing information gradually---we structure the agent's interaction with directives in stages.
Agents initially see only directive names and high-level descriptions.
When an agent selects a directive, it loads the full specification (detailed descriptions, instantiation schemas, examples) on-demand.
Additionally, the agent can invoke a \ttt{read\_document} tool at any point during instantiation to inspect sample data, grounding its decisions in actual document characteristics.}

\topic{New Search Algorithm} Another challenge is designing an {\em efficient} global search algorithm.
Since evaluating a pipeline requires executing it on sample data, we can only explore a limited number of pipelines.
Rewrites can be applied in sequence, creating a vast search space, as shown in \Cref{fig:moar-search-tree}.
For instance, the highest-accuracy pipeline (\#46, green) is discovered through four rewrites: first switching to a cheaper model (\#2), then applying data decomposition to process document chunks (\#6), followed by prompt improvements to the map (\#24) and reduce (\#46) operations.
While LLM agents can determine which directives to apply and how to instantiate them, it is unclear which pipelines will lead to high-accuracy, low-cost descendants.
{\bf \em Our insight is to learn which pipelines are promising to rewrite, by adapting multi-armed bandits.}
We use a variant of the UCB (Upper Confidence Bound) algorithm, adapted for the tree setting~\cite{auer2002finite}, to select which pipeline to rewrite.
However, we define a custom metric rather than using the typical hypervolume metric from multi-objective optimization~\cite{zitzler1999evolutionary}, which treats all Pareto frontier points as equally valuable.
In LLM-powered data processing, low-accuracy pipelines vastly outnumber high-accuracy ones, so optimizing for hypervolume would waste the evaluation budget exploring low-accuracy regions of the search space.
We instead introduce a metric based on {\em marginal accuracy contribution}---the vertical distance between a pipeline's accuracy and the best accuracy at comparable cost (shown as the red line in \Cref{fig:systemarch})---and score each pipeline by aggregating this metric across the pipeline and its descendants.

Then, with over 30 directives, each applicable to multiple operators, a single pipeline could spawn hundreds of children, exhausting the evaluation budget on variants of one pipeline rather than exploring deeper rewrite sequences.
We employ {\em progressive widening}~\cite{chaslot2008progressive}, a technique that limits how many edges (i.e., immediate rewrites) a node (i.e., pipeline) can have based on its visit count (i.e., total number of descendants of any depth).
As a node accumulates more visits, it is allowed to have more edges, but its edge growth is sublinear, forcing the search to explore other regions of the graph before returning to generate additional variants from any single pipeline.

Overall, the contributions of this paper include:
\begin{itemize}[nosep, leftmargin=*, wide=0pt]
\item {\bf An expanded and extensible library of rewrite directives (\Cref{sec:rewrite}).}
We design and implement an extensible library of 
18 new directives, 
greatly improving the expressive power of DocETL's original 13 directives.
These include directives targeted at cost reduction 
(e.g., model substitution, context truncation, operator fusion), 
as well as directives that replace LLM calls with synthesized code implementations.
\item {\bf A search algorithm for multi-objective optimization (\Cref{sec:search}).}
We introduce a new global search algorithm for 
discovering sequences of rewrites 
that improve both accuracy and cost 
while operating under a limited number of pipeline evaluations.
\item {\bf Empirical evaluation across six workloads (\Cref{sec:evaluation}).}  
We evaluate MOAR on six real-world workloads spanning 
legal, medical, and enterprise domains.
Compared to state-of-the-art systems and naive agentic baselines, 
{\bf \em MOAR achieves up to 11$\times$ higher accuracy
and up to 99\% lower inference cost}
at equivalent accuracy levels, 
while dominating the Pareto frontier of cost and accuracy {\bf \em on all workloads}.
On average, compared to the next best optimizer (ABACUS~\cite{russo2025abacus}), MOAR achieves 27\% higher accuracy, while matching its best accuracy at only 55\% of its cost.
\end{itemize}

MOAR is open-source: documentation is available at \href{https://ucbepic.github.io/docetl/optimization/moar/}{\color{blue!60!black}this link}.

\section{Background and Definitions}
\label{sec:background}

We build our optimizer on top of the DocETL~\cite{shankar2025docetl} system (i.e., DSL, parser, and execution engine), though our techniques can extend to other systems. 
We write DocETL-V1 when referring to contributions from \citet{shankar2025docetl}, such as the original query optimizer and rewrite directives.
\Cref{sec:background-dop,subsec:rewrite} describe background on DocETL and semantic operators. \Cref{subsec:problem} describes the optimization problem setup. \Cref{tab:notation} summarizes all notation. 

\subsection{Datasets, Operators, and Pipelines}
\label{sec:background-dop}

\topic{Datasets} A {\em dataset} $D$ is a collection of documents. \rone{Users can provide datasets in JSON or CSV formats,} where each document is a set of key--value pairs, each representing metadata or free-form text.  
In the workload from \Cref{ex:enhancement}, 
each document has a \ttt{case\_id} field and a \ttt{notes} field containing the report text, which can be tens or hundreds of pages.  

\topic{Semantic Operators} We refer to operators that transform data using natural language (NL) specifications as {\em semantic operators}, following~\citet{patel2025semantic}.
Each semantic operator in DocETL has four components:  
{\em (i)} an {\em operator type} such as \ttt{map}, \ttt{filter}, or \ttt{reduce} (e.g., \ttt{map} applies a transformation to each document, \ttt{reduce} aggregates groups of documents into one); 
{\em (ii)} a {\em prompt template} written in natural language that describes the operator's semantics, expressed in Jinja~\cite{jinja2};
{\em (iii)} an {\em output schema} that declares the structure of the operator's results; and  
{\em (iv)} a {\em model} specifying the LLM used to execute the operator 
(e.g., \ttt{gpt-4o-mini}).  
We denote $M$ as the set of models that the user makes available to the optimizer, 
such as GPT or Gemini variants, and by $m \in M$ a particular model.  
While the user selects a specific $m$ when authoring an operator, 
the optimizer is free to choose any model from $M$ for any semantic operator, 
or replace LLM execution with {\em code-powered} implementations---synthesized code that realizes the task specified by the operator's prompt template and output schema.

Formally, we denote a semantic operator generically by $o$.
When we need to make its configuration explicit,
we write $o_x$, where $x = (p, s, m)$ denotes its prompt template $p$,
output schema $s$, and model $m \in M$.
Applying $o_x$ to a dataset $D$ produces a new dataset $D' = o_x(D)$, with a schema following $s$.
When the operator type is important,
we write $\ttt{map}_x$, $\ttt{reduce}_x$, and so on.
Unless otherwise noted, $o_x$ (and typed forms like $\ttt{map}_x$, $\ttt{reduce}_x$) denote an LLM-powered semantic operator.
For code-powered implementations, $m$ is set as $\emptyset$ and $p$ contains synthesized Python code that obeys the NL specification, with output obeying schema $s$.
We may write $\ttt{code\_map}_x$, $\ttt{code\_reduce}_x$, etc., to emphasize the fact that an operator has a code-powered implementation.
Relational operators are a special case of code-powered operators. 

Consider the following example pipeline in \Cref{ex:enhancement}.
Let $\ttt{map}_x$ have configuration $x = (p, s, m)$, where
$p$ is the prompt template:
\op{"Given the text in \{\{ input.notes \}\}, return all the enhancement factors present, along with supporting evidence."} 
$s$ is the output schema:  
\op{enhancements: list[\{factor: str, evidence: str\}]},  
and $m$ is the model (e.g., \ttt{gpt-4o-mini}).
For a dataset $D$, the result $\ttt{map}_x(D)$ is a dataset where each document now includes a new key (or attribute) \ttt{enhancements}, derived using its \ttt{notes}.  

\topic{Pipelines} Given operators, either LLM or code-powered, a {\em pipeline} $P$ is then a sequence of $k$ semantic operators $(o^{(1)}, \ldots, o^{(k)})$; given an input dataset $D$, we write its execution as $o^{(1)} \to o^{(2)} \to \cdots \to o^{(k)}$, where $\to$ denotes function composition, following~\citet{shankar2025docetl}: $o^{(k)}\big(o^{(k-1)}(\cdots o^{(1)}(D)\cdots)\big)$.
While users specify operators via NL prompt templates, schemas, and model choices, these serve as a baseline specification. The optimizer may rewrite pipelines by substituting models or replacing LLM-powered implementations with code-powered ones (including relational operators), decomposing operators into multiple ones, or fusing multiple operators, to discover cheaper or more accurate alternatives. DocETL-V1 supports six semantic operator types (\ttt{map}, \ttt{parallel\_map}, \ttt{reduce}, \ttt{filter}, \ttt{resolve}, \ttt{equijoin}) and three \papertext{non-LLM} operators (\ttt{split}, \ttt{gather}, \ttt{unnest}).
\techreport{The auxiliary operators do not invoke LLMs.}
MOAR extends this with additional operator types detailed in \Cref{sec:rewrite}.
In DocETL, pipelines and operators are specified in YAML~\cite{yaml12}: a pipeline is a list of operator configurations, where each operator's {\em configuration} is represented as a dictionary of parameters (such as prompt template, output schema, and model).\footnote{\rone{For simplicity, we define pipelines as linear sequences, which cover the predominant use cases in our deployments and all workloads in our evaluation, but our subsequent rewriting techniques generalize to tree-structured pipelines.}}

\subsection{Rewrites and Directives}
\label{subsec:rewrite}

A {\em rewrite} transforms a pipeline $P$ into a new pipeline $P'$.  
Formally, if $P=(o_1,\ldots,o_k)$, a rewrite $r$ replaces a subsequence 
$(o_i,o_{i+1},\ldots,o_j)$ with $(o'_1,o'_2,\ldots,o'_\ell)$, yielding a new pipeline $P'$.  
Intuitively, rewrites modify one or more operators in $P$ to produce an 
alternative pipeline.

A {\em rewrite directive} is a transformation rule that induces rewrites, analogous to rewrite rules in traditional databases.
A directive $d$ consists of:  
{\em (i)} a {\em left-hand side (LHS)}, i.e., a pattern over operator 
types (and optional conditions on their configurations) that must match 
a subsequence of operators in the pipeline, and  
{\em (ii)} a {\em right-hand side (RHS)} that specifies the new operator 
sequence to substitute, and how their configurations (prompt templates, schemas, models, 
or code) are to be constructed.  
As in DocETL-V1, we call them rewrite {\em directives} instead of rewrite rules because, unlike traditional rules that are fully specified, directives are abstract patterns requiring concrete instantiation of operator configurations. LLM agents instantiate these directives. 

To apply a directive $d$, the optimizer selects a target subsequence 
that matches $d$'s LHS, then instantiates $d$ by generating concrete configurations for the RHS operators, yielding a specific rewrite $r$.  
$d$ is the general rule, and $r$ is one concrete instantiation. For example, one directive from DocETL-V1 has LHS $\ttt{map}_x$ and RHS $\ttt{code\_split} \to \ttt{code\_gather} \to \ttt{map}_{x'} \to \ttt{reduce}$, where $x'$ is a modified version of $x$ adapted to process chunks rather than full text. This directive splits the largest text field in each document into chunks (we colloquially refer to this text field as the ``document'' for simplicity; when we mean the full JSON object later, we will say ``document JSON object''), augments each chunk with ``peripheral'' context from elsewhere in the document, applies the map to augmented chunks, and aggregates results---improving accuracy when text is too long for the LLM's context window. The $\ttt{reduce}$ operator (subscripts omitted) is newly synthesized to aggregate chunk-level results.
For this directive, DocETL-V1 generates $\ttt{code\_split}$ and $\ttt{code\_gather}$ configurations non-agentically by trying different chunk sizes or peripheral contexts, and uses an LLM agent to generate the prompt template and schema for $\ttt{map}_{x'}$ and the new $\ttt{reduce}$ operator. \techreport{To illustrate how directives combine to transform a pipeline, \Cref{fig:rewrite-paths} shows the sequence of rewrites that leads
from the user-authored pipeline to the highest-accuracy pipeline in \Cref{fig:moar-search-tree}.}

\techreport{\begin{figure}
    \centering
    \vspace{-15pt}
    \includegraphics[width=0.95\linewidth]{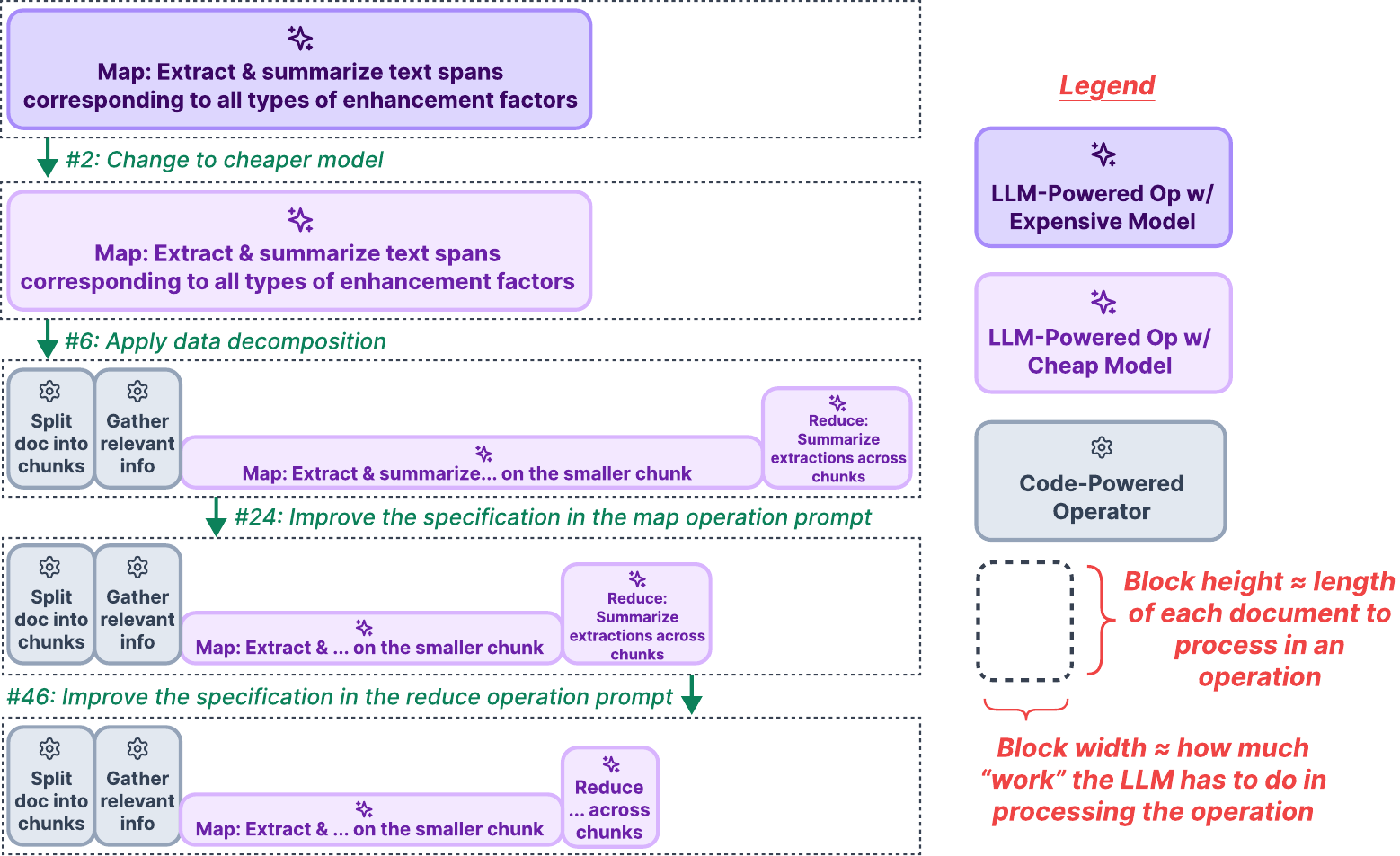}
     \vspace{-10pt}
    \caption{Illustration of the rewrite sequence that leads to the highest-accuracy pipeline in \Cref{fig:moar-search-tree}, qualitatively showing how rewrites restructure the pipeline so that LLMs perform less work on smaller portions of each document.}
    \label{fig:rewrite-paths}
    \vspace{-10pt}
\end{figure}}

\subsection{Optimization Problem Setup} 
\label{subsec:problem}

We now formalize our optimization problem. We focus on monetary cost as our primary cost metric, though our framework can be extended to capture other costs such as latency.

\topic{Cost Model} Each operator has an associated {\em cost}.  
For an LLM-powered operator $o_x$ with configuration $x = (p, s, m)$,  
the cost $c(o_x)$ is typically proportional to the number of input and output tokens in $p$ and $s$,  
multiplied by the per-token price of model $m$, and by the number of JSON documents $o_x$ processes.
For a code-powered operator $\mtt{code\_op}$, the cost is set to $c(\mtt{code\_op})=0$,  
since we want to minimize monetary cost. The cost of a pipeline $P=(o_1,\ldots,o_k)$ is the sum of individual operator costs: $c(P) = \sum_{i=1}^k c(o_i)$.
Each pipeline is evaluated by a user-defined scoring function $a(\cdot)$ applied to its final output on dataset $D$, i.e., $a(P) = a(P(D))$.  
The function $a$ may include accuracy, precision, recall, or other application-specific metrics\rone{; if $a$ requires ground truth labels, the user provides them, or $a$ can be an LLM-as-judge implementation~\cite{zheng2023judging}.}
For simplicity, we refer to this objective as ``accuracy,'' with $a(P) \in [0,1]$. Our goal, therefore, is to surface high-accuracy pipelines that offer trade-offs between accuracy and cost.
We formalize this goal using the notion of Pareto optimality.
\definition[Pareto set]
For any set of pipelines {\small $S$}, {\small 
$\mathrm{Pareto}(S)
= 
\{\, 
P \in S :
\{\, P' \in S \mid a(P') > a(P),\, c(P') \le c(P),\, P' \ne P \,\}
= \varnothing
\}$}.

Thus the Pareto set of $S$ is the set of pipelines that are not dominated (on both cost and accuracy) by any other member of $S$.

\topic{Objective} 
Let $\mathcal{P}$ denote the set of pipelines reachable from the initial user-authored pipeline $P_0$ through any sequence of rewrites.
$\mathcal{P}$ is clearly infinite, as there are an arbitrary number of ways to instantiate a given rewrite directive.
Let $B$ denote the evaluation budget, i.e., the maximum number of candidate pipelines that can be executed and scored.  
\rone{We define the budget in terms of pipeline evaluations because, in our experience, users can reason about ``how many pipeline variants to explore'' more easily than a monetary budget, which depends on unfamiliar factors like LLM pricing and data characteristics. However, the budget could also be extended to wall-clock time or total monetary cost.}
The optimizer selects a subset $\mathcal{P}_B \subseteq \mathcal{P}$ with 
$|\mathcal{P}_B|\le B$ and returns the approximate Pareto frontier $\hat{\mathcal{F}} = \mathrm{Pareto}(\mathcal{P}_B)$.
The optimization objective is then:
\begin{align*}
\max_{\mathcal{P}_B \subseteq \mathcal{P}}\; Q\!\bigl(\hat{\mathcal{F}}, \mathcal{F}\bigr) \quad \text{s.t.} \quad |\mathcal{P}_B|\le B,
\end{align*}
where $\mathcal{F} = \mathrm{Pareto}(\mathcal{P})$ denotes the true (but
unobservable) frontier. Here $Q(\hat{\mathcal{F}}, \mathcal{F})$ denotes the quality of the
approximation. In practice,
however, the true frontier $\mathcal{F}$ is unknown, so this
formulation serves as a conceptual benchmark.
Our evaluation in
\Cref{sec:evaluation} instead assesses our optimizer empirically, by comparing the
approximate frontiers $\hat{\mathcal{F}}$ it discovers against others.

In the following sections, we detail the two main components of the MOAR optimizer (depicted in \Cref{fig:systemarch}): an expanded library of rewrite directives (\Cref{sec:rewrite}) and a search algorithm that efficiently discovers high-quality pipelines within the evaluation budget (\Cref{sec:search}).

\begin{table*}[t]
\centering
\footnotesize
\vspace{-15pt}
\caption{\revision{Summary of new rewrite directives in MOAR\papertext{ {\em (the LLM-Centric category is omitted here but appears in our technical report~\cite{moartechreport})}}. Categories marked with $\dagger$ are novel to MOAR. Directives marked with $\ddag$ generate multiple candidate pipelines.} The ``visual'' column qualitatively depicts each rewrite\techreport{ (as in \Cref{fig:rewrite-paths})}: block height reflects the length of the document in an LLM call, block width reflects complexity of the task performed by the LLM, and block depth represents the number of documents processed (shown only when it changes). Darker purple blocks indicate more expensive operators; lighter purple blocks indicate cheaper ones. Blocks with \textcolor{rewrite}{green} borders are new or modified operators, and \textcolor{rewritered}{red} borders denote unchanged operators that do less ``work.''}
\vspace{-10pt}
\label{tab:rewrite_directives}
\setlength{\tabcolsep}{3pt}
\begin{tabular}{@{}l p{2.2cm} p{3cm} p{8cm} c@{}}
\toprule
\textbf{Category} & \textbf{Directive} & \textbf{Transformation Pattern} & \textbf{Description} & \textbf{Visual} \\
\midrule
\multirow{5}{*}{\shortstack[l]{Fusion and\\Reordering$^\dagger$}} 
  & \dnum{1}Same-type Fusion & 
    $\tinytt{map}_x \to \tinytt{map}_y \Rightarrow \textcolor{rewrite}{\tinytt{map}_z}$; similarly for $\tinytt{filter}$ and $\tinytt{reduce}$ &
    Fuses pairs of same-type operators (\tinytt{map--map}, \tinytt{filter--filter}, \tinytt{reduce--reduce}) into a single operator. &
    \raisebox{-0.5\height}{\includegraphics[height=0.65cm]{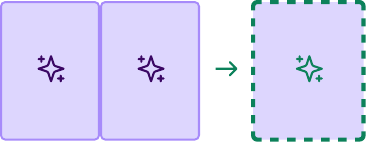}} \\
  & \dnum{2}Map--Reduce & 
    $\tinytt{map}_x \to \tinytt{reduce}_{K,y} \Rightarrow \textcolor{rewrite}{\tinytt{reduce}_{K,z}}$ &
   Combines the \tinytt{map} and \tinytt{reduce} into a single \tinytt{reduce}. Applicable only when the output schema in $x$ does not include the key(s) in $K$. &
    \raisebox{-0.5\height}{\includegraphics[height=0.65cm]{figures/rewrites/MapReduceFusion.pdf}} \\
  & \dnum{3}Map--Filter & 
    $\tinytt{map}_x \to \tinytt{filter}_y \Rightarrow \textcolor{rewrite}{\tinytt{map}_z} \to \textcolor{rewrite}{\tinytt{code\_filter}}$ &
    Expands the \tinytt{map} to also compute the predicate produced by the downstream operator $\tinytt{filter}_y$ (so $z$'s output schema is the union of those from $x$ and $y$), followed by a  \tinytt{code\_filter} that simply checks the boolean attribute generated (in $y$'s schema). &
    \raisebox{-0.5\height}{\includegraphics[height=0.65cm]{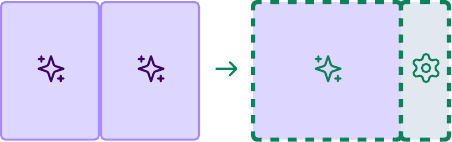}} \\
  & \dnum{4}Filter--Map & 
    $\tinytt{filter}_x \to \tinytt{map}_y \Rightarrow \textcolor{rewrite}{\tinytt{map}_z} \to \textcolor{rewrite}{\tinytt{code\_filter}}$ &
    Fuses \tinytt{filter} and \tinytt{map} logic into a single \tinytt{map}. As in Map--Filter, the fused operation is ``harder'' or requires the LLM to do more ``work;'' thus it is wider in the visual. &
    \raisebox{-0.5\height}{\includegraphics[height=0.65cm]{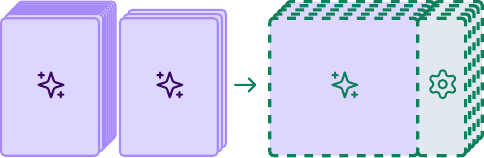}} \\
  & \dnum{5}Reordering & 
    $o_x \to o_y \Rightarrow o_y \to o_x$ &
    Reorders commuting operators so that cheaper operators run earlier, akin to traditional operator reordering. &
    \raisebox{-0.75\height}{\includegraphics[height=0.65cm]{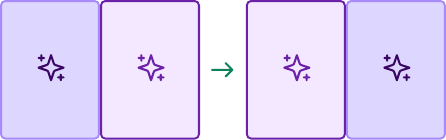}} \\
\midrule
\multirow{4}{*}{\shortstack[l]{Code\\Synthesis$^\dagger$}}
  & \dnum{6}Code Substitution & 
    $\tinytt{o}_x \Rightarrow \textcolor{rewrite}{\tinytt{code\_op}_{\hat{x}}}$ &
    Replaces an LLM-powered operator with synthesized Python code. &
    \raisebox{-0.5\height}{\includegraphics[height=0.65cm]{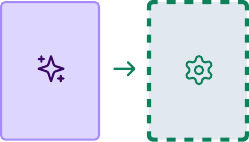}} \\
  & \dnum{7}Code Sub. (Reduce) & 
    $\tinytt{reduce}_x \Rightarrow \textcolor{rewrite}{\tinytt{code\_reduce}_{\hat{x}}} \to \textcolor{rewrite}{\tinytt{map}}$ &
    Splits a \tinytt{reduce} into code-based aggregation plus a \tinytt{map} that handles logic requiring an LLM and transforms the output into the schema specified in $x$.  For example, a \tinytt{reduce} that asks ``generate a report of the most common themes in the documents'' can be rewritten so \tinytt{code\_reduce} counts themes and concatenates relevant context, and the \tinytt{map} generates the natural-language report. &
    \raisebox{-1\height}{\includegraphics[height=0.65cm]{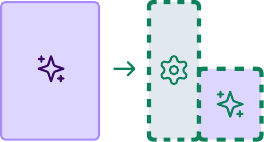}} \\
  & \dnum{8}Doc. Compression (Code)$^\ddag$ & 
    $\tinytt{o}_x \Rightarrow \textcolor{rewrite}{\tinytt{code\_map}} \to \textcolor{rewrite}{\tinytt{o}_{x'}}$ &
    Uses synthesized Python code (e.g., with regexes) to deterministically extract only the relevant portions of the document, producing a shorter input for the downstream operator. &
    \raisebox{-0.5\height}{\includegraphics[height=0.65cm]{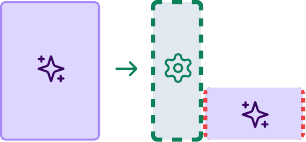}} \\
  & \dnum{9}Head/Tail Compression.$^\ddag$ & 
    $\tinytt{o}_x \Rightarrow \textcolor{rewrite}{\tinytt{code\_map}} \to \textcolor{rewrite}{\tinytt{o}_{x'}}$ &
    Retains only the first $h$ and last $\ell$ words (or lines) of each document via a synthesized \tinytt{code\_map}. Useful when key information typically appears at document boundaries (e.g., abstract, conclusion). &
    \raisebox{-0.5\height}{\includegraphics[height=0.65cm]{figures/rewrites/CodeCompression.pdf}} \\
\midrule
\multirow{2}{*}{\shortstack[l]{Data\\Decomposition}}
  & \dnum{10}Chunk Sampling$^\ddag$ & 
    $\tinytt{split} \to \tinytt{gather} \to \tinytt{map} \to \tinytt{reduce} \Rightarrow \tinytt{split} \to \tinytt{gather} \to \textcolor{rewrite}{\tinytt{sample}} \to \tinytt{map} \to \tinytt{reduce}$ &
    Samples relevant chunks using BM25, embeddings, or random sampling. Reduces cost by processing only relevant chunks when full documents contain mostly irrelevant content. &
    \raisebox{-0.5\height}{\includegraphics[height=0.8cm]{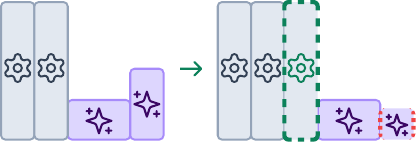}} \\
  & \dnum{11}Doc. Sampling$^\ddag$ &
    $\tinytt{reduce}_{K,x} \Rightarrow \textcolor{rewrite}{\tinytt{sample}_{K}} \to \tinytt{reduce}_{K,x}$ &
    Samples a subset of documents within each group (e.g., using BM25, embeddings, or random sampling) before the \tinytt{reduce}, reducing cost when groups contain many redundant or low-signal documents. &
    \raisebox{-0.5\height}{\includegraphics[height=0.8cm]{figures/rewrites/CodeCompression.pdf}} \\
  & \dnum{12}Cascade Filtering$^\ddag$ & 
    $\tinytt{filter}_x \Rightarrow \textcolor{rewrite}{\tinytt{code\_filter}^\ast} \to \textcolor{rewrite}{\tinytt{filter}_y^\ast} \to \tinytt{filter}_x$ &
    Inserts one or two cheaper ``pre-filters'' before $\tinytt{filter}_x$. $\tinytt{code\_filter}$ and $\tinytt{filter}_y$ (marked with $^\ast$ to indicate optionality; at least one is required) form a cascade in which each pre-filter removes additional documents before they reach $\tinytt{filter}_x$. &
    \raisebox{-0.5\height}{\includegraphics[height=0.75cm]{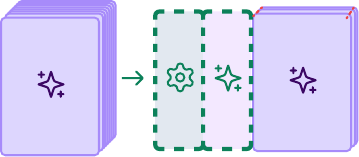}} \\
\midrule
\multirow{2}{*}{\shortstack[l]{Projection\\Synthesis}}
  & \dnum{13}Doc. Summarization & 
    $\tinytt{o}_x \Rightarrow \textcolor{rewrite}{\tinytt{map}} \to \textcolor{rewrite}{\tinytt{o}_{x'}}$ &
    Produces a shorter version of each document by generating an LLM-written summary (via \tinytt{map}) and passing that condensed text to the downstream operator. &
    \raisebox{-0.5\height}{\includegraphics[height=0.65cm]{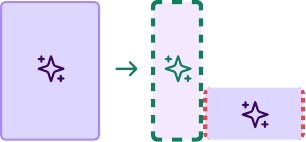}} \\
  & \dnum{14}Doc. Compression (LLM)$^\ddag$ & 
    $\tinytt{o}_x \Rightarrow \textcolor{rewrite}{\tinytt{extract}} \to \textcolor{rewrite}{\tinytt{o}_{x'}}$ &
   Produces a shorter version of each document by generating an \tinytt{extract} operator to return text spans from the original document; unlike the previous summarization rewrite, the output is a subset of the original text, not a transformation. &
    \raisebox{-0.5\height}{\includegraphics[height=0.65cm]{figures/rewrites/DocSumm.pdf}} \\
\techreport{\midrule
\multirow{4}{*}{LLM-Centric}
  & \dnum{15}Model Substitution & 
    $\tinytt{o}_x \Rightarrow \textcolor{rewrite}{\tinytt{o}_{x'}}$ where $x' = (p, s, \textcolor{rewrite}{m'})$ &
    Replaces an operator's LLM with a different model. &
    \raisebox{-0.5\height}{\includegraphics[height=0.65cm]{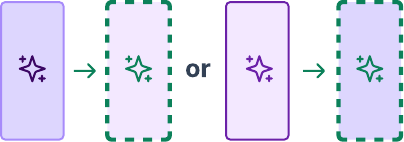}} \\
  & \dnum{16}Clarify Instructions$^\ddag$ & 
    $\tinytt{o}_x \Rightarrow \textcolor{rewrite}{\tinytt{o}_{x'}}$ where $x' = (\textcolor{rewrite}{p'}, s, m)$ &
    Rewrites the prompt template to be more specific and detailed, reducing ambiguity and thus making the task ``easier'' for the LLM. &
    \raisebox{-0.5\height}{\includegraphics[height=0.65cm]{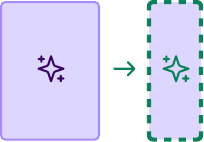}} \\
  & \dnum{17}Few-shot Examples & 
    $\tinytt{o}_x \Rightarrow \textcolor{rewrite}{\tinytt{o}_{x'}}$ where $x' = (\textcolor{rewrite}{p'}, s, m)$ &
    Adds few-shot examples to prompts (a standard strategy for improving accuracy~\cite{brown2020language}), thus making the task ``easier'' for the LLM. &
    \raisebox{-0.5\height}{\includegraphics[height=0.65cm]{figures/rewrites/ImprovePrompt.pdf}} \\
  & \dnum{18}Arbitrary Rewrite$^\dagger$ & 
    $P \Rightarrow \textcolor{rewrite}{P'}$ &
    Allows the agent to propose free-form pipeline transformations beyond the predefined directives. &
    \raisebox{-0.5\height}{\includegraphics[height=0.65cm]{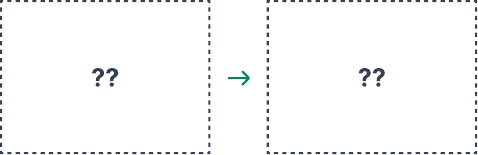}} \\}
\bottomrule
\end{tabular}
\vspace{-10pt}
\end{table*}

\section{New Operators and Directives}
\label{sec:rewrite}

\revision{The design space of rewrites for semantic operator pipelines is theoretically infinite, spanning rewrites that target cost reduction (e.g., replacing LLM calls with code, fusing operators), accuracy improvement (e.g., decomposing complex tasks, chunking long documents), or both simultaneously. As in traditional query optimizers~\cite{Graefe1995TheCF}, there is no guarantee that the transformation rules are complete or sufficient for optimality; what matters is that the framework is extensible. MOAR's optimization framework (\Cref{sec:search}) is designed independently of the specific directive library, so new directives can be incorporated as workloads evolve and models improve.}

DocETL-V1 proposed three categories of rewrite directives targeting accuracy: {\em projection synthesis} (i.e., creating \ttt{map} operators that decompose tasks), {\em data decomposition} (i.e., splitting documents into chunks or groups of many documents into smaller-size groups), and {\em LLM-centric improvements} (e.g., prompting strategies)~\cite{shankar2025docetl}. However, these directives typically increase cost.
MOAR extends DocETL's directive library in two ways. First, we introduce two new categories: {\em fusion and reordering} and {\em code synthesis}. \rfour{Fusion combines multiple sequential operators into fewer ones, reducing the number of LLM calls---for example, two adjacent maps can be merged into a single map whose prompt instructs the LLM to perform both tasks and whose output schema is the union of both original schemas, or a map followed by a reduce can be absorbed into a single reduce when the map's output does not generate the grouping keys. {\em Reordering} re-arranges commuting operators so that more selective operators run earlier. {\em Code synthesis} replaces LLM-powered operators with synthesized Python code, or inserts code-powered operations that preprocess data (e.g., regex-based extraction) before downstream LLM calls.}
Second, we adapt existing \rfour{rewrite directive} categories to also reduce cost---e.g., projection synthesis can compress documents to reduce tokens processed by downstream operators, data decomposition can limit processing to relevant chunks, and LLM-centric rewrites can substitute cheaper models. With DocETL-V1's 13 directives, MOAR's 18 new directives bring the total to over 30. \rfour{All directives in \Cref{tab:rewrite_directives} are new to MOAR}\papertext{; detailed descriptions \revision{and the typing and composability constraints they must satisfy} are in \Cref{sec:detailed-rewrite} in our technical report~\cite{moartechreport}}\techreport{; detailed descriptions \revision{and the typing and composability constraints they must satisfy} are in \Cref{sec:detailed-rewrite}}. We highlight a few examples below (directive numbers reference the corresponding row in \Cref{tab:rewrite_directives}):

\begin{itemize}[nosep, leftmargin=*, wide=0pt]
\item In {\bf \em Fusion and Reordering.} The \dnum{3}Map--Filter directive fuses a \ttt{map} followed by a \ttt{filter} into a single \ttt{map} whose prompt incorporates both the transformation and the filter predicate, producing a Boolean attribute that a downstream \ttt{code\_filter} checks---reducing cost by eliminating an LLM call per document. (\dnum{4}Filter--Map fusion into a Map is analogous but may not reduce cost when the filter has high selectivity.) These directives compose powerfully with others. For instance, in \Cref{ex:enhancement}, if public defenders are only interested in defendants where firearm-related enhancement factors appear in both the police report {\em and} the charging summary, and the defender adds a \ttt{filter} to the pipeline in \Cref{ex:enhancement}. MOAR can apply task decomposition (e.g., pipeline \#10 in \Cref{fig:moar-search-tree}) to extract each factor type in parallel, reorder operations as per \dnum{5} so the \ttt{filter} immediately follows the firearm-extraction \ttt{map}, and fuse them---avoiding downstream processing for irrelevant factors.

\item In {\bf \em Code Synthesis.} The \dnum{8}Code-based Document Compression directive inserts a new \ttt{code\_map} operator (in Python) to extract relevant portions of each document {\em before} the downstream LLM operator. Unlike \dnum{6}Code Substitution (which replaces the LLM entirely), this directive preprocesses data to reduce document size. Our key insight is that relevant content can often be identified via complex regular expressions or keyword matching---that LLM agents readily synthesize (e.g., 50--100 keyword variations for firearm-related content in \Cref{ex:enhancement}). The downstream operator then runs on shorter documents, thereby lowering cost.

\item In {\bf \em Data Decomposition.} DocETL-V1 introduced chunking to handle documents exceeding LLM context limits. MOAR adds the \dnum{10}Chunk Sampling directive: after splitting, a \ttt{sample} operator selects relevant chunks (via BM25, embeddings, or random sampling) before the \ttt{map}. Analogously, the \dnum{11}Document Sampling directive selects a subset of documents before a \ttt{reduce}. These directives work well for tasks where processing all data is unnecessary---e.g., identifying common themes across thousands of customer reviews does not require reading every review, just a representative sample.

\item In {\bf \em Projection Synthesis.} MOAR extends projection synthesis with directives that {\em compress} documents rather than decompose tasks. \dnum{13}Document Summarization inserts a \ttt{map} generating an LLM-written summary of each document before downstream operators, reducing downstream operator costs.
\end{itemize}

To support the new directives, we introduce three new operator types. The \ttt{sample} operator selects a subset of documents (or chunks) most relevant to the downstream operator. It may use BM25 keyword search~\cite{robertson2009probabilistic}, embedding-based similarity, random sampling, or stratified variants of these methods that ensure each subgroup (e.g., based on metadata keys in the document) is proportionally represented in the sample.
\techreport{For instance, when extracting enhancement factors from police reports (\Cref{ex:enhancement}), if a prior \ttt{split} operator has divided each report into chunks---each chunk now becoming a document---then \ttt{sample} can issue a BM25 query with terms like ``firearm'' and ``weapon'' or use embedding similarity to a query such as ``threatening with a weapon'' to select only the relevant chunks before the downstream \ttt{map}. By processing fewer documents, the subsequent LLM-powered operator incurs lower cost.}
Next, the \ttt{extract} operator presents a document's JSON representation with line numbers to an LLM, which returns ranges relevant to the operator's natural language specification (e.g., ``lines 45--67''); only those lines are retained, preserving the document's key--value structure.
Unlike a \ttt{map} that outputs text verbatim, \ttt{extract} guarantees exact subsets of documents and requires far fewer output tokens (thus reducing cost). Finally, code-powered operators (\ttt{code\_map}, \ttt{code\_reduce}, \ttt{code\_filter}) execute synthesized Python instead of invoking an LLM. Overall, the aforementioned operators enable directives to reduce costs by processing smaller document portions (\ttt{sample}, \ttt{extract}) or replacing LLM calls with code.
A complete operator list appears in \Cref{tab:docetl-operators}\papertext{ in our technical report~\cite{moartechreport}}. \papertext{\revision{Directives must also satisfy typing and composability constraints (formalized in \Cref{sec:detailed-rewrite} of our technical report~\cite{moartechreport}) that guarantee rewritten pipelines match the user's expected output schema. However, there is no guarantee that rewritten prompts match the intent of the original, so the search algorithm evaluates each rewritten pipeline on a held-out sample. We discuss the soundness and completeness of our directive library, contrasting with the relational setting, in our technical report~\cite{moartechreport}.}}

\techreport{\topic{Comparison with relational rewrite rules} \revision{Since rewrite directives for semantic operator pipelines are analogous to rewrite rules in relational query optimizers, we can ask whether they satisfy similar properties.}

\revision{Regarding {\em soundness}: in relational query optimizers, all rewrite rules are sound. In our setting, soundness splits into two notions. ``Type soundness'' requires that a rewritten pipeline preserves the input-output schema of the original; each directive's preconditions and postconditions (formalized in \Cref{sec:detailed-rewrite}) enforce this. ``Semantic soundness'' requires that the rewritten pipeline preserves the intent of the original prompts, e.g., that splitting a map into two maps via a projection synthesis directive produces prompts that together accomplish the same task as the original. Unlike type soundness, semantic soundness cannot be formally verified.}

\revision{Regarding {\em completeness}: like relational rewrite rules, our directive library is not complete. However, we take steps toward high coverage: the library spans three categories (cost reduction, accuracy improvement, and combinations of both), directives are composable (applying one does not prevent applying another), and we include an ``arbitrary rewrite'' directive that allows the agent to propose any change without directive scaffolding.}

\rfour{Then, we can also reflect on what is different about rewrite directives compared to relational rewrites. One observation is that, unlike in the relational setting, relational operators (i.e., what we call code-powered operators) are essentially ``free'' compared to LLM-powered operators. Offloading computation to such operators can reduce the amount of data processed by the LLM. In some special cases, we can entirely replace an LLM-powered operator with a code-powered operator (directive \dnum{6} in \Cref{tab:rewrite_directives}). More generally, we can fuse an LLM-powered map and LLM-powered filter (in either order) into one LLM-powered map followed by a (free) code-powered filter, where the new map performs the logic in the old map {\em and} prepares new information in a way that the code-based filter can use it, e.g., by creating a new 0/1 attribute that the code-powered filter can operate on, reducing the cost by half (directives \dnum{3}, \dnum{4} in \Cref{tab:rewrite_directives}). Or, we can prefix a code-powered operator to strip irrelevant tokens from, or reduce the size of, each document before it reaches an LLM operator (directives \dnum{8}, \dnum{9}, and \dnum{11} in \Cref{tab:rewrite_directives}, where \ttt{sample} is a special code-powered operator that performs sampling). Note, however, that in all of these cases, unlike the relational setting, equivalence is {\em not guaranteed}; rewriting could lead to a plan that has worse accuracy than the original. Thus, we still need to evaluate each rewritten plan for accuracy (and cost).}
}

\section{Search Algorithm}
\label{sec:search}

\begin{figure*}[t]
\vspace{-20pt}
    \begin{center}
\includegraphics[width=0.8\linewidth]{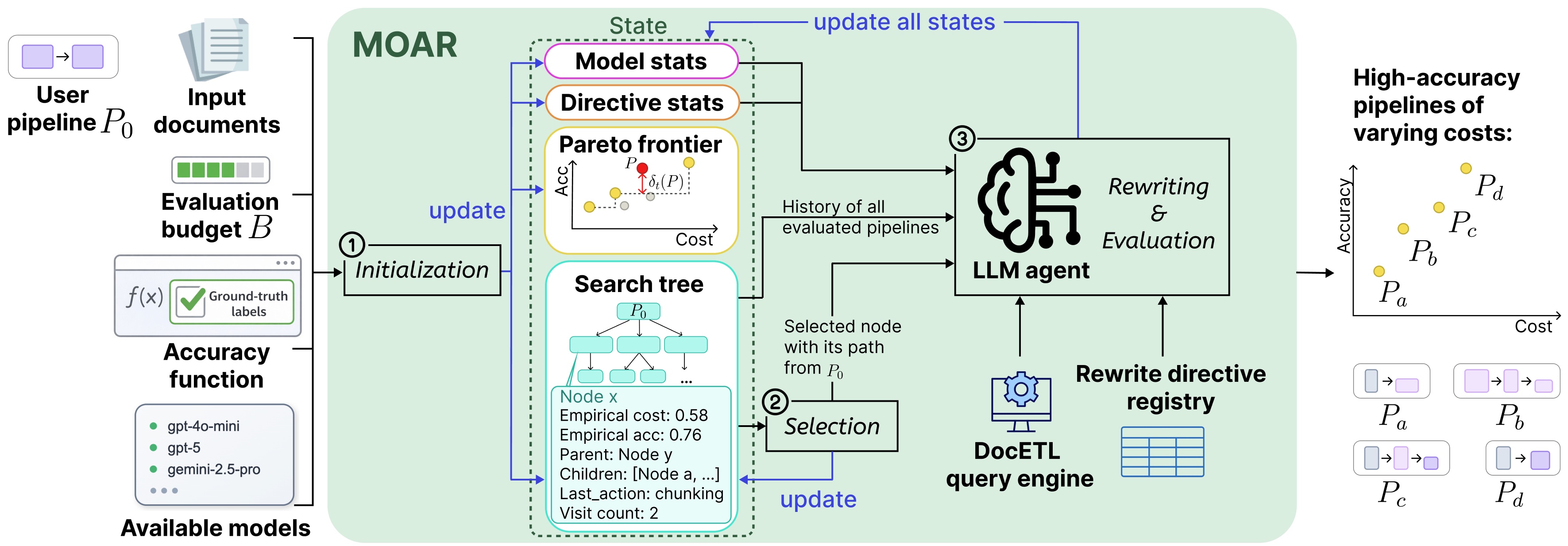}
    \end{center}
    \vspace{-10pt}
    \caption{\rone{System architecture of MOAR. The optimizer takes as input a user pipeline $P_0$, input documents, user-defined evaluation budget, user-authored accuracy function, and available models. In the initialization phase (\circled{1}), MOAR evaluates $P_0$ with all models to construct an initial frontier. The main search loop alternates between selection (\circled{2})---choosing which pipeline to rewrite based on contribution of it and its descendants to the Pareto frontier---and rewriting \& evaluation (\circled{3})---using an LLM agent to select and instantiate a rewrite directive from the registry, then executing the resulting pipeline. The output is a set of high-accuracy pipelines spanning different costs.}}
\label{fig:systemarch}
\vspace{-10pt}
\end{figure*}

\begin{table}[t]
\vspace{-15pt}
\centering
\scriptsize
\caption{Table of Notation}\label{tab:notation}
\vspace{-10pt}
\begin{tabular}{@{}l l p{4.5cm}@{}}
\toprule
\textbf{Category} & \textbf{Symbol} & \textbf{Description} \\
\midrule
\multirow{6}{*}{\textit{\begin{tabular}[c]{@{}l@{}}Datasets \&\\ Pipelines\end{tabular}}} 
& $D$; $D_o \subset D$ & Dataset; sample to evaluate candidate pipeline on \\
& $P$; $P_0$ & A candidate pipeline; the user-authored pipeline \\
& $o^x$ & Operator with config $x = (p, s, m)$ \\
& $p$; $s$; $m \in M$ & Prompt template; schema; model \\
& $P \xrightarrow{r} P'$ & Rewrite $r$ transforms $P$ into $P'$ \\
& $d$ & A rewrite directive \\
\midrule
\multirow{2}{*}{\textit{\begin{tabular}[c]{@{}l@{}}Cost \&\\ Accuracy\end{tabular}}} 
& $c(P)$; $a(P)$; $\hat{c}(P)$; $\hat{a}(P)$ & Cost and accuracy of $P$; empirical estimates on $D_o$ \\
& $B$ & Evaluation budget \\
\midrule
\multirow{4}{*}{\textit{\begin{tabular}[c]{@{}l@{}}Search\\ Tree\end{tabular}}} 
& $T_t = (V_t, E_t)$ & Search tree with vertices $V_t$, edges $E_t$ \\
& $V_t$; $\hat{\mathcal{F}}_t$; $\mathcal{F}$ & Evaluated pipelines; frontiers \\
& $\mathrm{children}(P)$; $\mathrm{parent}(P)$ & Child/parent nodes \\
& $\mathrm{desc}_t(P)$; $\mathrm{depth}(P)$ & Descendants; depth \\
& $last\_action(P)$ &Directive used to generate $P$ from $\mathrm{parent}(P)$\\
\midrule
\multirow{5}{*}{\textit{\begin{tabular}[c]{@{}l@{}}Pipeline\\Selection \&\\ Utility\end{tabular}}} 
& $A_t(P, c)$ & Max accuracy at cost $\leq c$ excluding $P$ \\
& $\delta_t(P)$ & Vertical gain: $\hat{a}(P) - A_t(P, \hat{c}(P))$ \\
& $n_t(P)$ & Visit count \\
& $\overline{\delta}_t(P)$; $U_t(P)$ & Avg improvement; utility \\
& $W(n_t(P))$ & Max children $P$ may generate \\
\midrule
\multirow{2}{*}{\textit{Rewriting}} 
& $\nu(P, d)$ & Usage count of $d$ from $P$ \\
& $k$ & Number of candidate pipelines generated for a rewrite \\
& $\text{rank}_t(P)$ & Accuracy rank of $P$ compared to all pipelines explored\\
\bottomrule
\end{tabular}
\vspace{-15pt}
\end{table}

In this section, we describe how MOAR efficiently searches over rewrite directives to discover high-quality and low-cost pipelines.
\Cref{fig:systemarch} shows an overview of MOAR's search 
algorithm. MOAR maintains 
state (shown in the dashed box) including model statistics, directive statistics, the 
current Pareto frontier, and a search tree rooted at the 
user-authored pipeline $P_0$.

\topic{Search Space Representation} Unlike traditional query optimizers that construct plans from optimal subplans~\cite{selinger1979access, Graefe1995TheCF}, MOAR performs global search over complete pipelines. This design reflects our observation from \Cref{sec:intro} that the best plan for a semantic operator pipeline may rely on subplans that are individually suboptimal, because the benefit of a subplan isn't independent of its outputs. To support this global search, we represent the search space as a tree $T = (V, E)$ with the user-authored pipeline $P_0$ at the root. Each node $P \in V$ represents a {\em complete} pipeline configuration. Each edge $P \xrightarrow{r} P'$ applies a single rewrite $r$ to produce child pipeline $P'$. Every node has exactly one parent and may have multiple children (shown as the search tree in \Cref{fig:systemarch}). The path from $P_0$ to any node captures the sequence of rewrites used to construct it.

At each iteration $t$ of the search algorithm, MOAR selects a node, applies a rewrite to generate a child pipeline, and {\em evaluates} it---that is, executes it on a small sample $D_o \subset D$ (e.g., 40 documents) to measure empirical cost $\hat{c}(P)$ and accuracy $\hat{a}(P)$. We write $V_t$ for the set of pipelines evaluated after $t$ iterations, $E_t$ for the corresponding edge set, and $\hat{\mathcal{F}}_t$ for the Pareto frontier of $V_t$. Throughout this section, $\hat{c}(P)$ and $\hat{a}(P)$ denote sample estimates on $D_o$, not the population-level quantities $c(P)$ and $a(P)$ defined in \Cref{sec:background}.

\begin{algorithm}
\scriptsize
\SetAlgoLined
\KwIn{User pipeline $P_0$, dataset $D$, sample $D_o \subset D$, model pool $M$, budget $B$}
\KwOut{Pareto frontier $\hat{\mathcal{F}}_B$}
\BlankLine
\SetKwFunction{FMain}{MOAR}
\SetKwFunction{FInit}{Initialize}
\SetKwFunction{FSelect}{Select}
\SetKwFunction{FRewrite}{RewriteAndEvaluate}
\SetKwProg{Fn}{Function}{:}{}
\Fn{\FMain{$P_0, D_o, M, B$}}{
    \tcp{Initialization: evaluate $P_0$ with all models}
    $T \gets$ \FInit{$P_0, D_o, M$}\;
    initialize directive usage map $\nu(P, d) \gets 0$ for all $P, d$\;
    $t \gets |V|$ \tcp*{Number of pipelines evaluated so far}
    
    \tcp{Main search loop}
    \While{$t < B$}{
        \tcp{Selection: traverse tree to find node to rewrite}
        $P^\star \gets$ \FSelect{$P_0, T_t$}\;
        
        \tcp{Rewriting and evaluation (parallelized)}
        $(P', r, \hat{c}(P'), \hat{a}(P'), k) \gets$ \FRewrite{$P^\star, G_t, \nu, D_o$}\;
        
        \tcp{Update tree with new pipeline and rewrite edge}
        add $P'$ to $V$ and edge $P^\star \xrightarrow{r} P'$ to $E$\;
        $t \gets t + k$ \tcp*{Increment by number of candidates evaluated}
        recompute $\hat{\mathcal{F}}_t$ from $V$\;
    }
    \Return $\hat{\mathcal{F}}_B$\;
}
\caption{MOAR Search}
\label{alg:main}
\end{algorithm}

\topic{Algorithm Overview} \rfour{MOAR searches for better pipelines by iteratively selecting, rewriting, and evaluating candidates, as described in Algorithm~\ref{alg:main} and illustrated in \Cref{fig:systemarch}.
The search process begins with {\em initialization}, where MOAR constructs an initial frontier that provides diverse starting points across the accuracy--cost spectrum. After initialization, each iteration of the search loop proceeds through two phases. 
First, in the {\em selection phase}, MOAR adapts a multi-armed bandit framework~\cite{auer2002finite} to decide which pipeline to rewrite next. Unlike traditional UCT, where each node's reward is independent, a node's reward in our setting depends on all other nodes because the Pareto frontier is a global function of every pipeline explored; we therefore define a custom utility metric based on marginal accuracy contribution that handles this interdependence. There are also infinitely many possible rewrites, so we limit the branching factor and delegate action selection to an LLM agent that reasons about pipeline semantics to prioritize which directives to try.
Second, in the {\em rewriting and evaluation phase}, the LLM agent chooses and instantiates a rewrite directive to produce a child pipeline. The child $P'$ is then executed on $D_o$ to obtain $(\hat{c}(P'), \hat{a}(P'))$, which informs selection in future iterations. 
Each iteration incurs significant latency due to LLM-guided rewriting and pipeline execution. MOAR therefore parallelizes search across multiple workers, with only the selection phase synchronized to ensure consistency. The loop repeats until the evaluation budget $B$ is exhausted, returning the final frontier $\hat{\mathcal{F}}_B$.}

Next, we describe each phase in detail: initialization (\Cref{sec:search:init}), selection (\Cref{sec:search:select}), and rewriting and evaluation (\Cref{sec:search:rewrite}).
\Cref{tab:notation} summarizes notation used.

\subsection{Initialization}
\label{sec:search:init}

Before the main search loop begins, MOAR initializes two components: the Pareto frontier and the state used throughout search.

\topic{Frontier Initialization}
Building an initial frontier prevents the search from getting trapped early in local minima and provides statistics that aid future selection decisions. Given the user-authored pipeline $P_0$ configured with a default model $m_0$ (e.g., gpt-4o-mini), MOAR evaluates $P_0$ with all models in the pool $M$ (e.g., gpt-4o-mini, gpt-4o, gemini-2.5-flash).\footnote{If $|M| > C_m$ (set to 12 in our implementation), we subsample up to 3 models per family (e.g., {\scriptsize\texttt{gpt-4.1-nano}, \texttt{gpt-4.1-mini}, \texttt{gpt-4.1}}) from randomly selected families.}
In other words, for each model $m_i \in M$, MOAR creates a pipeline variant by substituting $m_i$ into all operators of $P_0$, then measures its cost and accuracy $(\hat{c}, \hat{a})$ on $D_o$. These model variants become children of $P_0$ in the search tree, yielding frontier $\hat{\mathcal{F}}_{|M|}$. 
Next, for each pipeline in $\hat{\mathcal{F}}_{|M|}$, MOAR uses an LLM agent to generate exactly two rewrites---one targeting accuracy improvement and one targeting cost reduction---using our standard rewriting procedure (to be described in \Cref{sec:search:rewrite}). By spawning grandchildren only from frontier children ($\hat{\mathcal{F}}_{|M|}$) rather than all pipelines in $V_{|M|}$, we limit the budget consumption from initialization. 
At the end of initialization, we disable non-frontier model variants from future selection, ensuring subsequent iterations focus on promising regions of the search space.
At the end of initialization (step \circled{1} in \Cref{fig:systemarch}), the search tree contains $P_0$ at the root, model variants as children, and rewrites of frontier pipelines as grandchildren.

\topic{Other State Initialization} For each node $P$, MOAR maintains: 
\begin{enumerate}[label=(\roman*), nosep, left=0pt]
    \item $\hat{c}(P)$ and $\hat{a}(P)$: empirical cost and accuracy on $D_o$;
    \item $\mathrm{parent}(P)$ and $\mathrm{children}(P)$: encoding tree structure;
    \item $n_t(P)$: visit count, equal to $1 + |\mathrm{desc}_t(P)|$ where $\mathrm{desc}_t(P)$ denotes the descendants of $P$ in $V_t$; and
    \item $\mathit{last\_action}(P)$: the directive applied to generate $P$ from $\mathrm{parent}(P)$.
\end{enumerate}
Most statistics remain fixed throughout search; only $\mathrm{children}(P)$ and $n_t(P)$ evolve as new pipelines are evaluated. Additionally, MOAR initializes aggregate statistics. {\em (i) Model statistics} record the cost and accuracy achieved by each model variant of the original pipeline (i.e., each child of $P_0$), providing a controlled comparison across models on identical operators; this helps agents select models when synthesizing new operators. {\em (ii) Directive statistics} record, for each directive $d$, the average change in cost and accuracy induced by applying $d$, measured as the difference between a pipeline's metrics and those of its parent. Agents receive both model and directive statistics as context during rewriting (\Cref{sec:search:rewrite}).

\subsection{Selection}
\label{sec:search:select}

After initialization, the search tree contains $|V_t| = |M| + 2|\hat{\mathcal{F}}_{|M|}|$ pipelines (counting toward the budget $B$). Now, in each iteration, the selection phase identifies which pipeline $P^\star$ to rewrite---equivalently, which path in the search tree to extend by one more rewrite. The selector must balance exploitation of high-performing paths with exploration of under-tested ones. MOAR achieves this balance by assigning each pipeline a multi-armed bandit-style utility score~\cite{auer2002finite}.

\topic{Utility Function} We begin by formalizing utility. Let
{\small \begin{align*}
A_t(P) := \max \{\, \hat{a}(P') : P' \in \mathrm{Pareto}(V_t \setminus \{P\}),\; \hat{c}(P') \le \hat{c}(P) \,\}
\end{align*}}
be the highest accuracy achievable at cost $\hat{c}(P)$ or lower, excluding $P$ itself. The {\em contribution} of pipeline $P$ to the frontier is then $\delta_t(P) := \hat{a}(P) - A_t(P)$, measuring how much $P$ improves accuracy beyond other pipelines with comparable cost. When $\delta_t(P) > 0$, pipeline $P$ extends the frontier; otherwise it is dominated. This quantity $\delta_t(P)$ is visualized as the vertical distance between the red point and the frontier in the ``Pareto frontier'' panel of \Cref{fig:systemarch}.

One widely used strategy for multi-armed bandits is the Upper Confidence Bound (UCB) algorithm~\cite{auer2002finite}. However, we cannot apply UCB directly because the pipelines are not independent; each is derived from rewrites of its ancestors. 
Instead, we use UCT~\cite{kocsis2006bandit}, which extends UCB to a tree-structured search space. 
Following UCT, we define the utility score as:
{\small \begin{align}
U_t(P) = \underbrace{\frac{\delta_t(P) + \sum_{P' \in \mathrm{desc}_t(P)} \delta_t(P')}{n_t(P)}}_{\text{exploitation}} + \underbrace{\sqrt{\frac{2\ln n_t(\mathrm{parent}(P))}{n_t(P)}}}_{\text{exploration}},
\label{eq:utility}
\end{align}}
where $\mathrm{desc}_t(P)$ denotes all descendants of $P$ in $V_t$.

The {\em exploitation} term in \Cref{eq:utility} measures the average frontier contribution along the subtree rooted at $P$: we sum $\delta_t$ over $P$ and all its descendants, then divide by the visit count $n_t(P) = 1 + |\mathrm{desc}_t(P)|$. Unlike standard UCB, which tracks rewards for independent arms, UCT aggregates rewards across an entire subtree because a node's value depends on what rewrites become reachable after selecting it.
Then, the {\em exploration} term encourages trying under-visited nodes. Standard UCB uses the total number of iterations across all nodes in the numerator; UCT instead uses the parent's visit count $n_t(\mathrm{parent}(P))$ because this represents how many times we have had the opportunity to even select $P$ for rewriting. A child with few visits relative to $n_t(\mathrm{parent}(P))$ is under-explored compared to its siblings and receives a higher exploration bonus.

\topic{Hierarchical Traversal with Progressive Widening}
Because utility scores depend on $n_t(\mathrm{parent}(P))$, they are only comparable among siblings. We therefore traverse the tree top-down to select $P^\star$: starting at $P_0$, we repeatedly select the child with the highest utility until reaching a node $P^\star$ that can generate a new child. In standard UCT~\cite{kocsis2006bandit}, when a node is selected, {\em all} possible rewrites (i.e., actions) are applied before the next iteration of selection. This is feasible for small action spaces (e.g., board games), but problematic here since a single pipeline could spawn hundreds of children (30+ directives $\times$ multiple operators $\times$ multiple instantiations). To limit branching, we use {\em progressive widening}, a  technique from Monte Carlo Tree Search for large action spaces~\cite{chaslot2008progressive}. The idea is to cap the number of children based on visit count, allowing the action space to be explored gradually. We accordingly set the maximum number of children for node $P$ to $W(n_t(P)) = \max(2,\, 1 + \sqrt{n_t(P)})$, where the $\sqrt{n_t(P)}$ growth rate is typical~\cite{chaslot2008progressive}. For example, a node with four visits may have at most three children; only after nine visits may it produce a fourth. This sublinear growth forces exploration of deeper paths before generating more rewrites from any single node. Algorithm~\ref{alg:select} in \Cref{app:search}\papertext{ in our technical report~\cite{moartechreport}} specifies the complete procedure.

\subsection{Rewriting and Evaluation}
\label{sec:search:rewrite}

Given the selected pipeline $P^\star$, the rewriting phase generates a new child pipeline $P'$ by applying a directive. In DocETL-V1~\cite{shankar2025docetl}, optimization enumerates all applicable directives for each operator (or operator prefix) in the pipeline, then invokes an LLM agent to instantiate each one. The exhaustive approach is infeasible for MOAR, because the rewrite space is too large. 
Instead, MOAR delegates the entire rewriting decision to an LLM agent, which can reason about pipeline semantics to choose which directive to apply, which operators within the pipeline to target, and how to instantiate the directive.
Algorithm \ref{alg:rewrite-eval} in \Cref{app:search}\papertext{ in our technical report~\cite{moartechreport}} details the rewriting and evaluation procedure. Unlike selection, rewriting and evaluation are fully parallelized---multiple workers can simultaneously rewrite different selected pipelines and execute them on $D_o$. \techreport{Our implementation uses 3 workers (capped by LLM API rate limits).} We now describe how directives are encoded, how the agent chooses and instantiates them, and how resulting pipeline(s) are evaluated.

\subsubsection{How Directives Are Encoded}
Each rewrite directive is defined by a Python class that encapsulates documentation for the LLM agent and execution logic. The documentation includes:
\begin{itemize}[nosep, leftmargin=*, wide=0pt]
\item \textbf{Name and descriptions.} A unique identifier (e.g., \ttt{code\_sub}), the LHS/RHS pattern (e.g., $\ttt{reduce} \Rightarrow \ttt{code\_reduce} \to \ttt{map}$), and a plain-language explanation.
\item \textbf{Use case guidance.} A natural language explanation of when to apply the directive and what scenarios benefit most from it. \rone{These are manually curated based on experience from our deployment of DocWrangler, an IDE for authoring DocETL pipelines~\cite{shankar2025steering}, and do not overlap with pipelines in our evaluation (\Cref{sec:evaluation}). The full directive documentation is available in the \href{https://github.com/ucbepic/docetl/tree/2bf97c66/docetl/reasoning_optimizer/directives}{\color{blue!60!black}codebase}.}
\item \textbf{Instantiation schema.} A Pydantic model~\cite{pydantic} specifying parameters needed to apply the directive (e.g., \ttt{clarify\_instructions} requires a \ttt{clarified\_prompt}), with optional validators to enforce constraints (e.g., that the new prompt preserves all input variables).
\item \textbf{Example application.} A concrete example showing the original pipeline configuration, the instantiation parameters that would be generated, and the resulting transformed pipeline.
\techreport{\item \textbf{Test cases.} Scenarios specifying an input pipeline, target operators, expected transformation behavior, and whether the test should pass. These tests ensure that LLM agents can understand the directive specifications and instantiate rewrites correctly.}
\end{itemize}

Each directive also implements dynamic execution logic through two methods: \ttt{instantiate()} and \ttt{apply()}. The \ttt{instantiate()} method generates the parameters needed to apply the directive to a target pipeline, described in detail below. The \ttt{apply()} method takes the instantiated parameters (the structured object produced by \ttt{instantiate()}) and the current pipeline, and produces the rewritten pipeline to be parsed and executed by DocETL. 

\subsubsection{Choosing and Instantiating Directives}

Before the agent chooses a directive, we prune the set of directives to filter out redundant or trivial rewrites. First, we prune {\em cycles}---rewrites that reverse a transformation applied earlier on the path from $P_0$ to $P^\star$. 
Specifically, we prune: {\em (i)} a chaining directive immediately followed by a fusion directive, and {\em (ii)} applying a model substitution directive at a first-layer node, which only switches back to previously tried models. 
Second, we prune {\em no-ops}---rewrites that redundantly apply the same type of transformation. 
We prune: {\em (i)} applying a chunking directive to a pipeline that already uses chunking, and {\em (ii)} consecutive compression or summarization directives that attempt to reduce already condensed content.

Our agent-based rewriting proceeds in two steps. First, the agent chooses which directive to apply and which operators to target, seeing only directive names, descriptions, and use case guidance. Second, the agent loads the full instantiation schema and example application to generate concrete parameters. This {\em progressive disclosure}---a technique in user interface design for reducing cognitive load by revealing information gradually~\cite{CarrollRosson1987ProgressiveDisclosure}---avoids overwhelming the agent with details.\techreport{\footnote{Claude Code uses a similar approach for presenting documentation to agents~\cite{ZhangLazukaMurag2025}.}}

\topic{Choosing a Directive}
The agent receives as input (via its prompt):
\begin{itemize}[nosep, leftmargin=*, wide=0pt]
\item Pipeline $P^\star$ (as YAML).
\item For each directive after pruning: its name, description, and use case guidance.
\item A list of all rewrite paths explored so far, along with corresponding $\hat{c}$ and $\hat{a}$. One rewrite path might look like: {\footnotesize \texttt{ROOT $\to$ model\_sub(gpt-4.1) $\to$ doc\_chunking(size=1000) (cost: \$4.07, acc: 0.739)}}.
\item The path of rewrites from $P_0$ to $P^\star$ and its $\mathrm{depth}(P^\star)$.
\item Model and directive statistics, as defined in \Cref{sec:search:init}.
\item An objective---either ``reduce cost while preserving accuracy'' or ``improve accuracy''---determined by $P^\star$'s rank among evaluated pipelines. If $\text{rank}_t(P^\star) \leq |V_t|/2$ (where rank 1 is most accurate), the objective is cost reduction; otherwise, accuracy improvement. Providing different objectives to the agent helps discover a diverse Pareto frontier spanning different accuracy-cost trade-offs.
\end{itemize}

The agent chooses \revision{exactly one} directive $d$ whose LHS matches a subsequence of operators in $P^\star$ and identifies which  operators to rewrite.

\topic{Instantiating the Directive}
Once chosen, the agent loads the directive's full documentation (e.g., instantiation schema, example application) and generates concrete parameters to produce a specific rewrite $P^\star \xrightarrow{r} P'$.
MOAR invokes the directive's \ttt{instantiate()} method. Instantiation is an interactive loop. 
The agent receives: {\em (i)} a system prompt establishing its role (e.g., ``expert at optimizing LLM-powered data processing pipelines''), {\em (ii)} the directive's full documentation (instantiation schema and example application) along with the target pipeline $P^\star$ and target operators, and {\em (iii)} access to sample documents from $D_o$. At each step, the agent can call \ttt{read\_next\_doc()} to inspect samples---e.g., to identify appropriate chunk sizes or detect patterns informing prompt refinements---or output a structured object matching the instantiation schema. 
MOAR validates outputs against the schema; if validation fails, the error is returned for refinement. The loop continues until valid or a retry limit is reached\techreport{ (3 in our implementation)}. Validated parameters are passed to \ttt{apply()}, producing $P'$.
For directives with parameters that are difficult for LLMs to select---e.g., chunk sizes in document splitting---MOAR generates multiple instantiations with different parameter values, evaluates each on $D_o$, and picks the highest-accuracy instantiation.
\rfour{Six of the 18 new directives are parameter-sensitive (marked with $\ddag$ in \Cref{tab:rewrite_directives}); \Cref{sec:detailed-rewrite}\papertext{ in our technical report~\cite{moartechreport}} details the number of candidates for each and references the agent prompts used for directive selection and instantiation.}

\papertext{\subsubsection{Evaluation and Error Handling}

Once a valid rewrite $P'$ is generated, MOAR executes it on $D_o$ to obtain $(\hat{c}(P'), \hat{a}(P'))$, reusing cached measurements for identical pipelines. MOAR handles errors---e.g., rewrites producing invalid YAML or malformed operator configurations, or LLM API failures---by retrying or discarding the pipeline and adjusting visit counts accordingly (details in our technical report~\cite{moartechreport}). Once the budget $B$ is exhausted, MOAR outputs the final Pareto frontier $\hat{\mathcal{F}}_{B}$.}

\techreport{\subsubsection{Evaluation and Error Handling}

Once a valid rewrite $P'$ is generated, MOAR executes it on the evaluation sample $D_o$ to obtain empirical cost and accuracy $(\hat{c}(P'), \hat{a}(P'))$. If an identical pipeline was evaluated previously, MOAR reuses the cached measurements. The measured statistics are recorded in the tree along with the pipeline's depth and parent pointer.

Any agentic query optimizer must handle errors that arise when delegating decisions to LLM agents and evaluating generated pipelines. MOAR encounters three types of errors. 
First, the agent may choose a directive whose applicability signature does not match $P^\star$, or generate a rewrite that cannot be parsed by the DocETL query engine; in both cases, MOAR retries by invoking the agent again. 
Second, LLM API errors may occur during pipeline execution on $D_o$ (e.g., rate limits, service outages); MOAR discards these pipelines without retry since they indicate transient infrastructure issues rather than problems with the rewrite. If a retry also fails, the pipeline is discarded. 
When any pipeline is discarded, MOAR decrements the visit count of $P^\star$ to ensure that failed attempts do not artificially inflate visit counts. 

Once the evaluation budget $B$ has been exhausted, MOAR outputs the final Pareto frontier $\hat{\mathcal{F}}_{B}$ constructed from all evaluated pipelines.}

\section{Evaluation}
\label{sec:evaluation}

We evaluate MOAR across a diverse set of workloads spanning legal, biomedical, government, consumer, and corporate domains. \rone{Compared to baselines,} {\bf \em MOAR  discovers the highest-accuracy pipeline on every workload.} On average, MOAR achieves 27\% higher accuracy than the next-best optimizer, ABACUS~\cite{russo2025abacus}, while matching its best accuracy at only 54.5\% of its cost.
We first describe our setup (\Cref{sec:setup}). We then present the accuracy improvements and cost savings achieved by MOAR (\Cref{sec:evaluation:results}). Finally, we examine the characteristics of high-accuracy pipelines discovered by MOAR (\Cref{sec:evaluation-insights}). Our experiment artifacts are released at \href{https://drive.google.com/drive/folders/1Meebo7J8bhHg9b1MOP14MjQorx0OgIcF?usp=drive_link}{\color{blue!60!black}this Google Drive link}. \rfour{The full agent prompts used for directive selection and instantiation are available in our codebase: the \href{https://github.com/ucbepic/docetl/blob/2bf97c66/docetl/reasoning_optimizer/agent.py}{\color{blue!60!black}main agent logic} and \href{https://github.com/ucbepic/docetl/tree/2bf97c66/docetl/reasoning_optimizer/directives}{\color{blue!60!black}individual directive prompts}.}

\subsection{Setup}
\label{sec:setup}

We run MOAR with a budget of 40 pipeline evaluations per workload, using \ttt{gpt-5} as the agent for instantiating rewrite directives. MOAR selects models from a pool of 11 LLMs, including \ttt{gpt-4o-mini}, \ttt{gpt-4o}, 3 \ttt{gpt-4.1} variants, 3 \ttt{gpt-5} variants, and 3 \ttt{gemini-2.5} variants. All models are available to all optimizers. Unless otherwise noted, the user-specified or initial pipelines prior to optimization use \ttt{gpt-4o-mini}, as in prior work~\cite{russo2025abacus, shankar2025docetl, patel2025semantic}.

DocETL comprises over 30,000 lines of Python code, of which MOAR accounts for approximately 16,000 lines. The directive library alone requires over 9,000 lines (each directive requires 300--600 lines). We use LiteLLM as a wrapper around Google Gemini and Azure OpenAI APIs. All experiments were executed on Modal, a cloud computing platform.

\subsubsection{Baselines}
\label{subsec:baselines}

\rfour{We compare MOAR against 4 baseline systems and 1 ablation.}
We compare against open-source systems that support a semantic \ttt{map} operator.
For each baseline system, we express the pipeline using the baseline's interface, minimally modifying the DocETL operators so the pipeline can be parsed by the baseline system.
All baselines except DocETL-V1 support selecting from multiple models; DocETL-V1 uses only \ttt{gpt-4o-mini}.

\begin{itemize}[nosep, leftmargin=*, wide=0pt]
  \item \textbf{DocETL-V1~\cite{shankar2025docetl}.}
We run the original DocETL optimizer, which optimizes only for accuracy and returns a single plan. Unlike all other baselines, DocETL-V1 does not access ground-truth labels; it relies solely on an LLM judge to select plans.
  \item \textbf{LOTUS~\cite{patel2025semantic}.}
  We express each pipeline using LOTUS's semantic operators. LOTUS does not support structured output schemas, so we minimally edit prompts to request JSON-formatted responses and write custom Python code to parse LOTUS outputs. LOTUS performs cost reduction for filters, joins, and group-bys by swapping in cheaper models (i.e., \ttt{gpt-5-nano}). Like DocETL-V1, LOTUS always returns a single optimized plan.
  \item \textbf{Palimpzest/ABACUS~\cite{liu2024declarative, russo2025abacus}.}
  We express each pipeline using Palimpzest (PZ)'s~\cite{liu2024declarative} operators and use its ABACUS optimizer~\cite{russo2025abacus}. ABACUS does not directly return a Pareto frontier; instead, it allows users to specify a cost budget and returns the maximum-accuracy plan within that budget.
  Following discussion with the authors, we construct a Pareto frontier by running PZ with: (i) no budget constraint (to obtain the highest-accuracy plan), and (ii) budgets set to 50\%, 25\%, and 10\% of the unconstrained plan's cost. As of writing this paper (October 15, 2025), PZ does not support LLM-powered reduce operators, so we omit it from workloads requiring the DocETL \ttt{reduce} operator.
  \item \textbf{Simple agent (SA).}  \rfour{We test whether an LLM agent, that can use DocETL as a tool and read DocETL-V1's rewrite directives in the codebase, can discover effective pipelines.} We provide a \ttt{gpt-5} agent with three tools: (i) reading sample documents, (ii) reading the original\techreport{ DocETL} paper~\cite{shankar2025docetl}\rfour{, codebase,} and \href{https://ucbepic.github.io/docetl/}{\color{blue!60!black}DocETL documentation}, and (iii) executing any pipeline on samples and observing its accuracy and cost. The agent proposes rewrites until its context window is exhausted ($\approx$400k tokens) or it calls a ``done'' tool to indicate completion. Then, of {\em all} pipelines generated by the agent, we retain the Pareto frontier.
  \item \rfour{\textbf{MOAR (no search).} Here, we run SA but also provide descriptions of all 18 new MOAR rewrite directives---to serve as a true ablation of MOAR's search algorithm. As with SA, we retain the Pareto frontier over all pipelines produced.}
\end{itemize}

\techreport{PZ and LOTUS each accept a \ttt{num\_samples} parameter for optimization (PZ defaults to 100, LOTUS to 200); we set both to 200 to provide each system ample budget for exploration (unless mentioned otherwise in \Cref{sec:evaluation-workloads}).}

\subsubsection{Workloads}
\label{sec:evaluation-workloads}

\techreport{
We evaluate across six workloads. We take text processing workloads identified in prior published work~\cite{shankar2025docetl, patel2025semantic, russo2025abacus}, and introduce two new workloads from medical analysis and enterprise sustainability domains, inspired by real DocETL users. The Sustainability workload is inspired by the \href{https://www.climateintelligenceservice.scot/}{\color{blue!60!black}Scottish Climate Intelligence Service}, who uses DocETL to identify common sustainability initiatives across organizations and regional authorities. The Medec workload is inspired by \href{https://www.ankihub.net/}{\color{blue!60!black}AnkiHub}, who builds study tools for medical students (e.g., personalized quizzes and automated error detection in their reasoning). For privacy, we use public datasets with similar task structures rather than proprietary user data.

For each workload, we describe the task, initial pipeline, and metric. The initial pipeline for each workload uses a minimal number of semantic operators, representing what a user would naturally write when first encountering the problem~\cite{shankar2025steering}. Each workload uses a dataset $D$ sampled from larger source datasets, split into $D_o$ (40 documents per workload, for optimization) and a held-out test set $D_T = D \setminus D_o$ (100 documents per workload).
Optimizers search over candidate pipelines using $D_o$, and all reported results reflect execution on the held-out test set $D_T$.

\begin{itemize}[nosep, leftmargin=*, wide=0pt]
\item {\bf CUAD (Legal Analysis)~\cite{hendrycks2021cuad}.} This dataset contains 510 legal contracts (each averaging 7,727 words) annotated with 41 clause categories, used as a benchmark in prior work~\cite{shankar2025docetl, russo2025abacus}. The task is to extract text spans for each clause type present in a contract. The initial pipeline consists of a single \ttt{map} that prompts for all clause types at once and outputs a list of \{\ttt{clause\_type}, \ttt{text\_span}\} objects. For PZ, we use the pipeline provided by the ABACUS authors, doubling the number of samples (as mentioned in \Cref{subsec:baselines}) to aid PZ's exploration. The metric is F1 score, counting an extraction as correct if the clause type matches and the span has Jaccard similarity $>0.15$ with the ground truth.

\item {\bf Game Reviews.} This dataset, from prior work~\cite{shankar2025docetl}, involves documents from the Steam reviews dataset~\cite{sobkowicz2016steam}, where each document contains 300 reviews for a single game (averaging 97,696 words). The task is to identify ten positive and ten negative reviews per game, presented in chronological order. The initial pipeline, as in ~\cite{shankar2025docetl}, is a single \ttt{map} over each document that attempts to extract and order the required reviews directly. The metric is an average of hallucination rate (fraction of extracted reviews not present in the source text), sentiment accuracy (agreement with ground-truth ratings for non-hallucinated reviews), and Kendall's $\tau$ between predicted and correct orderings. Since documents exceed context windows of most LLMs, we use only
\ttt{gpt-4.1} and \ttt{gemini} series models, with \ttt{gpt-4.1-mini} as default.
We note that PZ crashed during optimization with a 200-sample budget ($\sim$4 hours in), so we reduced PZ's budget to 50 samples.

\item {\bf BlackVault (Declassified Articles).} This dataset from~\cite{shankar2025docetl} contains 733 articles (each averaging 7,351 words) describing international paranormal events. The task is to classify each document's event type and aggregate distinct location mentions across documents of the same type. The initial pipeline, as in \cite{shankar2025docetl}, has two operators: a \ttt{map} that extracts an event type per article (e.g., UFO sighting), followed by a \ttt{reduce} that aggregates locations by type. The metric is average recall of distinct locations per event type, normalized by the maximum recall achieved across all methods in \Cref{subsec:baselines} (to get a score between 0 and 1).

\item {\bf Biodex (Biomedical Classification).} This benchmark~\cite{d2023biodex} evaluates biomedical paper classification (each averaging 71,151 words). Each paper must be linked to the adverse drug reactions it discusses, drawn from over 24,000. Prior work expressed this task using different pipelines (code provided by the respective paper authors). LOTUS-r\&r (retrieve-and-rerank) implements a join (to find all reactions for each document) followed by a semantic aggregation (to group by document and rerank reactions with an LLM)~\cite{patel2025semantic}. PZ-r\&r implements a map, retrieve, and map pipeline~\cite{russo2025abacus}. However, now that LLMs support sufficiently long context windows (all models in our pool exceed 128k tokens), we implement the simplest initial pipeline for MOAR: a single \ttt{map} operation where the prompt contains the full list of reactions and the output is a ranked list of reactions relevant to the document.
For fair comparison, we also express this single-map initial pipeline in LOTUS and PZ, called LOTUS-d and PZ-d (direct). We report results for both pipeline formulations (d and r\&r) for LOTUS and PZ, taking the union of plans found for their Pareto frontiers. The metric is rank-precision@5 (RP@5), measuring how often the correct reactions appear among the top-5 predictions.

\item {\bf Medec (Medical Error Detection and Correction); new.}  The Medec dataset~\cite{abacha2024medec} contains 3,848 clinical notes (each averaging 147 words) with labeled medical errors across five categories. The task is to detect whether an error is present in each note, identify the sentence containing the error, and generate a corrected version. The initial pipeline is a single \ttt{map} that produces three outputs: \ttt{error\_flag}, \ttt{error\_sentence}, and \ttt{corrected\_sentence}. The metric is an average of the error-detection F1 scores and the corrected sentence Jaccard similarities with the reference corrections.

\item {\bf Sustainability (Corporate ESG Reports); new.} This dataset~\cite{datanneed2024sustainability} contains 5,436 enterprise reports (each averaging 38,668 words) spanning annual reports, sustainability reports, financial reports, and others. The task is to (i) filter to retain only sustainability reports, (ii) classify each company's economic sector (e.g., health, real estate), and (iii) for each sector, produce a summary listing each company and its key sustainability initiatives (e.g., for the technology sector: ``Apple: carbon neutrality by 2030; Microsoft: 100\% renewable energy; Google: water replenishment programs''). The initial pipeline applies a \ttt{filter} to select sustainability reports, a \ttt{map} to classify the economic sector, and a \ttt{reduce} to group reports by sector and generate a summary per sector containing all companies and their initiatives. The metric is the average of sector classification accuracy (the fraction of reports assigned to the correct sector) and company name accuracy (the fraction of company names in the per-sector summaries that match ground-truth companies for that sector).
\end{itemize}
}

\begin{table}[t]
\vspace{-5pt}
\centering
\footnotesize
\setlength{\tabcolsep}{2pt}
\caption{Best accuracy by method. Highest per workload is bolded; next-best is underlined. Last row shows MOAR's average relative gain. For Biodex, LOTUS and PZ include two variants (d, r\&r). \rfour{For BlackVault, the first row reports raw recall values, and the second row reports normalized recall.}}
\vspace{-10pt}
\label{tab:results}
\begin{tabular}{l C{1.35cm} cccccc C{2cm}}
\toprule
\textbf{Workload} & \textbf{DocETL-V1} & \textbf{LOTUS} & \textbf{PZ} & \textbf{SA} & \rfour{\makecell{\textbf{MOAR} \\ \textbf{(no search)}}} & \textbf{MOAR}  \\
\midrule
CUAD            & 0.471 & 0.402 & \underline{0.694} & 0.521 & \rfour{0.495} & \textbf{0.762} \\
Game Reviews    & 0.608 & 0.522 & \underline{0.683} & 0.467 & \rfour{0.703} & \textbf{0.804} \\

BlackVault      & 5.339 & 3.020 & -- & 7.230 & \rfour{4.055} & \textbf{37.333} \\
(normalized)    & (0.143) & (0.081) & -- & \underline{(0.194)} & \rfour{(0.109)} & \textbf{(1.000)} \\

Biodex & 0.247 &
\makecell[l]{0.260 {\scriptsize(d)}\\ 0.202 {\scriptsize(r\&r)}} &
\makecell[l]{0.260 {\scriptsize(d)}\\ 0.296 {\scriptsize(r\&r)}} &
\underline{0.333} & \rfour{0.259} & \textbf{0.370} \\
Medec           & 0.534 & 0.538 & 0.536 & 0.726 & \rfour{\textbf{0.755}} & \rfour{\underline{0.742}} \\
Sustainability  & \underline{0.632} & 0.516 & -- & 0.543 & \rfour{0.560} & \textbf{0.646} \\
\midrule
\textbf{Avg Gain (\%)}
                & +130.71\% & +209.53\% & +26.65\% & +94.36\% & \rfour{+157.04\%} & -- \\
\bottomrule
\end{tabular}
\end{table}

\subsection{Results}
\label{sec:evaluation:results}

We present main results on accuracy (\Cref{sec:evaluation:sb:acc}), cost savings and Pareto frontier quality (\Cref{sec:evaluation:sb:cost}), and optimization overhead (\Cref{sec:evaluation:sb:overhead}), and additional experiments and ablations in \Cref{sec:evaluation:sb:additional}.

\subsubsection{Accuracy Improvement}
\label{sec:evaluation:sb:acc}

\Cref{tab:results} summarizes MOAR's accuracy improvements across all baselines. {\bf \em MOAR achieves the highest accuracy on every workload among all baselines, with gains over the next most accurate baseline ranging from +2.2\% (Medec, Sustainability) to +415.5\% (BlackVault).} MOAR's pipelines are also structurally more complex than those produced by the baselines. {\bf \em MOAR's highest-accuracy pipelines use, on average, 2.3$\times$ as many operators as the baseline pipelines.}
\rfour{MOAR achieves the highest accuracy on every workload except Medec, where MOAR (no search) edges it out (0.755 vs.\ 0.742) by applying five rewrites: four of our predefined directives (model substitution, code-based document compression, gleaning, and prompt rewriting) and one the agent wrote from scratch (a code map to programmatically clean the output). SA also performs well by rewriting the prompt and switching to \ttt{gpt-5}. Just switching to \ttt{gpt-5} achieves 8\% lower accuracy at 1.25$\times$ the cost and does not appear on the Pareto frontier.\papertext{ On Biodex, RP@5 is low across all methods; an error analysis in our technical report~\cite{moartechreport} finds that most of the gap is attributable to errors in the ground truth rather than model performance.}}

\rfour{\techreport{On Biodex, RP@5 is low across all methods. To understand whether further improvement is possible, we classified all mismatches from MOAR's best pipeline on 50 sampled documents using an LLM judge (\ttt{gpt-5}) that read each source biomedical paper and determined whether each mismatch reflects a model error or an error in the ground truth. For each false positive (predicted by MOAR but absent from the ground truth), the judge classified the prediction as either a clinically relevant ADR described in the paper but missing from the ground truth, or a true model error. For each false negative (in the ground truth but not in MOAR's top-5), the judge classified whether the label is explicitly described in the paper, only implicitly inferable, or not present in the paper at all. We manually verified a random subset of 50 judge outputs and found 100\% agreement; all judge outputs are available at \href{https://ucbepic--biodex-labeling-web.modal.run/}{\color{blue!60!black}this interactive web app}. Of the 192 false positives, only 8 (4.2\%) are true model errors; the rest are clinically relevant reactions described in the paper but absent from the ground truth. Of the 132 false negatives, 37 (28\%) are ground truth labels not found in the source paper. Correcting the ground truth raises RP@5 from 37.9\% to 97.5\%, suggesting that most of the gap is attributable to errors in the ground truth rather than model performance.}}

\begin{figure}
    \centering
    \vspace{-15pt}
    \captionsetup[subfigure]{skip=0pt}
    \setlength{\abovecaptionskip}{2pt}
    \setlength{\belowcaptionskip}{0pt}
    
    \begin{subfigure}{0.49\columnwidth}
        \centering
        \includegraphics[width=\linewidth]{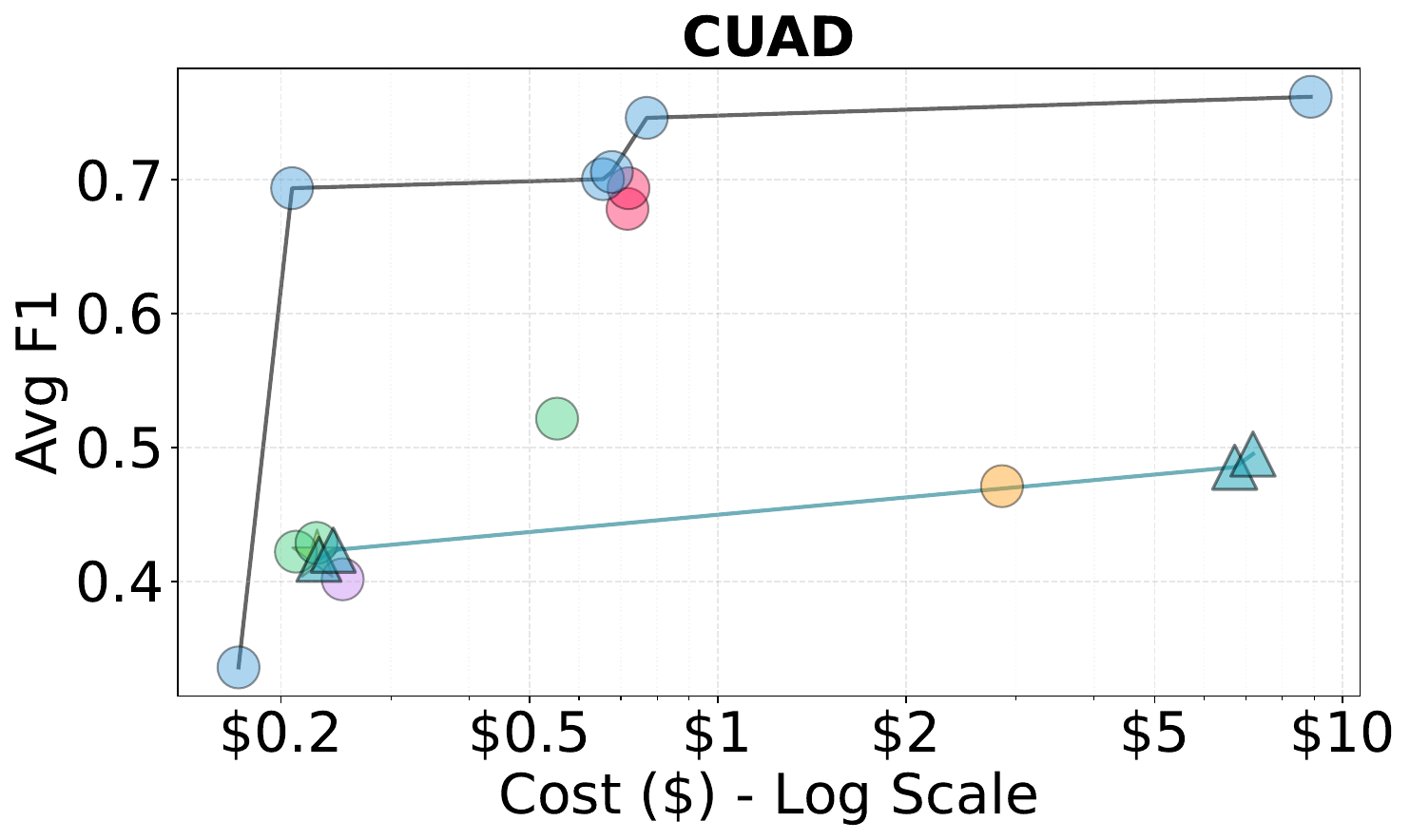}
    \end{subfigure}%
    \begin{subfigure}{0.49\columnwidth}
        \centering
        \includegraphics[width=\linewidth]{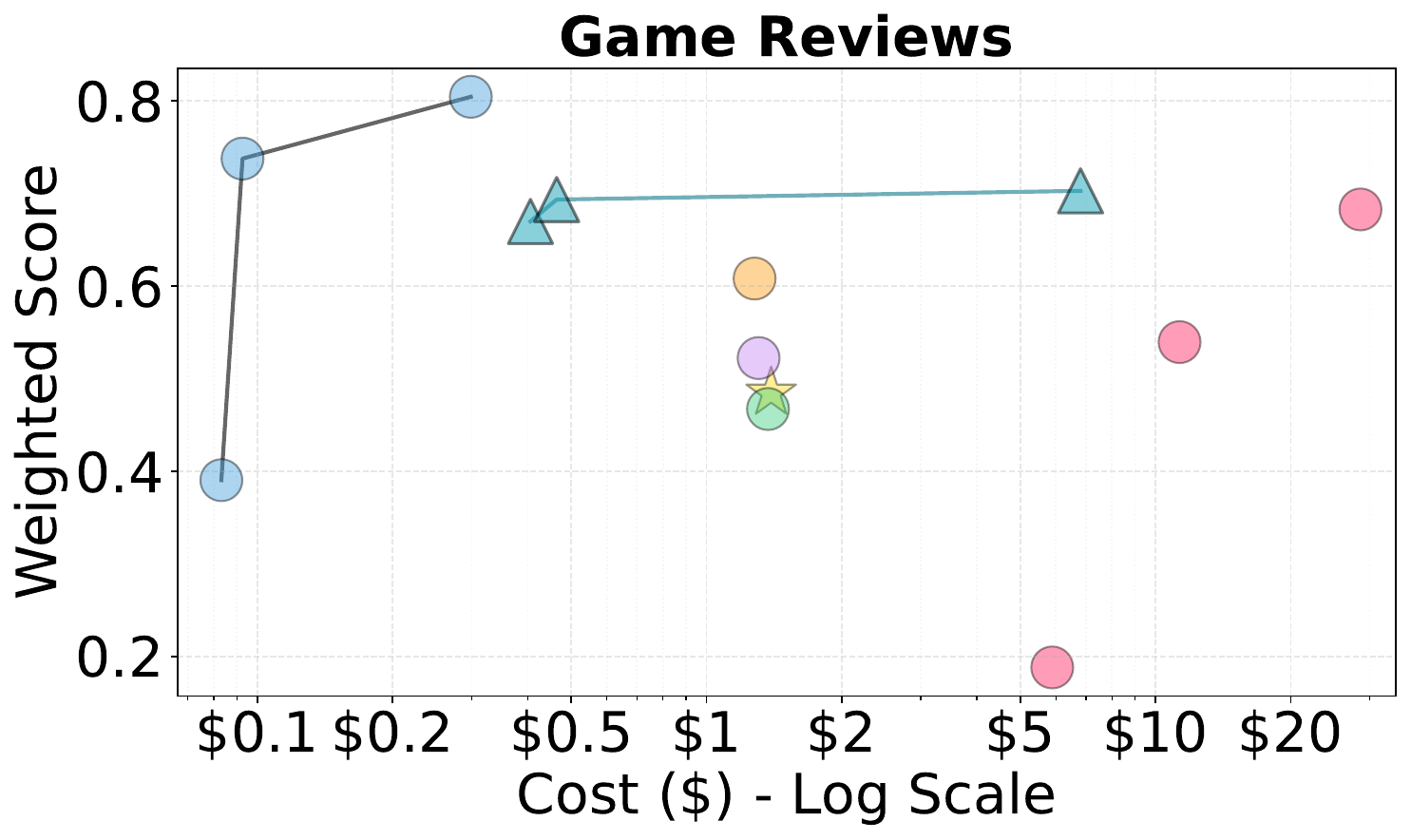}
    \end{subfigure}\\[-1mm]
    \begin{subfigure}{0.49\columnwidth}
        \centering
        \includegraphics[width=\linewidth]{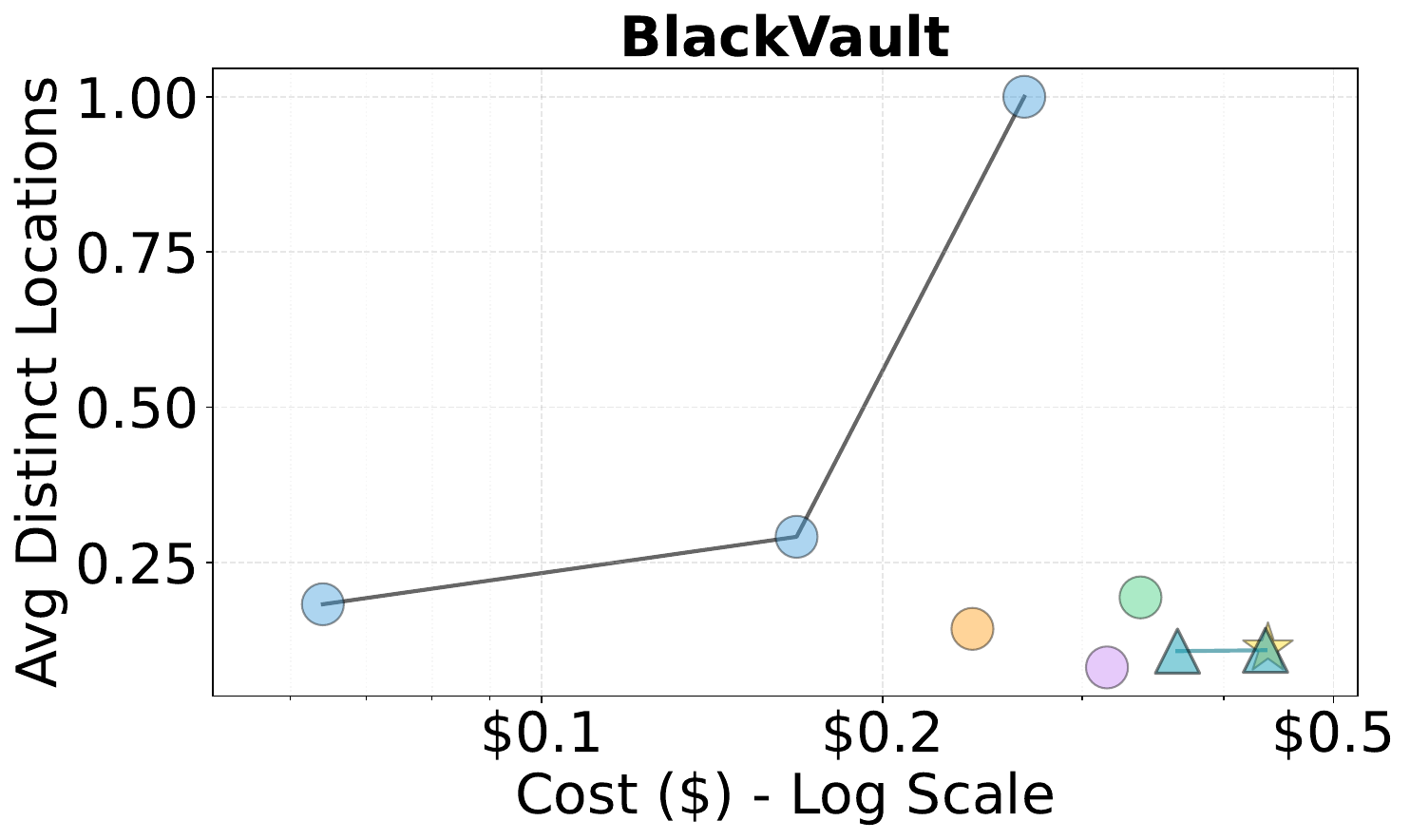}
    \end{subfigure}%
    \begin{subfigure}{0.49\columnwidth}
        \centering
        \includegraphics[width=\linewidth]{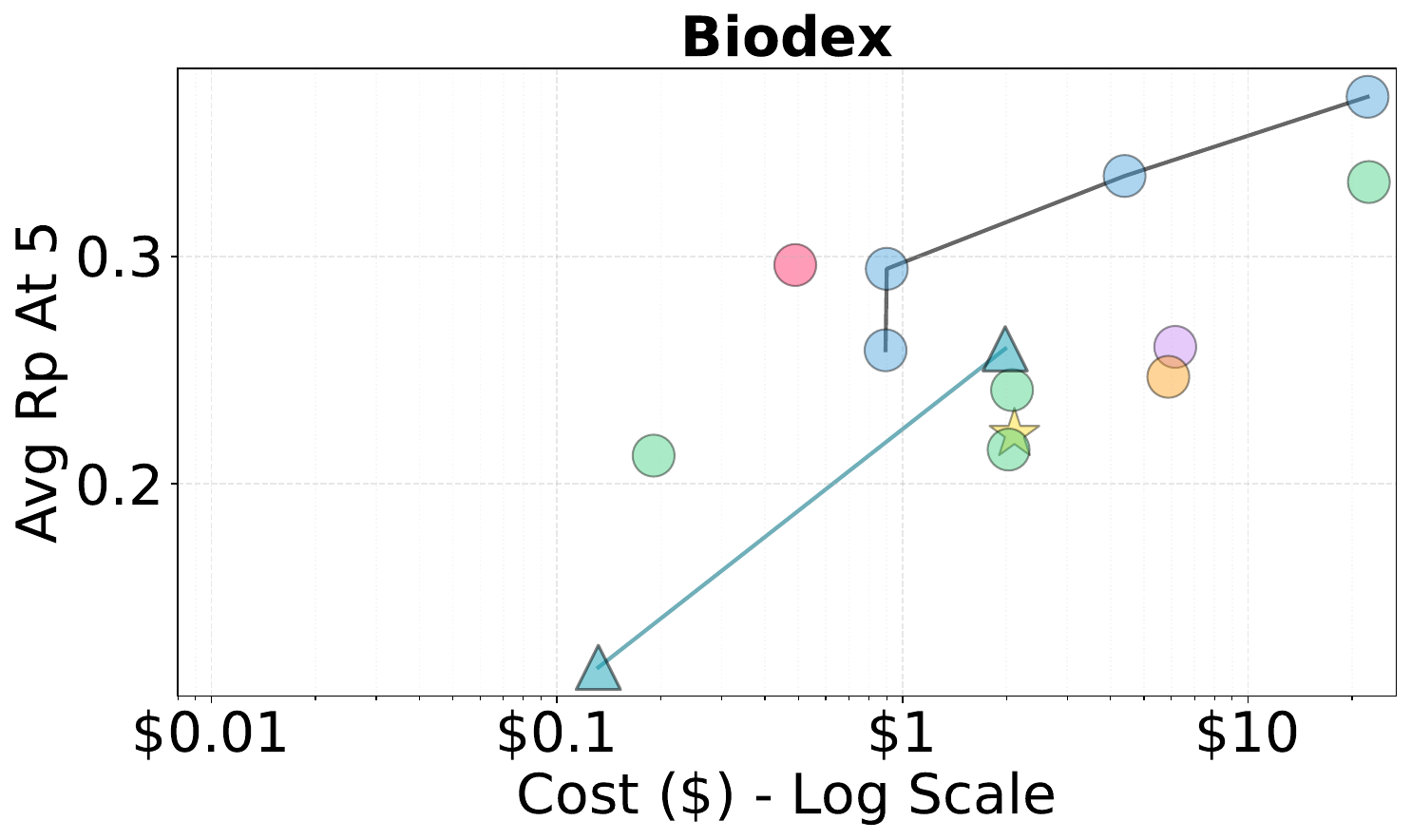}
    \end{subfigure}\\[-1mm]
    \begin{subfigure}{0.49\columnwidth}
        \centering
        \includegraphics[width=\linewidth]{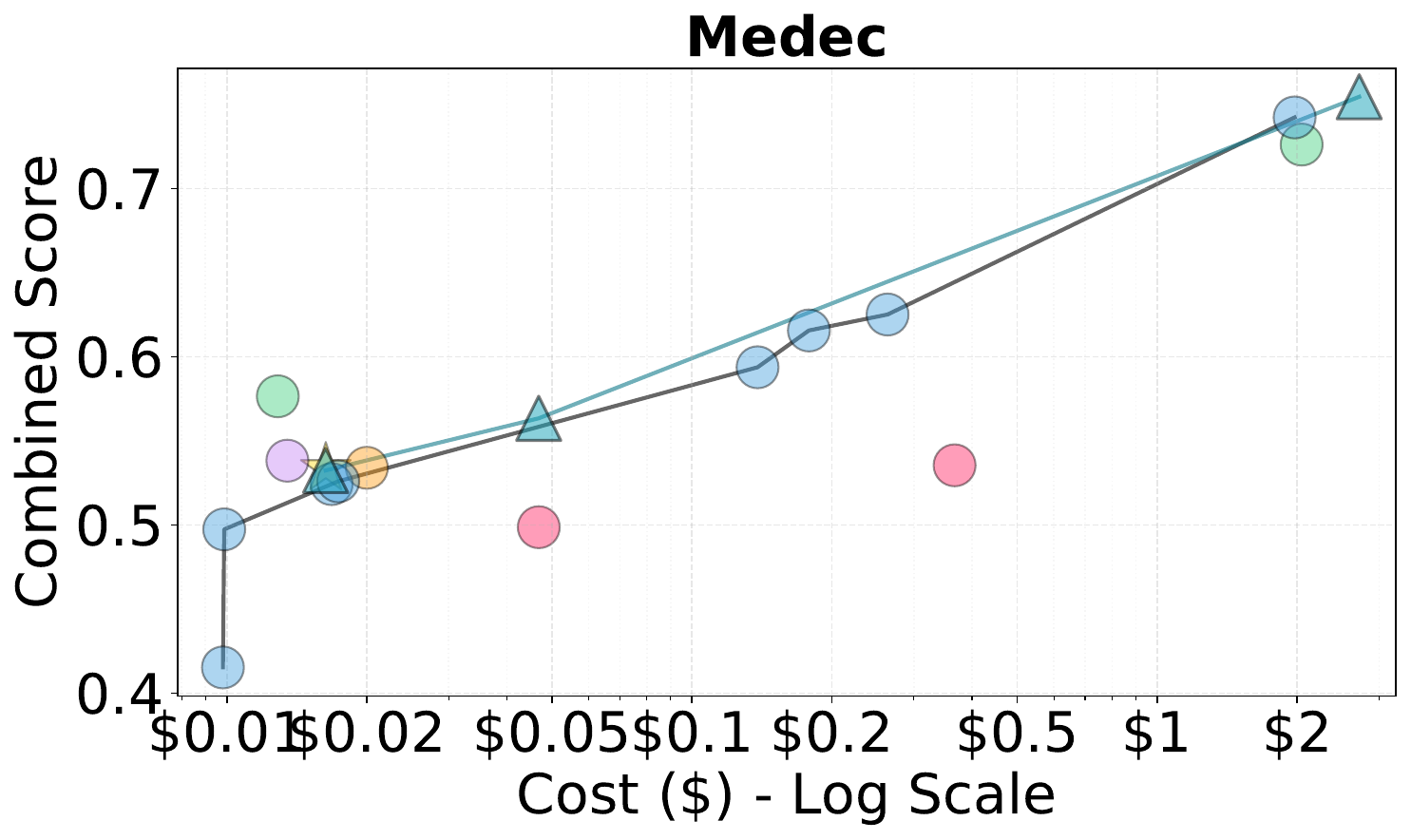}
    \end{subfigure}%
    \begin{subfigure}{0.49\columnwidth}
        \centering
        \includegraphics[width=\linewidth]{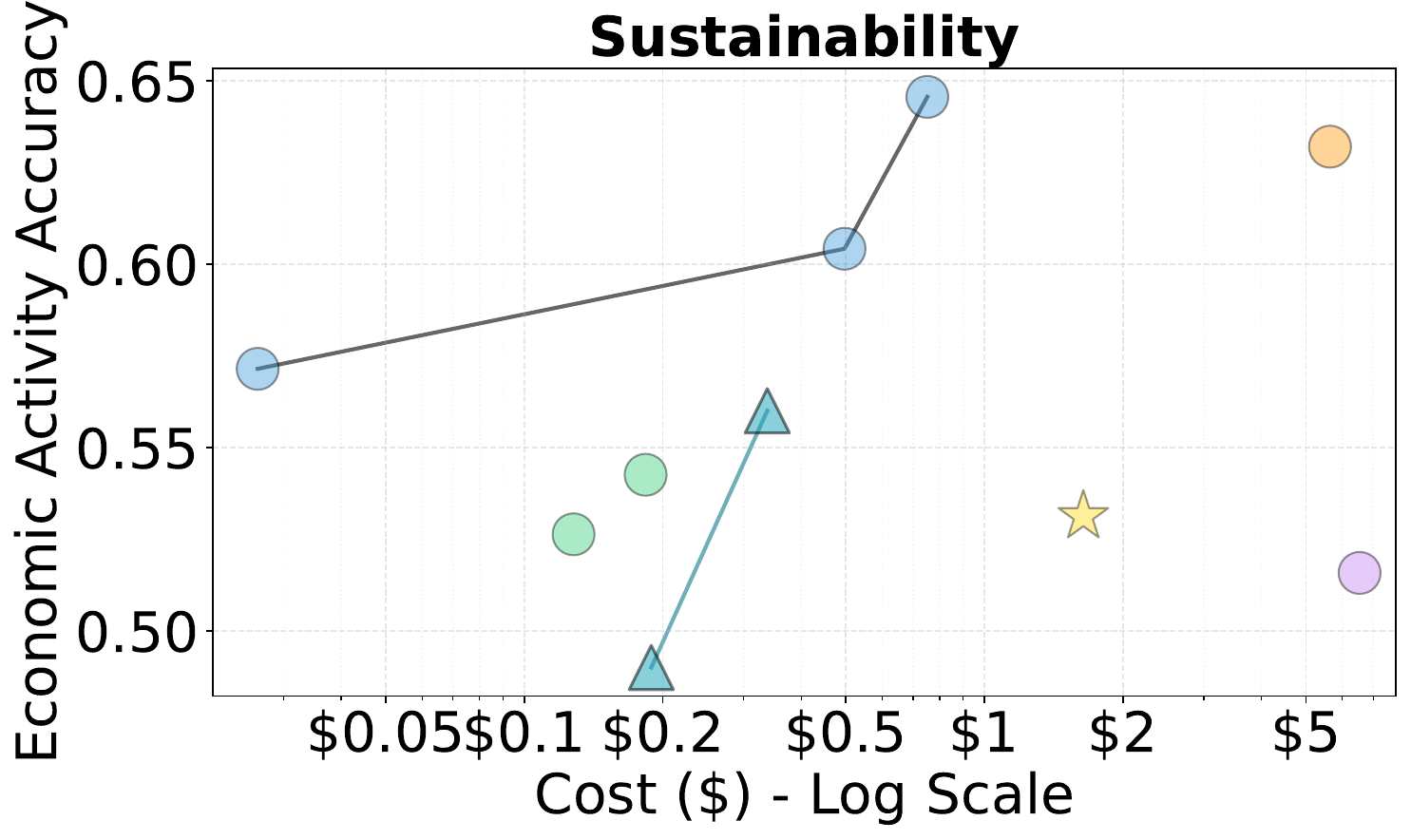}
    \end{subfigure}
    \includegraphics[width=\columnwidth]{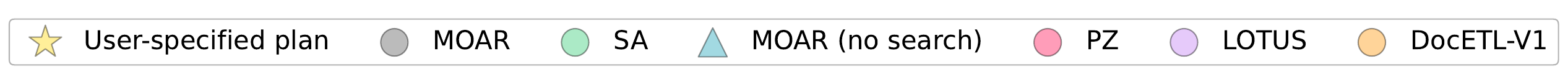}
    \vspace{-10pt}
    \caption{\revision{Pareto frontiers for each method, for each held-out test set. ``User-specified plan'' is the initial pipeline authored by the user.}}
    \vspace{-20pt}
    \label{fig:frontier}
\end{figure}

\subsubsection{Cost Savings and Pareto Frontier Quality}
\label{sec:evaluation:sb:cost}

\Cref{tab:cost-savings} shows MOAR's cost to match each baseline's highest accuracy.
\Cref{app:results}\papertext{ in our technical report~\cite{moartechreport}} provides pairwise cost comparisons between all methods. 
On average, MOAR finds plans that {\bf \em match the simple agent's best accuracy with $0.436\times$ the cost, LOTUS's with $0.487\times$, PZ's with $0.545\times$, and DocETL-V1's with $0.256\times$ the cost}. The largest cost savings occur on Game Reviews, where MOAR achieves PZ's highest accuracy with $0.003\times$ the cost.

There are two cases where MOAR does not achieve cost savings for top-accuracy baseline plans. First, for Medec, MOAR costs $1.245\times$ LOTUS to match its accuracy. However, LOTUS's accuracy on Medec is 53.8\%---nearly 20 percentage points worse than MOAR's maximum (74.2\%)---so returning plans at such low accuracy may not be useful in practice.
Second, for Biodex, MOAR costs $1.840\times$ to match PZ-r\&r's accuracy. 
However, MOAR still finds a plan that is more accurate, albeit more expensive.

Biodex also illustrates how different logical plans yield unpredictable performance: LOTUS-d (direct) outperforms LOTUS-r\&r (retrieve-and-rerank), while PZ-r\&r outperforms PZ-d. MOAR discovers its highest-accuracy pipeline through chunking and sampling rewrites, which resembles PZ-r\&r's retrieval-based approach. Interestingly, the simple agent achieves the second-best accuracy on Biodex by simply using \ttt{gpt-5} on the single-map pipeline, outperforming all LOTUS and PZ variants.

Finally, \Cref{fig:frontier} shows the Pareto frontier for each method. MOAR completely dominates all other methods\rfour{, and the MOAR (no search) ablation,} on CUAD, Game Reviews, BlackVault, and Sustainability. On Biodex and Medec, only two baseline pipelines in each of these two workloads are not dominated. For Medec, both non-dominated pipelines achieve lower accuracy than the original user-specified pipeline, limiting their practicality.
For Biodex, only one baseline pipeline (PZ's highest-accuracy plan) is not dominated by MOAR.

\subsubsection{Pareto Optimization Overhead}
\label{sec:evaluation:sb:overhead}

\revision{In \Cref{tab:overhead}, we report optimization costs and latencies, as well as expected execution costs at 1-million document scale.} MOAR and ABACUS (PZ) are the only optimizers that construct a Pareto frontier. Relative to PZ, MOAR discovers substantially more accurate pipelines (27\% higher accuracy on average) and more cost-efficient pipelines (achieving PZ's accuracy at 0.545$\times$ the cost). This comes with a higher optimization cost: MOAR's optimization cost is 2.11$\times$ that of ABACUS, though its optimization latency is only 0.562$\times$ that of ABACUS. 
Users are often willing to incur substantial optimization costs to achieve highest accuracies~\cite{khattab2024dspy}. 
Moreover, since optimization is a one-time cost that amortizes over repeated pipeline executions, a high cost can be acceptable if it lowers execution costs at scale~\cite{aggarwal2025cortex, Shankar2026TaskCascades}.

\subsubsection{Additional Experiments}
\label{sec:evaluation:sb:additional}

\revision{We conduct four additional experiments, with full details in \Cref{app:results}\papertext{ of our technical report~\cite{moartechreport}}; we summarize the results here.}

\begin{itemize}[nosep, leftmargin=*, wide=0pt]
\item \rfour{To determine whether MOAR's gains come primarily from selecting better models, we run the user-specified pipeline with every model in our pool and construct the resulting Pareto frontier for each workload. MOAR achieves higher accuracy on every workload: on Game Reviews, BlackVault, and Sustainability, MOAR's best accuracy is 1.3--8.6$\times$ higher; on CUAD, Biodex, and Medec, the accuracy is within five percentage points. Regarding cost, MOAR matches the model-substitution frontier's best accuracy at 0.009--0.435$\times$ the cost on four workloads, and still at lower cost on the remaining two (Biodex, Medec).}

\item \rfour{To understand how much the new search algorithm and the expanded directive library each contribute to MOAR's performance, we evaluate MOAR with only DocETL-V1's directives (which we call MOAR-V1; see \Cref{fig:directive-library}\papertext{ in our technical report~\cite{moartechreport}}). On four of six workloads, MOAR produces strictly better Pareto frontiers than MOAR-V1; on the remaining two, MOAR-V1 dominates DocETL-V1. Overall, the search algorithm and the expanded library are both necessary for good performance on all workloads.}

\item \rtwo{To evaluate whether open-source LLMs can replace gpt-5 as the optimization agent, we evaluate four open-source LLMs on CUAD (see \Cref{fig:cuad-os}\papertext{ in our technical report~\cite{moartechreport}}) and find that two (Kimi K2 Thinking~\cite{kimi_k2}, Llama-4-Maverick~\cite{llama4}) produce comparable Pareto frontiers while two smaller LLMs fail due to insufficient context or instruction-following capabilities.}

\item \rtwo{To understand whether accuracies on the optimization sample generalize, we report accuracy differences between $D_o$ and $D_T$ across all Pareto-optimal pipelines (see \Cref{tab:generalization}\papertext{ in our technical report~\cite{moartechreport}}); the average difference is less than 5 percentage points.}
\end{itemize}

\begin{table}
\centering
\footnotesize
\vspace{-15pt}
\caption{\rfour{Cost of the cheapest MOAR plan that matches or exceeds each baseline's best accuracy, as a multiple of that baseline's cost. ``--'' denotes that the baseline does not reach the original accuracy; ``n/a'' that it is not evaluated on the workload. Reported savings exclude one-time optimization costs.}}
\vspace{-10pt}
\label{tab:cost-savings}
\begin{tabular}{lcccccc}
\toprule
\textbf{Workload} & \textbf{DocETL-V1} & \textbf{LOTUS} & \textbf{PZ} & \textbf{SA} & \rfour{\makecell{\textbf{MOAR} \\ \textbf{(no search)}}}\\
\midrule
CUAD & \cellcolor{green!50}0.073$\times$ & — & \cellcolor{green!40}0.290$\times$ & \cellcolor{green!35}0.377$\times$ & \cellcolor{green!55} \rfour{0.029x} \\
Game Reviews & \cellcolor{green!50}0.072$\times$ & \cellcolor{green!50}0.071$\times$ & \cellcolor{green!65}0.003$\times$ & — & \cellcolor{green!55} \rfour{0.014x} \\
BlackVault & \cellcolor{green!40}0.267$\times$ & — & n/a & \cellcolor{green!40}0.497$\times$ & \rfour{—} \\
Biodex & \cellcolor{green!45}0.152$\times$ & \cellcolor{green!45}0.145$\times$ & \cellcolor{red!20}1.840$\times$ & \cellcolor{green!40}0.196$\times$ & \cellcolor{green!35}\rfour{0.451x} \\
Medec & \cellcolor{green!20}0.840$\times$ & \cellcolor{red!20}1.245$\times$ & \cellcolor{green!50}0.046$\times$ & \cellcolor{green!15}0.966$\times$ & \cellcolor{green!10}\rfour{0.999x} \\
Sustainability & \cellcolor{green!40}0.133$\times$ & — & n/a & \cellcolor{green!40}0.143$\times$ & \cellcolor{green!50} \rfour{0.078x} \\
\bottomrule
\end{tabular}
\end{table}

\begin{table}[t]
\vspace{-10pt}
\caption{\revision{Optimization Overhead and Expected Execution Costs. MOAR's cost multiple relative to the initial plan ($\times$ Init.) is shown in parentheses. Expected execution costs are computed by scaling the most accurate pipeline's cost to the full dataset and to 1M documents. ``--'' indicates the method is not evaluated on that dataset.}}
\label{tab:overhead}
\centering
\scriptsize
\setlength{\tabcolsep}{2pt}
\renewcommand{\arraystretch}{0.9}
\vspace{-10pt}
\begin{tabular}{p{0.7cm}cccc ccc cc}
\toprule
& \multicolumn{4}{c}{\textbf{Opt. Cost (\$)}}
& \multicolumn{3}{c}{\textbf{Opt. Latency (s)}}
& \multicolumn{2}{c}{\textbf{Exec. Cost (\$)}} \\
\cmidrule(lr){2-5}\cmidrule(lr){6-8}\cmidrule(lr){9-10}
& \textbf{MOAR}
& \textbf{SA}
& \textbf{PZ}
& \makecell{\textbf{DocETL} \\ \textbf{V1}}
& \textbf{MOAR}
& \textbf{SA}
& \textbf{PZ}
& \makecell{\textbf{Full} \\ \textbf{Data}}
& \textbf{1M} \\
\midrule
CUAD & \makecell{44.0 \\ {\tiny (36.0$\times$)}} & 0.34 & 54.9 & 1.58 & 5,536 & 157 & 11,579 & 45.3 & 88.9k \\
\makecell[l]{Game\\Reviews} & \makecell{57.3 \\ {\tiny (44.0$\times$)}} & 0.29 & 28.8 & 6.60 & 8,382 & 2,502 & 19,718 & 28.0 & 2.99k \\
\makecell[l]{Black-\\Vault} & \makecell{35.4 \\ {\tiny (11.0$\times$)}} & 0.24 & -- & 1.84 & 2,353 & 1,092 & -- & 1.95 & 2.67k \\
Biodex & \makecell{188 \\ {\tiny (0.45$\times$)}} & 8.94
& \makecell{93.2 (d) \\ 6.26 (r\&r)}
& 14.4
& 3,908 & 756
& \makecell{9,711 (d) \\ 1,768 (r\&r)}
& 4.31k & 222k \\
Medec & \makecell{16.7 \\ {\tiny (21.7$\times$)}} & 0.27 & 4.59 & 0.01 & 2,889 & 166 & 3,066 & 72.9 & 19.0k \\
\makecell[l]{Sustain-\\ability} & \makecell{79.2 \\ {\tiny (0.88$\times$)}} & 2.59 & -- & 43.7 & 2,070 & 547 & -- & 40.9 & 7.52k \\
\bottomrule
\end{tabular}
\papertext{\vspace{-15pt}}
\end{table}

\newcommand{\yes}{\cellcolor{gray!15}{\checkmark}}

\techreport{\begin{table}[t]
\centering
\footnotesize
\papertext{\vspace{-10pt}}
\caption{Model usage across 29 top-accuracy pipelines. Task types: Ext. = extraction, Class. = classification, Summ. = summarization.}
\label{tab:model-task}
\vspace{-10pt}
\begin{tabular}{l ccc cc c}
\toprule
& \multicolumn{3}{c}{\textbf{Task Type}} & \multicolumn{2}{c}{\textbf{Doc Length}} & \\
\cmidrule(lr){2-4} \cmidrule(lr){5-6}
\textbf{Model} & Ext. & Class. & Summ. & Short & Long & \textbf{Frac.} \\
\midrule
gpt-5-nano            & \yes & \yes &        & \yes & \yes   & 41\% \\
gemini-2.5-flash-lite &      & \yes & \yes   &      & \yes   & 17\% \\
gpt-4.1-mini          & \yes &      &        & \yes &        & 14\% \\
gpt-4o-mini           &      &      & \yes   &      & \yes   & 10\% \\
gpt-5                 &      & \yes &        & \yes &        & 10\% \\
\bottomrule
\end{tabular}
\vspace{-15pt}
\end{table}}

\subsection{Insights from MOAR's Pipelines}
\label{sec:evaluation-insights}

To understand what characterizes accurate pipelines, we analyze the 5 most accurate Pareto-optimal pipelines per workload (29 total, since Game Reviews has only 4 pipelines on the frontier).

\begin{itemize}[nosep, leftmargin=*, wide=0pt]

\item \textbf{86\% use a modified logical plan.} Top pipelines exhibit different logical structures—adding, removing, or restructuring operators. For example, in BlackVault, the initial pipeline first extracted event types via a \ttt{map}, then aggregated locations per event type using a \ttt{reduce} that reprocesses all documents. MOAR rewrote the initial \ttt{map} to extract both event types and locations, so the downstream \ttt{reduce} simply combines and deduplicates pre-extracted lists rather than re-analyzing full documents.

\techreport{\item \textbf{79\% use projection synthesis.} These strategies reduce document size before LLM operations, which not only reduces cost but also helps LLMs focus on relevant information. Among them, 55\% use deterministic methods (regex, full-text search, or code-based pruning), 17\% use embedding-based pruning, and 14\% use LLM-powered summarization.}

\item \textbf{48\% use agent-authored code.} These pipelines incorporate agent-authored code operations to replace or complement LLM-powered components. In a top-accuracy CUAD pipeline, a code operation using regex pattern matching was inserted before the LLM-powered map to identify clause-relevant sections and extract fixed-size context windows around each match. This reduced cost by 48.76\% while improving accuracy by 3.1\%.

\item \revision{\textbf{11 distinct rewrite directives are selected, but a few dominate.} Model substitution is the most frequent (25 usages) but {\em never} appears alone. Every pipeline combines it with at least one other rewrite directive. The most common non-model-substitution directives are clarify instructions (13 usages), chunking (11), code-based document compression (9), and head/tail compression (8), spanning LLM-centric improvements, data decomposition, and code synthesis categories. Fusion and reordering directives appear 4 times.}

\techreport{\item \textbf{Optimal models are workload-dependent.} All 29 pipelines switch from the default \ttt{gpt-4o-mini}, even though it is OpenAI's ``most cost-efficient small model''~\cite{openai_gpt4o_mini}. Each workload's most accurate pipeline uses a different model. The model usage patterns reveal specializations (\Cref{tab:model-task}): \ttt{gpt-5-nano} is the most prevalent (41\%), used for extraction and classification on both long and short documents, while smaller models like \ttt{gemini-2.5-flash-lite} and \ttt{gpt-4o-mini} are used for summarization on long documents.}

\item \textbf{High-accuracy pipelines are discovered late.} Among the 29 pipelines, 51.72\% were found after iteration 20 (the second half of the 40-iteration search), and 34.48\% were discovered after iteration 30---demonstrating that MOAR avoids premature convergence and maintains strong exploration throughout.

\item \rfour{\textbf{Most explored pipelines never contribute to the Pareto frontier.} Across workloads, 56--79\% of explored pipelines are neither on the final Pareto frontier nor ancestors of a frontier point in the search tree (58\% on CUAD, 79\% on Game Reviews, 76\% on BlackVault, 64\% on Biodex, 56\% on Medec, and 59\% on Sustainability). We discuss strategies for reducing this wasted budget in \Cref{sec:conclusion}\papertext{ of our technical report~\cite{moartechreport}}.}
\end{itemize}

\rfour{The empirical patterns above suggest some guidelines for pipeline authors. First, {\em invest in specification quality}---the logical plan will likely change during optimization, but a well-specified prompt helps the optimizer produce rewrites that preserve the user's intent. For example, on Medec, the carefully specified prompt from \citet{abacha2024medec}---which explicitly defined error categories and correction criteria---allowed MOAR to find a near-optimal plan early in search, whereas workloads with vaguer prompts required more iterations to converge.
Second, {\em examine the Pareto frontier rather than defaulting to the highest-accuracy plan}: the accuracy-cost trade-offs are steep (on CUAD, a 2.07\% accuracy drop reduced cost by 91.34\%). Users with budget constraints can often find plans that are nearly as accurate at a fraction of the cost.
}

\section{Related Work}
\label{sec:related}

We discuss semantic data processing systems and optimization.

\topic{Semantic Data Processing Systems} Systems that expose AI-powered data processing capabilities have proliferated over the past two years. ThalamusDB~\cite{jo2024thalamusdb} was among the earliest to support natural language filters and joins in SQL. LOTUS~\cite{patel2025semantic} coined the term ``semantic operators'' for this type of LLM-powered data processing; other semantic data processing systems include Palimpzest~\cite{liu2024declarative}, Aryn~\cite{anderson2024design}, DocETL~\cite{shankar2025docetl}, FlockMTL~\cite{flockdb}, ELEET~\cite{urban2024eleet}, Unify~\cite{unify}, DocDB~\cite{docdb}, \revision{and Stretto~\cite{sanmartino2026stretto}}. Other systems target specific applications, e.g., templatized documents~\cite{lin2025querying}, data extraction~\cite{sun2025quest, arora2023language}, and social science~\cite{hu2024leap}. Industrial systems from Databricks~\cite{databricks-llm}, DuckDB~\cite{duckdb-llm}, Snowflake~\cite{snowflake-llm, aggarwal2025cortex}, Google BigQuery~\cite{HeSethi2025BigQueryAIFunctions}, and Google AlloyDB~\cite{alloydb-llm} also now support LLM-powered UDFs.

\newcommand{\smallcite}[1]{{\scriptsize\cite{#1}}}

\begin{figure}[t]
\centering
\footnotesize
\vspace{-15pt}
\setlength{\tabcolsep}{3pt}
\begin{tabular}{@{}p{1cm} p{3.8cm} p{3.2cm}@{}}
\toprule
 & \textbf{Rule-based} & \textbf{Agentic} \\
\midrule
\textbf{Data-indep.} & 
  e.g., model substitution, ensembling, operator reordering \newline
  {\scriptsize\textcolor{violet}{\textsc{abacus}~\smallcite{russo2025abacus}, \textsc{flock-mtl}~\smallcite{flockdb}, \textsc{unify}~\smallcite{unify}, \textsc{docdb}~\smallcite{docdb}, \textsc{moar}}} & 
  e.g., operator fusion, task decomposition \newline
  {\scriptsize\textcolor{violet}{\textsc{docetl-v1}~\smallcite{shankar2025docetl}, \textsc{caesura}~\smallcite{urban2024demonstrating}, \textsc{unify}~\smallcite{unify}, \textsc{aop}~\smallcite{wang2025aop}, \textsc{moar}}} \\
\addlinespace
\textbf{Data-dep.} & 
  e.g., model cascades, fine-tuning, RAG \newline
  {\scriptsize\textcolor{violet}{\textsc{abacus}~\smallcite{russo2025abacus}, \textsc{lotus}~\smallcite{patel2025semantic}, \textsc{thalamusdb}~\smallcite{jo2024thalamusdb}, \textsc{cortex-aisql}~\smallcite{aggarwal2025cortex}, \textsc{eleet}~\smallcite{urban2024eleet}, \textsc{unify}~\smallcite{unify}, \textsc{docdb}~\smallcite{docdb}, \textsc{moar}}} & 
  e.g., code synthesis, other rewrite directives \newline
  {\scriptsize\textcolor{violet}{\textsc{docetl-v1}~\smallcite{shankar2025docetl}, \textsc{cortex-aisql}~\smallcite{aggarwal2025cortex}, \textsc{moar}}} \\
\bottomrule
\end{tabular}
\vspace{-10pt}
\caption{Space of rewrites for semantic operator pipelines. Data-independent rewrites can be instantiated without sample data; data-dependent rewrites require samples to learn configurations or synthesize transformations. Systems shown in \textcolor{violet}{purple}.}
\label{fig:plan-space}
\vspace{-15pt}
\end{figure}

\topic{Semantic Query Optimization} 
Query optimizers have three components: a plan space, a cost model, and a search algorithm.

\begin{enumerate}[nosep, leftmargin=*, wide=0pt]
\item \textbf{Plan space.} Classical relational database optimizers search over rewrites such as filter and join reordering~\cite{chaudhuri1998overview, hellerstein1993predicate, selinger1979access}.
These are rule-based, algebraic transformations.
Semantic operators admit a richer space of rewrites---including transformations over the meaning of the task and data---which we organize along two axes (\Cref{fig:plan-space}):
whether rewrites are {\em data-dependent} (requiring samples to instantiate) or {\em data-independent}, and whether rewrites are {\em agentic} (synthesized by an LLM, as in DocETL) or {\em rule-based} (not requiring an LLM agent to instantiate).
\techreport{Examples of rule-based, data-independent rewrites include physical implementations like model substitution and ensembling in ABACUS~\cite{russo2025abacus}.
Examples of rule-based, data-dependent rewrites---those that require samples to learn configurations but not an LLM agent to instantiate---include model cascades~\cite{patel2025semantic, viola2001rapid, kang2017noscope, zeighami2025cut}, context reduction~\cite{russo2025abacus}, and fine-tuning small LLMs~\cite{urban2024eleet}.
An example of a rewrite that requires an LLM agent to instantiate, but not sample data, is operator fusion.
Examples of agentic, data-dependent rewrites include most MOAR rewrite directives.}
These quadrants vary in instantiation cost---from cheap (rule-based, data-independent) to expensive (agentic, data-dependent)---but also in expressiveness: e.g., agentic rewrites can synthesize novel transformations.
MOAR is the only query optimizer with rewrites that span all four quadrants.
\item \textbf{Cost model.} \techreport{A cost model estimates the quality of candidate plans, guiding the search algorithm toward good plans.
Selinger's cost model, for example, estimates I/O and CPU costs using statistics like cardinality and index availability~\cite{selinger1979access}.} 
For semantic operators, the new challenge is estimating accuracy in addition to monetary cost or latency: there is no closed-form expression for how well an LLM will perform a task~\cite{semanticic, spade}. 
Approaches to accuracy estimation vary in cost and precision.
The simplest approach naively executes candidate plans on samples (as in LOTUS, DocETL-V1, and MOAR).
\rfour{ABACUS uses multi-armed bandits to adaptively sample plans and estimate their accuracy, tightening the accuracy bounds for promising (i.e., Pareto frontier) candidate plans.}
Some rewrites guarantee accuracy by construction, e.g., model cascades bound accuracy relative to the unrewritten plan~\cite{patel2025semantic, jo2024thalamusdb, jo2024smart, zeighami2025cut}.
All approaches require some ground truth---either a user-specified ``oracle'' LLM-powered implementation~\cite{patel2025semantic, russo2025abacus, jo2024smart}, a user-defined accuracy function~\cite{russo2025abacus, moartechreport}, or LLM-as-judge evaluation approaches~\cite{shankar2025docetl}.

\item \textbf{Search algorithm.} Given a plan space and cost model, a search algorithm finds good plans.
Classical optimizers use dynamic programming, composing optimal subplans into optimal plans~\cite{graefe1993volcano, Graefe1995TheCF}.
For semantic operators, two new challenges arise: plans may lack optimal substructure (as explained in \Cref{sec:intro}), and with accuracy as an objective, optimizers should return a Pareto frontier of plans, so users can choose their preferred cost-accuracy tradeoff.
ABACUS adapts Cascades~\cite{Graefe1995TheCF} to return a Pareto frontier; DocETL-V1 introduces a top-down search algorithm designed for LLM-as-judge evaluation; MOAR uses UCT-based search~\cite{kocsis2006bandit} over complete pipelines, avoiding optimal substructure assumptions.
ThalamusDB and LOTUS search for optimal parameters within specific rewrites (e.g., model cascade thresholds) but do not search across different rewrites.
Other solutions provide interactive interfaces to help users rewrite pipelines themselves~\cite{liu2025palimpchat, shankar2025steering}.
\end{enumerate}

Overall, MOAR spans the broadest rewrite space of any semantic query optimizer, returns a frontier of low-cost, high-accuracy plans, and handles the lack of optimal substructure in query plans.

\section{Conclusion\techreport{ and Future Work}}
\label{sec:conclusion}

We introduce MOAR (Multi-Objective Agentic Rewrites), a novel optimizer for LLM-powered data processing pipelines that jointly optimizes for both accuracy and cost. Building on DocETL's foundation of agentic rewrite directives, MOAR introduces {\em (i)} an expanded library of over 30 rewrite directives, {\em (ii)} a multi-armed bandit-based search algorithm that efficiently discovers sequences of rewrites that lead to good plans, and {\em (iii)} comprehensive empirical validation across six real-world workloads, demonstrating substantial improvements over state-of-the-art systems.
MOAR achieves the highest accuracy on all workloads. Compared to ABACUS~\cite{russo2025abacus}, the next-best optimizer, MOAR achieves 27\% higher accuracy on average while matching its best accuracy at only 55\% of its cost.

\techreport{\revision{Looking forward, several directions could improve the efficiency of MOAR's search process. First, {\em early pruning of unpromising pipelines}: across our six workloads, only 22--44\% of explored pipelines (varying by workload) are either on the Pareto frontier or are ancestors of a frontier point in the search tree---the remaining 56--78\% never contribute to the final frontier. Running fewer samples on these pipelines or terminating their evaluation early could substantially reduce wasted optimization budget. Second, {\em estimating accuracy and cost without executing pipelines on samples}: modeling how the accuracy of individual operators composes into end-to-end accuracy could allow pruning rewrites without running the full pipeline, though modeling this composition remains an open challenge given unpredictable operator interactions (as described in \Cref{sec:intro}). Third, {\em reducing reliance on frontier (e.g., gpt-5) LLM agents} for rewrite instantiation, e.g., by using open-source models or heuristics. Fourth, {\em user-specified accuracy lower bounds}: MOAR's search is currently biased toward high-accuracy regions of the Pareto frontier, which may under-explore cheap, moderate-accuracy plans. Allowing users to specify a minimum accuracy threshold would let the optimizer skip low-accuracy regions entirely, spending its budget on plans users actually care about.}}

\techreport{More broadly, MOAR demonstrates the promise of agentic approaches to query optimization: by delegating both the discovery and instantiation of rewrites to LLM agents guided by structured directives and intelligent search, we can effectively navigate the vast space of possible query plans for LLM-powered data processing.}

\bibliographystyle{ACM-Reference-Format}
\bibliography{sample}

\clearpage
\appendix
\section*{Appendix}

\noindent In this appendix, we provide the complete DocETL operator library (\Cref{app:operators}), detailed descriptions of all new rewrite directives introduced in MOAR (\Cref{sec:detailed-rewrite}), pseudocode for the pipeline selection and rewriting algorithms (\Cref{app:search}), and additional experimental results (\Cref{app:results}).

\begin{table*}
\centering
\footnotesize
\setlength{\tabcolsep}{4pt}
\vspace{-15pt}
\caption{DocETL operator library. Semantic operators invoke an LLM; operators marked with * do not. Operators marked with $\dagger$ are new in MOAR.}
\label{tab:docetl-operators}
\vspace{-10pt}
\begin{tabular}{l p{4cm} p{11cm}}
\toprule
\textbf{Operator} & \textbf{User configuration} & \textbf{Description} \\
\midrule
\texttt{map} &
prompt, output schema &
Uses an LLLM to execute a per-document transformation, adding new keys to the schema
(and optionally omitting existing ones). \\

\texttt{parallel-map} &
multiple prompts, output schemas &
Runs multiple independent \texttt{map} operations in parallel on each document, merging all
resulting fields into the schema. \\

\texttt{reduce} &
group-by keys, prompt, output schema &
Uses an LLM to aggregate groups of documents sharing the same key values, producing
one output document per group. \\

\texttt{filter} &
boolean prompt &
Uses an LLM to evaluate a boolean condition per document, retaining only documents
for which the condition is true. \\

\texttt{resolve} &
comparison prompt, resolution prompt &
Uses an LLM to identify fuzzily matching values across documents and replace them with
canonicalized versions through a two-step compare–resolve process. \\

\texttt{equijoin} &
comparison prompt &
Uses an LLM to semantically compare pairs of documents and determine whether they
should be joined on fuzzy or contextual matching of key values. \\
\midrule
\texttt{unnest}* &
array/dict field &
Flattens nested array or dictionary fields: arrays create multiple documents, while nested
dicts are merged into parent documents. \\

\texttt{split}* &
split key, chunk size &
Divides documents into token-limited or rule-based chunks, producing one document per
chunk. \\

\texttt{gather}* &
context-window configuration &
Augments each chunk with surrounding context (preceding and following chunks), without
changing the number of documents. \\

\texttt{sample}*${\dagger}$ &
sampling method; sample size; optional query; stratification keys &
Selects a subset of documents or chunks before downstream processing. The optional query
is a text template (provided by the user or synthesized by the agent) that the sampler
instantiates to assess relevance under BM25 or embedding-based sampling. Sampling can
also be stratified on user-provided keys. \\

\texttt{extract}${\dagger}$ &
prompt returning line ranges &
Uses an LLM to output only the relevant line spans (``lines 45–67, 103–120''), returning a
``compressed'' version or subset of the original document. \\

\texttt{code-map}*${\dagger}$ &
Python code, output schema &
A code-powered version of \texttt{map}; runs a user- or agent-generated Python function over each document and produces outputs matching the specified schema. \\

\texttt{code-reduce}*${\dagger}$ &
Python code, output schema &
A code-powered version of \texttt{reduce}; performs grouping and aggregation in Python, often followed by a lightweight \texttt{map} operator to generate narrative or structured summaries. \\

\texttt{code-filter}*${\dagger}$ &
Python code that returns true or false &
A code-powered version of ``filter``: evaluates a Python boolean function on each document and discards those for which the function returns false.\\
\bottomrule
\end{tabular}
\end{table*}

\section{Operators in DocETL}
\label{app:operators}

DocETL provides a library of semantic and auxiliary (i.e., not parameterized by natural language) operators used to construct document processing pipelines. \Cref{tab:docetl-operators} summarizes all operators supported in our implementation.

\section{Detailed Descriptions of New Rewrite Directives}
\label{sec:detailed-rewrite}

Throughout this section, we adopt the notation from \citet{shankar2025docetl}: given operators $A$ and $B$, we denote their composition as $A \to B$, where $(A \to B)(D) = B(A(D))$. For independent execution, we use $A \parallel B$. We may drop arguments (e.g., $\text{Map}_x(D)$ becomes $\text{Map}_x$) and omit subscripts except when the same operator appears multiple times. We color new or modified operators introduced by a rewrite in \textcolor{rewrite}{green}. The arrow $\Rightarrow$ denotes a rewrite of the operator (or operator sequence) on the left into the form on the right. \rtwo{Each rewrite directive is also accompanied by two kinds of validity checks. First, each newly generated operator must conform to the typing rules of its operator type (\Cref{tab:docetl-operators}). Second, {\em composability conditions} ensure that the rewritten pipeline produces outputs matching the user's expected schema. We distinguish two kinds: {\em preconditions}, which must hold for the rewrite to be applied, and {\em postconditions}, which the rewritten pipeline's output schema must satisfy. Both are checkable before execution and are denoted \emph{pre} and \emph{post} beneath each rewrite directive. Whether a rewritten prompt preserves the intent of the original cannot be verified automatically; this is validated empirically by evaluating each rewritten pipeline on $D_o$ (\Cref{sec:search}).}

\rtwo{For brevity, we write $s_x$ for the output schema of operator $o_x$. Formally, $s_x = \{(k_1, \tau_1), \ldots, (k_n, \tau_n)\}$ is a set of key--type pairs, where each $k_i$ is a key name and $\tau_i$ is its type. We say $s_x \subseteq s_y$ ({\em containment}) if every key--type pair in $s_x$ also appears in $s_y$. The {\em union} of two schemas, $s_x \cup s_y$, is the schema containing all key--type pairs from both (with shared keys required to have the same type). We write keys $(s_x) = \{k : (k, \tau) \in s_x\}$ for the key names in a schema.}

\topic{New Directive Categories} Categories marked with \newcat{} are entirely new to MOAR: operator fusion, approximation, reordering, and arbitrary rewrites.

\topic{Parameter-Sensitive Directives} Some directives are parameter-sensitive and marked with \paramsens{}: they generate multiple candidate rewrites with different parameter values, evaluate all candidates on the sample $D_o$, and select the one achieving highest accuracy. For these directives, we instruct the LLM agent to synthesize multiple distinct configurations exploring different trade-offs (e.g., precision vs.\ recall, cost vs.\ context length). Parameter-sensitive directives either enumerate discrete parameter values (e.g., chunk sizes) or use the agent to generate diverse configurations. The number of candidates generated is denoted by $k$ in Algorithm \ref{alg:rewrite-eval}.

\topic{Agent and Directive Prompts} The full agent prompts used for directive selection and instantiation are available in our codebase: the \href{https://github.com/ucbepic/docetl/blob/2bf97c66/docetl/reasoning_optimizer/agent.py}{\color{blue!60!black}main agent logic} and \href{https://github.com/ucbepic/docetl/tree/2bf97c66/docetl/reasoning_optimizer/directives}{\color{blue!60!black}individual directive prompts} (including documentation, use case guidance, instantiation schemas, and example applications).

\subsection{Fusion and Reordering\newcat{}}
\label{subsec:fusion}

Operator fusion combines multiple sequential operators into fewer operators, reducing the number of LLM calls and avoiding redundant passes over the same document. Reordering re-arranges commuting operators so that cheaper or more selective operators run earlier, reducing the amount of work done by expensive operators. This category is new in MOAR and primarily targets cost reduction, subject to preserving pipeline semantics.

When instantiating these directives, the LLM agent synthesizes merged prompts that unify the semantics of fused operators, combined output schemas, and, when necessary, auxiliary logic to maintain compatibility with downstream operators. It also verifies that reordering preserves semantics (e.g., ensuring a filter does not depend on outputs from an operator it is moved before).

In general, any pair of adjacent operators of the same type (e.g., \ttt{Map}--\ttt{Map} or \ttt{Filter}--\ttt{Filter}) can be fused into a single operator. Special cases arise when the operators differ in type but have dependent semantics, such as \ttt{Map}--\ttt{Reduce}, \ttt{Map}--\ttt{Filter}, or \ttt{Filter}--\ttt{Map}.

\subsubsection{Same-type Fusion}

This directive fuses adjacent operators of the same type (e.g., \ttt{Map}--\ttt{Map}, \ttt{Filter}--\ttt{Filter}, or \ttt{Reduce}--\ttt{Reduce}) into a single operator that implements the combined semantics:
\[
\begin{gathered}
\text{Map}_x \to \text{Map}_y \Rightarrow \textcolor{rewrite}{\text{Map}_z} \\
\rtwo{\text{pre: } s_x \cap s_y = \emptyset; \;\; \text{post: } s_z = s_x \cup s_y}
\end{gathered}
\]
and similarly for filters and reduces. The LLM agent rewrites the prompt template in $z$ to cover both tasks, and synthesizes an output schema that is the union of the original schemas (dropping intermediate fields that are not needed downstream). This reduces the number of LLM calls without requiring additional passes over the data.

\subsubsection{Map--Reduce Fusion}

This directive fuses a map with a downstream reduce, eliminating the need to materialize intermediate results:

\begin{equation}
\label{eq:mapreducefusion}
\begin{gathered}
\text{Map}_x \to \text{Reduce}_{K,y} \Rightarrow \textcolor{rewrite}{\text{Reduce}_{K,z}} \\
\rtwo{\text{pre: } K \cap \text{keys}(s_x) = \emptyset; \;\; \text{post: } s_z = s_y}
\end{gathered}
\end{equation}

The LLM agent rewrites the prompt of $\text{Reduce}_{K,y}$ to also perform the logic described by the preceding $\text{Map}_x$. This allows the reduce to compute per-document transformations and aggregate their results within a single LLM call. 
For example, in \Cref{ex:enhancement}, if police report documents already have a key or attribute representing case type, and the pipeline first maps each report to a list of enhancement factors ($\text{Map}_x$) and then summarizes them by case type ($ \text{Reduce}_{K,y}$), the agent can rewrite the reduce prompt in $z$ to both extract and summarize enhancement factors in one pass of the documents.

Note that a precondition for \Cref{eq:mapreducefusion} to be invoked is that when the output schema of $\text{Map}_x$ does not generate any of the grouping attributes in $K$; otherwise, the groupby keys would not exist prior to aggregation.

\subsubsection{Map--Filter Fusion}

This directive fuses an LLM-powered map followed by an LLM-powered filter that depends only on the map's outputs:

\begin{equation}
\label{eq:mapfilterfusion}
\begin{gathered}
\text{Map}_x \to \text{Filter}_y \Rightarrow \textcolor{rewrite}{\text{Map}_{z}} \to \textcolor{rewrite}{\text{CodeFilter}} \\
\rtwo{\text{pre: none}; \;\; \text{post: } s_z = s_x \cup s_y}
\end{gathered}
\end{equation}

The $\textcolor{rewrite}{\text{Map}_{z}}$ operator has a rewritten prompt that incorporates the logic of both the original map and filter, and an extended output schema that incorporates the boolean attribute in $y$'s output schema, indicating whether the document should be retained. The downstream $\text{CodeFilter}$ operator is then programmatically synthesized to drop documents where this attribute is false, ensuring that the pipeline preserves the original filter semantics. 

For example, in \Cref{ex:enhancement}, if $\text{Map}_x$ extracts snippets of text describing instances of excessive force and the $\text{Filter}_y$ identifies those involving a firearm, the agent rewrites the map prompt in $z$ to both extract the snippets {\em and} predict whether each involves a weapon, producing a boolean flag in addition to the output schema attributes requested in $x$. The code filter $\text{CodeFilter}$ then removes entries where the flag is false.

\subsubsection{Filter--Map Fusion}

This directive fuses an LLM-powered filter followed by an LLM-powered map, replacing the two-step evaluation with a single LLM call followed by a lightweight deterministic filter:

\begin{equation}
\label{eq:filtermapfusion}
\begin{gathered}
\text{Filter}_x \to \text{Map}_y \Rightarrow \textcolor{rewrite}{\text{Map}_{z}} \to \textcolor{rewrite}{\text{CodeFilter}} \\
\rtwo{\text{pre: none}; \;\; \text{post: } s_z = s_x \cup s_y}
\end{gathered}
\end{equation}

The $\textcolor{rewrite}{\text{Map}_{z}}$ operator has a rewritten prompt that combines the logic of both the original filter and map, and extends the output schema with a boolean attribute indicating whether the document satisfies the filter condition in $x$. Like in \Cref{eq:mapfilterfusion}, the downstream $\text{CodeFilter}$ operator is programmatically synthesized. For example, in \Cref{ex:enhancement}, if $\text{Filter}_x$ identifies police reports describing violent incidents and $\text{Map}_y$ extracts snippets of text describing the use of excessive force, the agent rewrites the map prompt to, at the same time, both extract these snippets and predict whether the incident qualifies as violent, outputting a boolean flag. $\text{CodeFilter}$ then simply removes entries where the flag is false, avoiding a separate LLM call for filtering. 

Note that Filter--Map fusion effectively ``pulls up'' the filter into the map stage. As a result, the rewrite may not always be optimal---especially when $\text{Filter}_x$ can be executed with a cheaper model than $\text{Map}_y$, or when the selectivity of $\text{Filter}_x$ is low enough that performing it separately would substantially reduce the number of documents processed by the more expensive map.

\subsubsection{Reordering}

Inspired by classical query optimization techniques, this directive reorders commuting operators to improve efficiency by moving selective or shrinking operations earlier in the pipeline:

\begin{equation}
\label{eq:reordering}
\begin{gathered}
o_x \to o_y \Rightarrow o_y \to o_x \\
\rtwo{\text{pre: } s_x \cap s_y = \emptyset; \;\; \text{post: none}}
\end{gathered}
\end{equation}

MOAR applies this rewrite only when the LLM agent verifies that reordering preserves pipeline semantics, e.g., the rewritten filter does not depend on attributes produced by the operator it is moved before. In practice, user-authored pipelines are short (2--3 operators)~\cite{shankar2025steering}, so reordering becomes more valuable as MOAR's search generates longer, more complex pipelines through sequences of rewrites.

\subsection{Code Synthesis\newcat{}}
\label{subsec:codesynthesis}

This new category directly targets cost reduction by replacing LLM calls with custom Python code intended to approximate the task described in the semantic operator.

\subsubsection{Code Substitution}

This directive replaces an LLM-powered operator with synthesized Python code:

\begin{equation}
\label{eq:codesubst}
\begin{gathered}
o_x \Rightarrow \textcolor{rewrite}{\text{Code}_{\hat{x}}} \\
\rtwo{\text{pre: none}; \;\; \text{post: } s_{\hat{x}} = s_x}
\end{gathered}
\end{equation}

where $\text{Code}$ is the code-powered version of the same operator type as $o_x$, and $\hat{x}$ contains Python code synthesized by an LLM agent to approximate the task specified in the original prompt template in $x$. The output schema remains unchanged---$\text{Code}_{\hat{x}}$ must produce outputs conforming to the same schema $s$ as the original operator. 
For \Cref{ex:enhancement}, if the task is to extract any mention of a firearm, the LLM agent might synthesize a regular expression that matches common firearm-related terms (e.g., gun, pistol, rifle, weapon, firearm, armed) and extracts surrounding sentences as context, avoiding LLM inference costs entirely while producing outputs in the same format as the original LLM-powered extraction.

\subsubsection{Code Substitution (Reduce)}

A specialized version of code substitution targets reduce operations where parts of the task are better suited to code than LLMs:

\begin{equation}
\label{eq:swapcode}
\begin{gathered}
\text{Reduce}_x \Rightarrow \textcolor{rewrite}{\text{CodeReduce}_{\hat{x}}} \to \textcolor{rewrite}{\text{Map}_y} \\
\rtwo{\text{pre: none}; \;\; \text{post: } s_y \subseteq s_x}
\end{gathered}
\end{equation}

This directive splits the reduce task into two stages: deterministic aggregations handled by code, followed by text generation or reasoning handled by an LLM. Both operators are synthesized by the LLM agent. For \Cref{ex:enhancement}, suppose the original reduce ($\text{Reduce}_x$) generates a detailed report summarizing enhancement factors per officer, including total counts and breakdowns by type. The agent might split this into: {\em (i)} a $\textcolor{rewrite}{\text{CodeReduce}}$ that groups documents by officer name, counts total enhancement factors, and computes counts per type (firearm, injury, kidnapping), producing structured data; and {\em (ii)} a $\textcolor{rewrite}{\text{Map}}$ that generates narrative report text from this data (e.g., ``Person X had 2 total enhancement factors: one firearm-related, one injury-related'').

\subsubsection{Document Compression (Code-based)\paramsens{}}

When the relevant content can be identified through deterministic rules rather than semantic understanding, this directive replaces the LLM-powered extraction with synthesized code, following the template:
\begin{equation}
\label{eq:doccompressiondet}
\begin{gathered}
o_x \Rightarrow \textcolor{rewrite}{\text{CodeMap}} \to \textcolor{rewrite}{o_{x'}} \\
\rtwo{\text{pre: none}; \;\; \text{post: } s_{x'} = s_x}
\end{gathered}
\end{equation}

The $\textcolor{rewrite}{\text{CodeMap}}$ operator executes Python code (using only standard libraries and the regular expression library \texttt{re}) that returns a compressed version of the document. This document compression approach avoids LLM calls entirely, unlike the previous approaches in \Cref{eq:docsummarization,eq:doccompression}. The operator $\textcolor{rewrite}{o_{x'}}$ is a modified version of $o_x$ whose prompt references the compressed content. For \Cref{ex:enhancement}, if enhancement factors are always mentioned in sections with specific headers (e.g., ``Incident Details,'' ``Evidence Collected''), the LLM agent might synthesize a code map that uses regular expressions to extract only paragraphs under these headers. The agent synthesizes both the code for $\textcolor{rewrite}{\text{CodeMap}}$ and the modified prompt for $\textcolor{rewrite}{o_{x'}}$.

This directive is parameter-sensitive. We instruct the LLM agent to synthesize two entirely different code implementations exploring different trade-offs: one optimizing for precision (stricter pattern matching) and one optimizing for recall (broader pattern matching). For example, when extracting firearm mentions, the precision-focused implementation might match only explicit weapon terms with exact regular expressions, while the recall-focused implementation might include broader contextual patterns and proximity-based matching. Both implementations are evaluated on $D_o$ and the higher-accuracy variant is selected.

\subsubsection{Head/Tail Compression\paramsens{}}

One specific instantiation of code-based document compression extracts only the first $h$ words (head) and last $\ell$ words (tail) from a document, which we refer to as \emph{head/tail} compression. While head/tail compression follows the same template in \Cref{eq:doccompressiondet}, we provide it as an explicit directive because agents do not reliably discover it independently, even though the pattern is broadly applicable. For example, classifying a document's genre or identifying its author may only require examining the opening paragraphs.

This directive is parameter-sensitive. In the directive description provided to the agent (as explained in \Cref{sec:rewrite}), we instruct the agent to generate two different configurations with different head/tail lengths: one using shorter context windows (e.g., $h=100, \ell=50$) optimizing for cost efficiency, and another using longer windows (e.g., $h=300, \ell=150$) optimizing for higher recall. For tasks where key information appears in opening paragraphs (e.g., document classification or author identification), the directive description (as explained in \Cref{sec:rewrite}) suggests the agent may allocate more words to the head. Both configurations are evaluated on $D_o$ and the higher-accuracy variant is selected.

As LLM agents improve, or we train our own agents based on known rewrite directive patterns, explicit instantiations of directives like head/tail compression may become unnecessary. But, for now, they provide valuable guidance for discovering common optimization patterns.

\subsection{Data Decomposition}
\label{subsec:data_decomp}

DocETL introduced directives for document chunking, to improve accuracy when processing long documents, and for multi-level aggregation, to combine results across groups of documents. MOAR extends the ``data decomposition'' category with additional chunking strategies that provide more fine-grained control over {\em which} portions of documents to process.

\subsubsection{Chunk Sampling\paramsens{}}

When documents are split into many chunks, processing all chunks may be unnecessary if only a subset contains relevant information. The chunk sampling directive introduces a sampling step after gathering context:

\begin{align}
\label{eq:chunksampling}
\text{Split} \to \text{Gather} \to \text{Map} \to \text{Reduce} \Rightarrow \; & \text{Split} \to \text{Gather} \to \textcolor{rewrite}{\text{Sample}} \nonumber \\
& \to \text{Map} \to \text{Reduce} \nonumber \\
\rtwo{\text{pre: none}; \;\; \text{post: none}} &
\end{align}

The $\textcolor{rewrite}{\text{Sample}}$ operator can be instantiated with random sampling, keyword search (based on BM25 retrieval~\cite{robertson2009probabilistic}), or embedding-based similarity, selecting the top-$k$ document chunks relevant to a query~\cite{lewis2020retrieval}. When instantiating the directive, the LLM agent synthesizes the sampling method, the query (if not random sampling), and the value of $k$. 

For example, in \Cref{ex:enhancement}, the agent might choose keyword-based sampling after splitting police records into chunks. It could synthesize the keyword list \ttt{["firearm", "injury", "kidnapping", "weapon", "harm"]} and set $k=20$ to select the 20 chunks with highest BM25 scores before applying the extraction map. Alternatively, for identifying cases involving excessive force, the agent might choose embedding-based sampling with the query ``excessive use of force by police officer,'' computing embeddings for all chunks and selecting the $k$ chunks with highest cosine similarity to this query. These selected chunks are then processed by the map and their results reduced into a final output.

This directive is parameter-sensitive. We instruct the LLM agent to generate two entirely different sampling configurations: one optimizing for precision (using stricter sampling criteria with lower $k$) and one optimizing for recall (using broader sampling criteria with higher $k$). For example, the precision-focused configuration might use BM25 with $k=10$ and strict keyword matching, while the recall-focused configuration might use embedding-based sampling with $k=30$ and a broader text query. Both configurations are evaluated on $D_o$ and the higher-accuracy variant is selected.

\subsubsection{Document Sampling\paramsens{}}

When a reduce operation aggregates over many documents within each group (where the grouping is defined by the reduce keys), it may be unnecessary to process every document if many contribute little or no signal to the final aggregation. The document sampling directive inserts a sampling stage before the reduce:

\begin{equation}
\label{eq:docsampling}
\begin{gathered}
\text{Reduce}_{K,x} \Rightarrow \textcolor{rewrite}{\text{Sample}_{K}} \to \text{Reduce}_{K,x} \\
\rtwo{\text{pre: none}; \;\; \text{post: none}}
\end{gathered}
\end{equation}

The operator $\textcolor{rewrite}{\text{Sample}_{K}}$ selects a subset of documents from each group defined by $K$, using random sampling, BM25 keyword search, or embedding-based similarity. The LLM agent synthesizes both the sampling method and the per-group sample size $k$, selecting the documents most relevant to the downstream aggregation logic.

For example, in \Cref{ex:enhancement}, suppose the pipeline aggregates enhancement factors per precinct, where $K = \{\texttt{precinct\_id}\}$. Some precincts may contain hundreds of reports, many of which have no enhancement-related content. The agent might synthesize an embedding-based sampler that, for each precinct, selects the $k=30$ reports most similar to the query ``mentions of injuries, weapons, or threats,'' forwarding only these to the reduce. A precision-oriented configuration might instead select the $k=10$ reports containing explicit weapon or injury keywords (``firearm'', ``injury'', ``harm'', ``weapon'') using BM25.

This directive is parameter-sensitive. We instruct the agent to generate at least two distinct sampling configurations—one emphasizing precision (smaller $k$, stricter criteria) and one emphasizing recall (larger $k$, broader retrieval). Both variants are evaluated on $D_o$, and the higher-accuracy configuration is selected.

\subsubsection{Cascade Filtering\paramsens{}}
This directive optimizes filtering costs by injecting a cascade of cheaper ``pre-filters'' before an expensive LLM-powered filter:
\begin{equation}
\label{eq:cascadefiltering}
\begin{gathered}
\text{Filter}_x \Rightarrow \textcolor{rewrite}{\text{CodeFilter}} \to \textcolor{rewrite}{\text{Filter}_y} \to \text{Filter}_x \\
\rtwo{\text{pre: none}; \;\; \text{post: none}}
\end{gathered}
\end{equation}

where one or more code filters and LLM filters may be synthesized. The cascade consists of two stages of pre-filters, ordered by increasing cost: first, deterministic Python code (using regular expressions, keyword matching, or simple logic) that quickly eliminates documents failing obvious criteria; second, cheap LLM-powered filters $\text{Filter}_y$ with simplified prompts and inexpensive models (e.g., gpt-5-nano), ordered by prompt length (shortest first). The pre-filters prioritize high recall (rarely rejecting documents that would pass the main filter) but may have lower precision (allowing through documents that will eventually be filtered out). This design ensures the final filter produces the same results as the original, while reducing cost by eliminating many documents before expensive evaluation.

When instantiating this directive, the LLM agent examines sample documents from $D_o$ to identify patterns distinguishing documents that pass versus fail the main filter. The agent then synthesizes code filters for patterns observable through keyword presence, regular expressions, or document structure, and LLM filters with short prompts that evaluate simple semantic properties difficult to capture with code.
For \Cref{ex:enhancement}, if the original filter identifies police reports describing violent incidents with firearms, the agent might synthesize: (i) a code filter checking for weapon-related keywords (``gun'', ``pistol'', ``firearm'', ``weapon''); (ii) a gpt-5-nano filter checking ``Does this report describe a violent incident?''; followed by (iii) the original expensive filter performing nuanced interpretation of what constitutes a violent firearm incident.

This directive is parameter-sensitive: we instruct the agent to generate two cascade configurations exploring different combinations of code filters and LLM pre-filters, evaluating each on $D_o$ to select the highest-accuracy pipeline.

\subsection{Projection Synthesis}
\label{subsec:proj_synth}

DocETL introduced projection synthesis directives that decompose complex tasks into simpler subtasks (e.g., chaining multiple maps, isolating independent projections). MOAR extends this category by identifying a sub-class of projection synthesis: rather than decomposing the {\em task} described in an operation's prompt, these directives reduce the {\em data} processed by the operation. By making documents smaller before applying an LLM-powered operator, these directives can improve cost while preserving the information needed for accurate results. We provide various directives to make the documents smaller.

\subsubsection{Document Summarization}

This directive inserts a map operation at the beginning of the pipeline to summarize each document, preserving all information needed for downstream operations, following the template:

\begin{equation}
\label{eq:docsummarization}
\begin{gathered}
o_x \Rightarrow \textcolor{rewrite}{\text{Map}} \to \textcolor{rewrite}{o_{x'}} \\
\rtwo{\text{pre: none}; \;\; \text{post: } s_{x'} = s_x}
\end{gathered}
\end{equation}

Here, $\textcolor{rewrite}{\text{Map}}$ produces a summary of the document, and $\textcolor{rewrite}{o_{x'}}$ is a modified version of the original operator $o_x$ whose prompt references the summary instead of the full document. The LLM agent synthesizes both the summarization operator $\textcolor{rewrite}{\text{Map}}$ and the modified prompt for $\textcolor{rewrite}{o_{x'}}$ to ensure all information needed by the downstream extraction is preserved.

\subsubsection{Document Compression (LLM-based)}

This directive inserts an LLM-powered {\em extraction} operation at the beginning of the pipeline to retain only content relevant for downstream operations, following the template:

\begin{equation}
\label{eq:doccompression}
\begin{gathered}
o_x \Rightarrow \textcolor{rewrite}{\text{Extract}} \to \textcolor{rewrite}{o_{x'}} \\
\rtwo{\text{pre: none}; \;\; \text{post: } s_{x'} = s_x}
\end{gathered}
\end{equation}

The $\textcolor{rewrite}{\text{Extract}}$ operator asks the LLM to output line ranges that are relevant (e.g., ``lines 45-67, 103-120''), which are then converted back into a subset of the original document. This differs from summarization in \Cref{eq:docsummarization}, where the map operation generates entirely new text---which might not be a subset of the original document. Since $\textcolor{rewrite}{\text{Extract}}$ outputs only line ranges rather than narrative text, it is typically cheaper to execute. The operator $\textcolor{rewrite}{o_{x'}}$ is a modified version of $o_x$ whose prompt references the extracted content. The LLM agent synthesizes both the extraction prompt and the modified prompt for $\textcolor{rewrite}{o_{x'}}$.

\subsection{LLM-Centric Rewrites}
\label{subsec:llm-centric}

DocETL introduced directives that improve LLM output quality by refining how tasks are specified to the LLM. For example, the gleaning directive uses a validator LLM to check outputs and provide feedback for iterative refinement. Similar strategies have been explored in ABACUS, which implements a ``critique-and-refine'' physical implementation of map operations~\cite{russo2025abacus}. Inspired by prompting strategies~\cite{liuprompting2023} and optimization techniques~\cite{ramnath2025systematic}, MOAR adds directives that improve prompt quality and provide examples to guide LLM behavior.

\subsubsection{Model Substitution}
This directive replaces the model used by an operator:
\begin{equation}
\label{eq:modelsubst}
\begin{gathered}
o_x \Rightarrow \textcolor{rewrite}{o_{x'}} \\
\rtwo{\text{pre: none}; \;\; \text{post: } s_{x'} = s_x}
\end{gathered}
\end{equation}
where $x = (t, s, m)$ and $x' = (t, s, m')$ with $m' \neq m$. When instantiating this directive, the LLM agent receives context about model performance: for each model in the available pool $M$, the agent sees the cost and accuracy achieved by the original pipeline when executed with that model on a sample of data. The agent also has access to each model's performance on MRCR (a long-context benchmark that evaluates an LLM's ability to retrieve and distinguish between multiple similar requests hidden in long contexts~\cite{openai_mrcr, vodrahalli2024michelangelo}), as well as context window size and pricing details.

Using this information, the agent can reason about model capabilities and select $m'$ based on the operator's complexity and position in the pipeline. For \Cref{ex:enhancement}, the agent might substitute GPT-4o-mini for operators extracting explicit mentions of weapons, while keeping GPT-4o for operators requiring more complex interpretation (e.g., determining if force was excessive given the circumstances).

\subsubsection{Clarify Instructions in Prompt\paramsens{}}

This directive rewrites an operator by making its prompt template more specific and detailed:

\begin{equation}
\label{eq:clarifyinstructions}
\begin{gathered}
o_x \Rightarrow \textcolor{rewrite}{o_{x'}} \\
\rtwo{\text{pre: none}; \;\; \text{post: } s_{x'} = s_x}
\end{gathered}
\end{equation}

where $x = (t, s, m)$ and $x' = (t', s, m)$, with $t'$ being a clarified version of prompt template $t$. The LLM agent analyzes the original prompt and a sample of documents, identifies ambiguous instructions, and generates a more detailed prompt that reduces the likelihood of misinterpretation. For \Cref{ex:enhancement}, an original prompt might say {\em ``extract evidence of threatening with a firearm.''} 
After examining sample police reports, the agent might observe that reports use varied terminology (``weapon,'' ``gun,'' ``pistol,'' ``armed'') and describe threats in different ways (``pointed at,'' ``brandished,'' ``displayed'').
The agent might then clarify the prompt to {\em ``extract evidence of threatening with a firearm. This includes any instance where: {\em (i)} the report mentions a firearm, weapon, gun, pistol, rifle, or other projectile weapon; AND {\em (ii)} the report describes the weapon being pointed at, brandished, displayed, or used to intimidate someone. Extract the complete sentence(s) containing both elements.''}

This directive is parameter-sensitive. We instruct the LLM agent to generate two  different clarified prompts exploring different clarification strategies, and both clarified prompts are evaluated on $D_o$ to determine which clarification strategy yields higher accuracy.

One might wonder why such clarifications help---why can't the LLM executing the operator simply reason through these ambiguities? In practice, the LLM agent performing optimization (e.g., gpt-5) is typically more powerful than the models used in operations (i.e., $m$, which might be gpt-4o-mini for cost efficiency). The more capable agent is therefore well-suited to identify ambiguities and aspects requiring additional reasoning, then encode that reasoning directly into the prompt so that cheaper models can execute the task reliably.

\subsubsection{Find Few-shot Examples}

A popular prompt engineering technique is to include few-shot examples in the prompt to demonstrate desired behavior~\cite{brown2020language}. This directive synthesizes such examples to improve operator accuracy:

\begin{equation}
\label{eq:fewshot}
\begin{gathered}
o_x \Rightarrow \textcolor{rewrite}{o_{x'}} \\
\rtwo{\text{pre: none}; \;\; \text{post: } s_{x'} = s_x}
\end{gathered}
\end{equation}

where $x' = (t', s, m)$ and $t'$ is a modified prompt template that incorporates few-shot examples. The agent examines sample documents, generates input-output pairs that demonstrate the desired behavior, and constructs the modified prompt with these examples embedded. Note that tools like DSPy~\cite{khattab2024dspy} could automate few-shot example generation when this directive is chosen. However, DSPy's iterative optimization process (which evaluates multiple example sets to find the best one) takes significantly longer than our agent's ``single-pass'' instantiation. Supporting such iterative optimization would require modifying our search algorithm to account for varying directive instantiation costs---we leave this to future work.

\subsubsection{Arbitrary Rewrite}

Beyond the structured directives described above, MOAR supports a ``meta-directive'' that allows LLM agents to propose custom rewrites without any directive scaffolding.  This flexibility is important because the space of possible rewrites is unbounded, and even a large directive library cannot anticipate all beneficial transformations. For \Cref{ex:enhancement}, an agent might propose adding a new map operation that extracts the reporting officer's precinct and experience level from metadata fields, then uses this information in downstream extraction prompts to adjust interpretation of what constitutes ``excessive force'' based on departmental policies and officer training. This would require edits to multiple different operators spread across the pipeline, and does not fit cleanly into any existing directive category.

To implement arbitrary rewrites, we pass the entire pipeline as YAML code to the agent and ask it to propose edits using a find-and-replace approach inspired by how coding agents work~\cite{aider_edit_formats}. Specifically, the agent returns a list of search-and-replace blocks, where each block specifies: {\em (i)} a unique string to search for in the original pipeline YAML, and {\em (ii)} the replacement text. 
After the agent produces an arbitrary rewrite, we verify that the resulting pipeline can be parsed and executed by DocETL. If parsing or validation fails, we provide the error message to the agent and retry up to 3 times before discarding the rewrite.

\section{Search Algorithm Details}

This appendix provides pseudocode for two of MOAR's search procedures. Algorithm \ref{alg:select} describes the selection phase, which traverses the search tree using UCT-based utility scores with progressive widening to choose which pipeline to rewrite next. Algorithm \ref{alg:rewrite-eval} describes the rewriting and evaluation phase, which uses an LLM agent to choose and instantiate a directive, then evaluates the resulting pipeline(s) on the sample $D_o$.

\label{app:search}
\begin{algorithm}
\scriptsize
\SetAlgoLined
\KwIn{Root pipeline $P_0$, current search tree $T_t$}
\KwOut{Pipeline $P^\star$ to rewrite}
\BlankLine
\SetKwFunction{FSelect}{Select}
\SetKwProg{Fn}{Function}{:}{}
\Fn{\FSelect{$P_0, G_t$}}{
    $P \gets P_0$\;
    \While{true}{
        \uIf{$|\mathrm{children}(P)| < W(n_t(P))$ \textbf{or} $P$ has no evaluated children \label{line:fanout-check}}{
            \textbf{break}\;
        }
        \tcp{Descend to child with highest utility}
        $P \gets$ child $P' \in \mathrm{children}(P)$ with highest $U_t(P')$\;
    }
    \tcp{Increment visit count for $P$ and all ancestors}
    $P' \gets P$\;
    \While{$P' \neq \mathrm{null}$}{
        $n_t(P') \gets n_t(P') + 1$\;
        $P' \gets \mathrm{parent}(P')$\;
    }
    \Return $P$\;
}
\caption{Selecting the pipeline to rewrite}
\label{alg:select}
\end{algorithm}

\begin{algorithm}
\caption{Rewriting and evaluation}
\label{alg:rewrite-eval}
\scriptsize
\SetAlgoLined
\KwIn{Selected pipeline $P^\star$, current search tree $T_t = (V_t, E_t)$, directive usage map $\nu$, evaluation sample $D_o$}
\KwOut{Evaluated child pipeline $P'$, applied rewrite $r$, statistics $(\hat{c}(P'), \hat{a}(P'))$, and candidate count $k$}
\BlankLine
\SetKwFunction{FRewriteEval}{RewriteAndEvaluate}
\SetKwProg{Fn}{Function}{:}{}
\Fn{\FRewriteEval{$P^\star, G_t, \nu, D_o$}}{
    \tcp{Determine objective based on frontier position}
    $\text{rank} \gets$ rank of $P^\star$ by accuracy among pipelines in $V_t$\;
    \uIf{$\text{rank} \leq |V_t|/2$}{
        objective $\gets$ ``reduce cost while preserving accuracy''\;
    }
    \Else{
        objective $\gets$ ``improve accuracy''\;
    }
    \tcp{Step 1: Prune registry and choose directive}
    allowed\_directives $\gets$ PruneRegistry($P^\star$, registry, $E_t$)\;
    $(d, \text{target\_ops}) \gets$ ChooseDirective($P^\star$, allowed\_directives, $T_t$, $\nu$, $\{\mu_t(\cdot)\}$, objective)\;
    $\nu(P^\star, d) \gets \nu(P^\star, d) + 1$ \tcp*{Soft-prevent other concurrent workers choosing this directive}
    \tcp{Step 2: Instantiate directive (may generate multiple candidates for parameter-sensitive directives)}
    $\{r_1, \ldots, r_k\} \gets$ InstantiateDirective($d$, target\_ops, $P^\star$, objective, $D_o$)\;
    \tcp{Evaluate all candidates and select most accurate}
    \For{$i = 1$ to $k$}{
        $P'_i \gets$ apply rewrite $r_i$ to $P^\star$\;
        execute $P'_i$ on $D_o$ to obtain $(\hat{c}(P'_i), \hat{a}(P'_i))$\;
    }
    $(r, P') \gets \arg\max_{(r_i, P'_i)} \hat{a}(P'_i)$ \tcp*{Select rewrite with highest accuracy}
    \Return $(P', r, \hat{c}(P'), \hat{a}(P'), k)$\;
}
\end{algorithm}

\newpage
\section{Additional Experimental Results}
\label{app:results}

\begin{figure}[h]
\centering
\includegraphics[width=0.6\columnwidth]{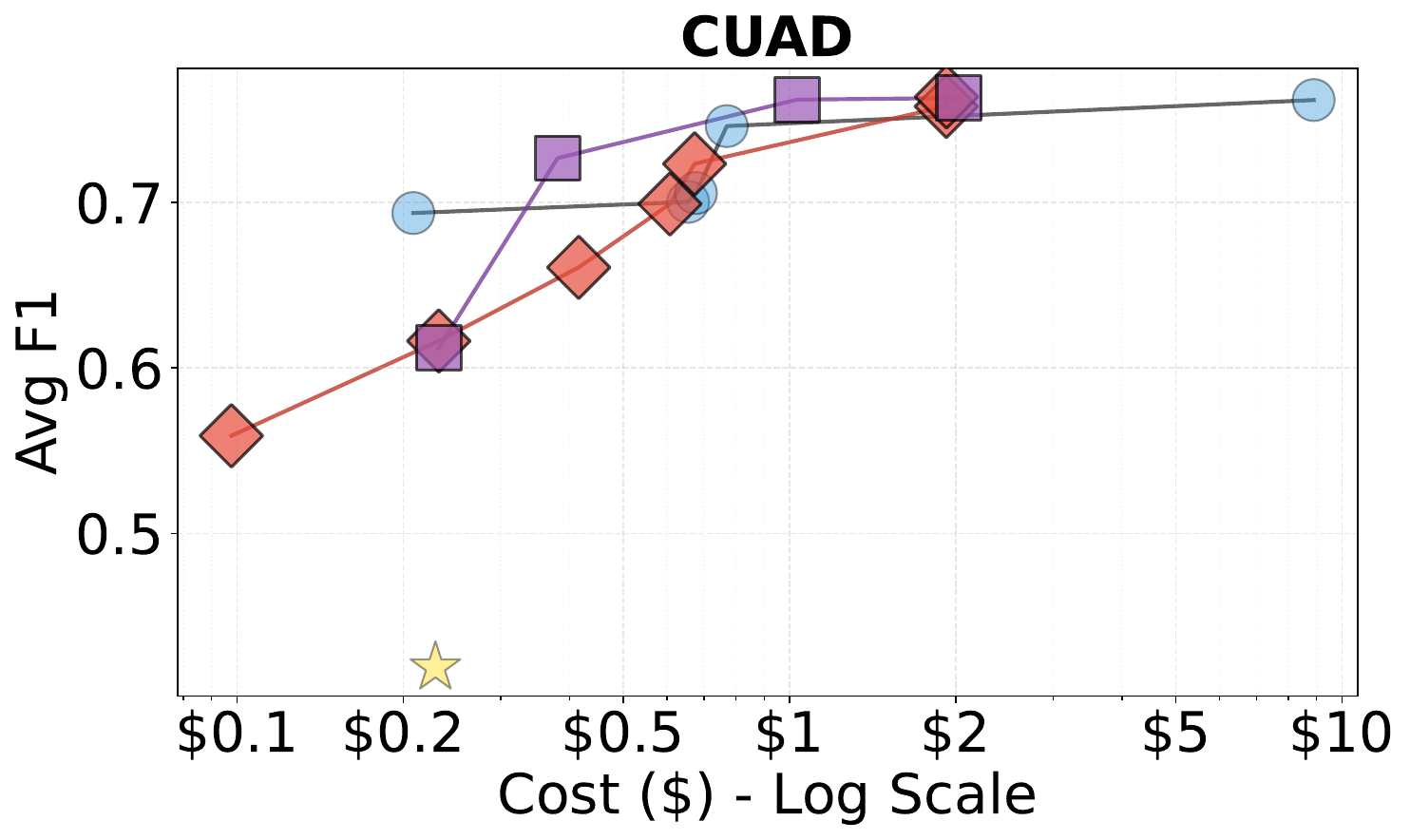}
\includegraphics[width=0.9\linewidth]{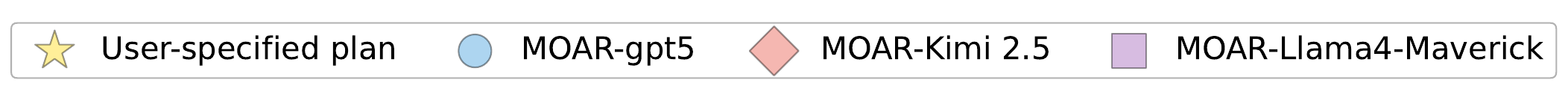}
\caption{\rtwo{Pareto frontiers on CUAD when varying the LLM used to select and instantiate rewrite directives. gpt-5 (default) shown as reference.}}
\label{fig:cuad-os}
\end{figure}

\topic{Open-Source Agent LLMs} \rtwo{We evaluate MOAR with four open-source LLMs as the agent on the CUAD workload: Kimi K2.5~\cite{team2026kimi}, Llama-4-Maverick-17B~\cite{llama4}, Qwen3-8B~\cite{qwen3}, and Llama-3.1-8B~\cite{llama3}. \Cref{tab:agent-models} summarizes the results. Kimi K2.5 and Llama-4-Maverick produce Pareto frontiers comparable to gpt-5 (\Cref{fig:cuad-os}). Qwen3-8B and Llama-3.1-8B fail to produce competitive pipelines. Qwen3-8B's context window (32k tokens) is too small for the agent to read document samples alongside directive schemas and instantiation examples, causing all directive instantiations to fail. Llama-3.1-8B-Instruct-Turbo has a sufficiently large context window (128k tokens) but cannot follow the instructions in the agent prompt for directive instantiation: it selects directives inapplicable to target operator types (e.g., applying \ttt{reduce\_gleaning} to a \ttt{map} operator), generates responses with invalid control characters that cannot be parsed, and produces malformed instantiations that cause runtime errors (\Cref{tab:failure-logs}).}

\begin{figure}
    \centering
    \vspace{-10pt}
    \captionsetup[subfigure]{skip=0pt}
    \setlength{\abovecaptionskip}{2pt}
    \setlength{\belowcaptionskip}{0pt}
    
    \begin{subfigure}{0.49\columnwidth}
        \centering
        \includegraphics[width=\linewidth]{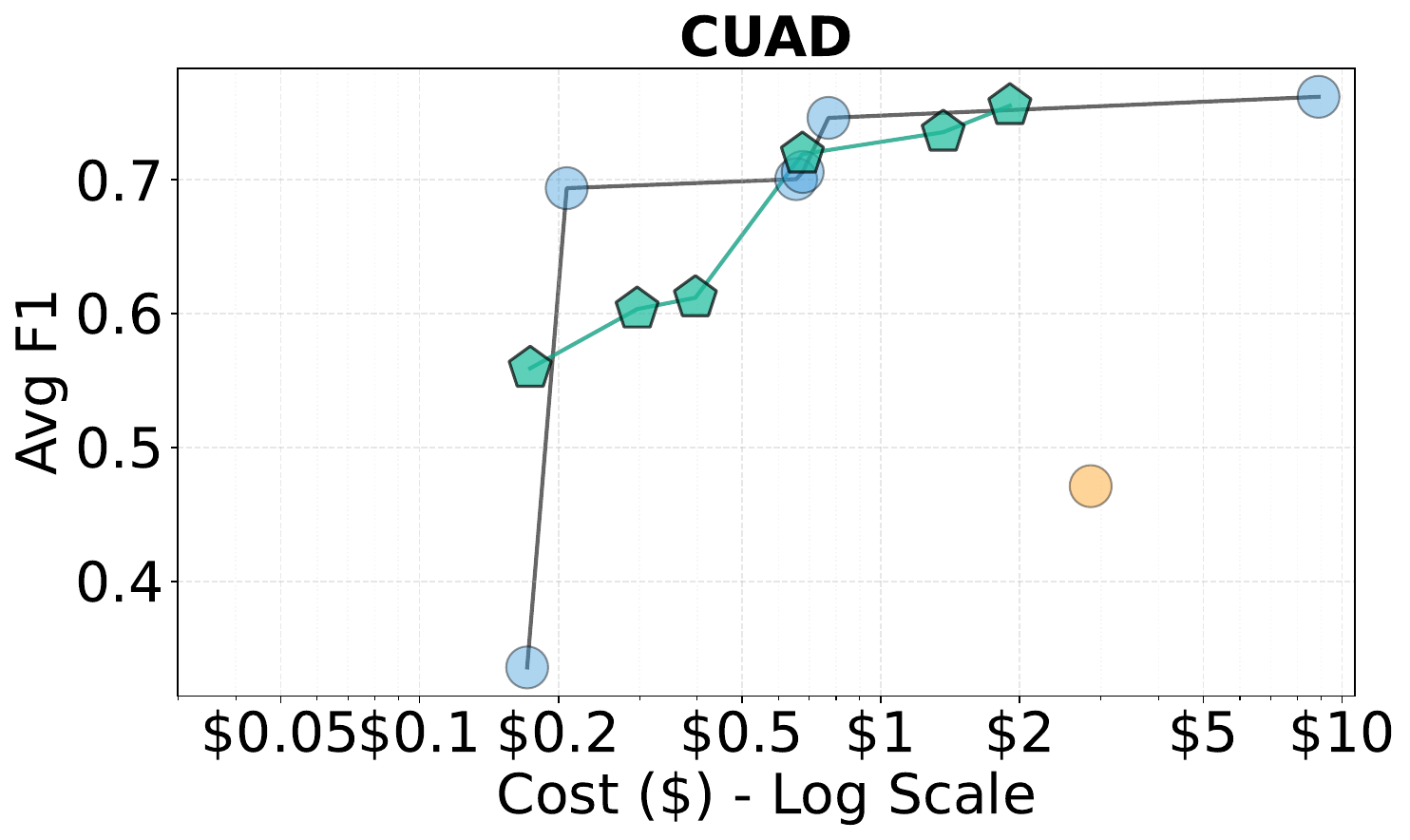}
    \end{subfigure}%
    \begin{subfigure}{0.49\columnwidth}
        \centering
        \includegraphics[width=\linewidth]{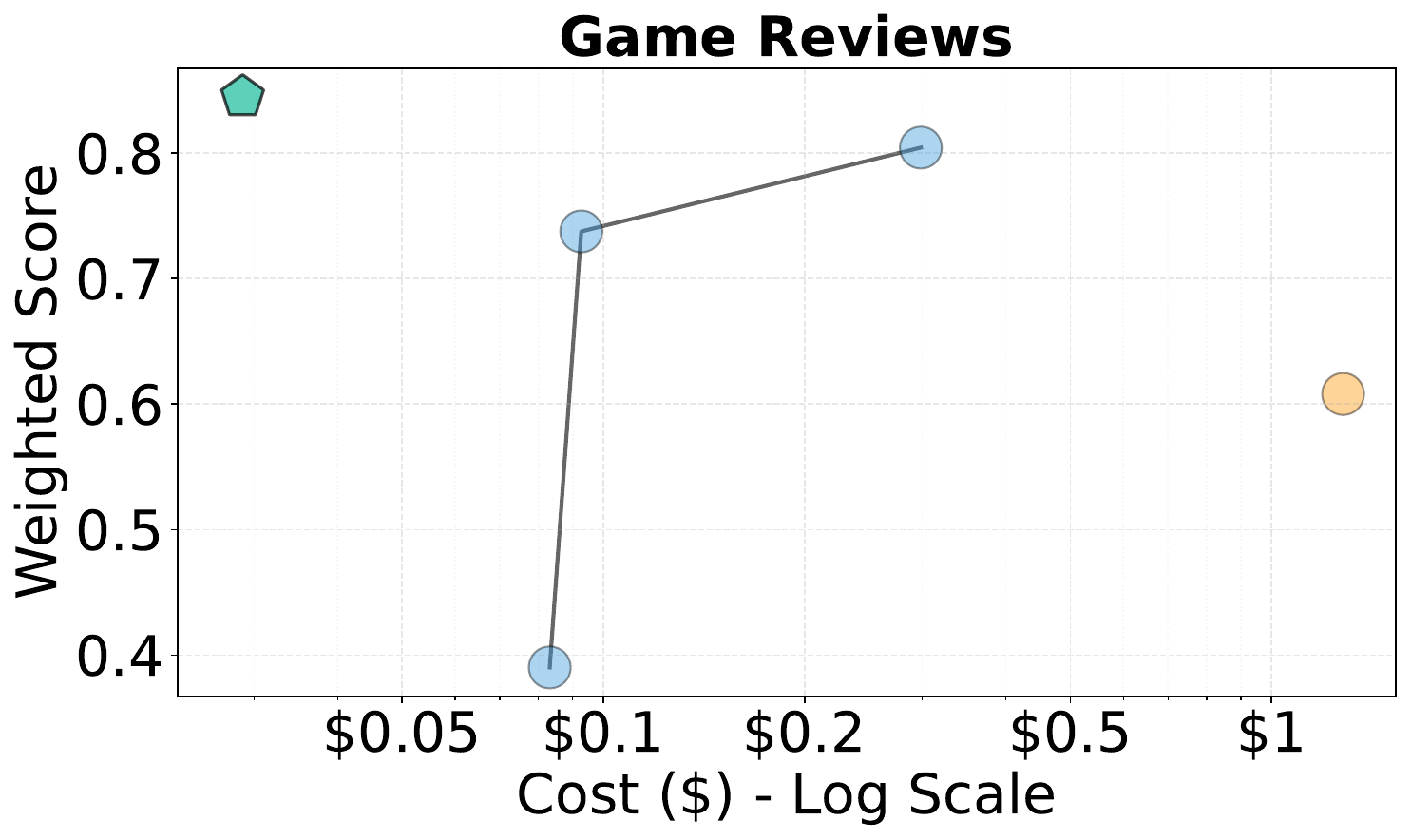}
    \end{subfigure}\\[-1mm]
    \begin{subfigure}{0.49\columnwidth}
        \centering
        \includegraphics[width=\linewidth]{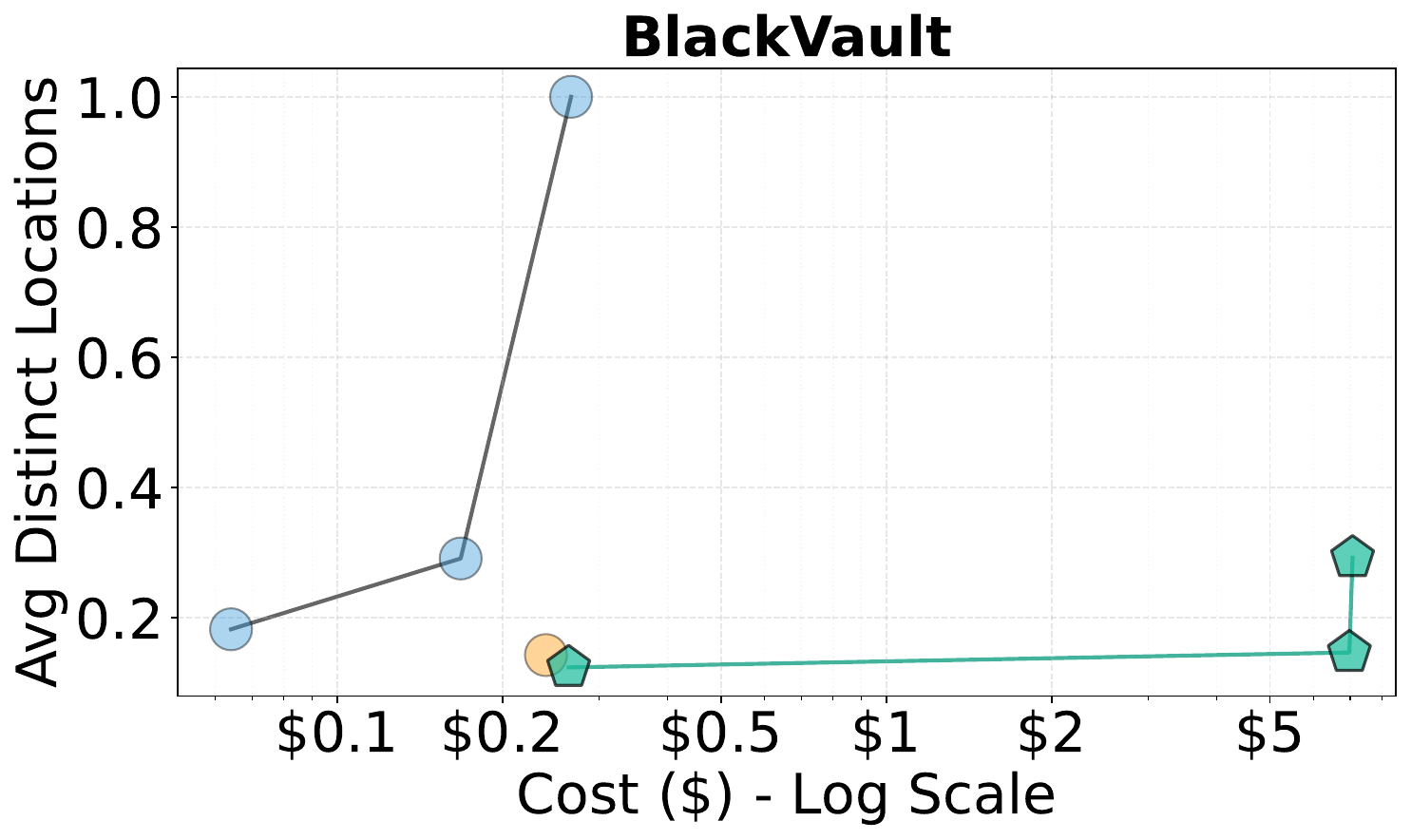}
    \end{subfigure}%
    \begin{subfigure}{0.49\columnwidth}
        \centering
        \includegraphics[width=\linewidth]{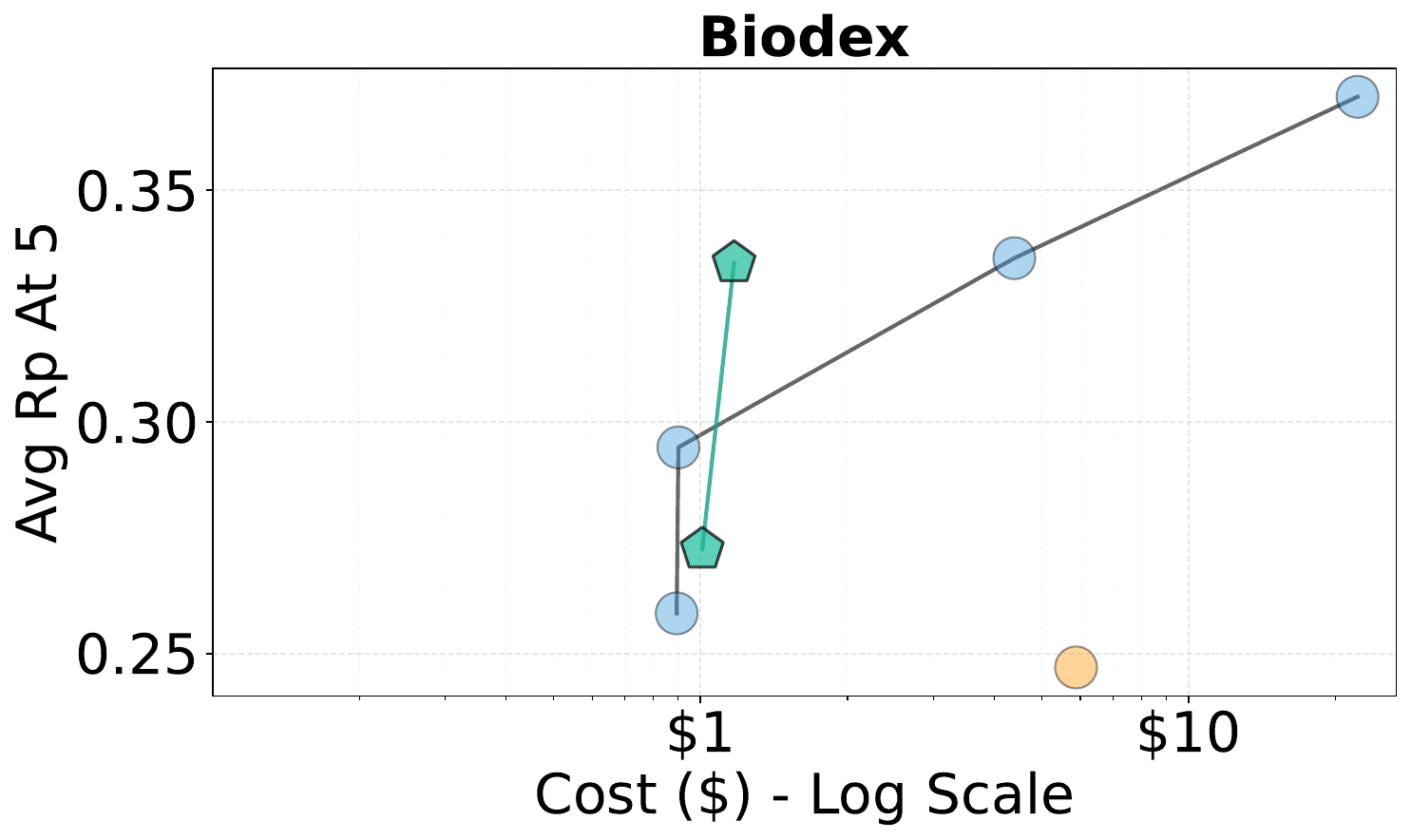}
    \end{subfigure}\\
    \begin{subfigure}{0.49\columnwidth}
        \centering
        \includegraphics[width=\linewidth]{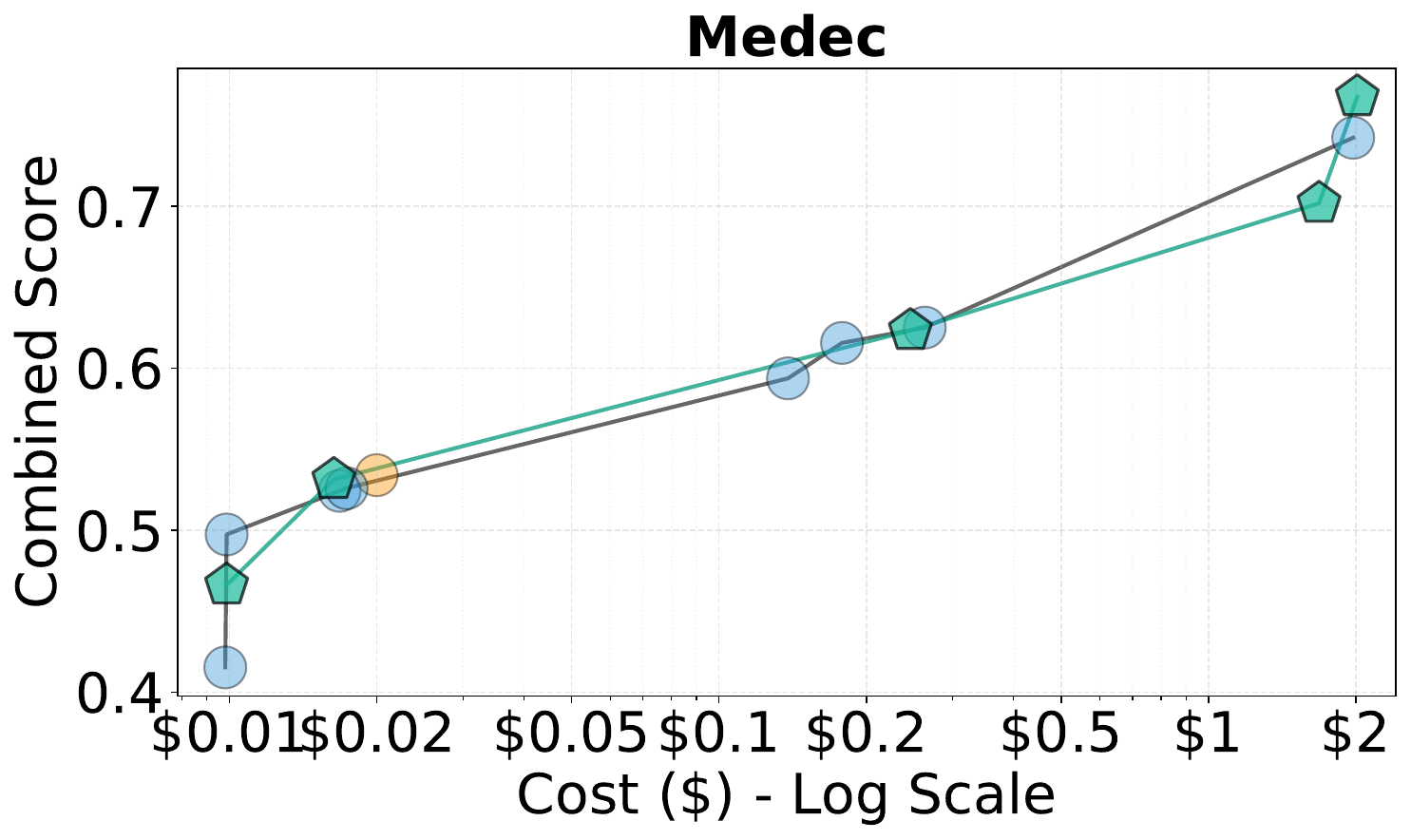}
    \end{subfigure}%
    \begin{subfigure}{0.49\columnwidth}
        \centering
        \includegraphics[width=\linewidth]{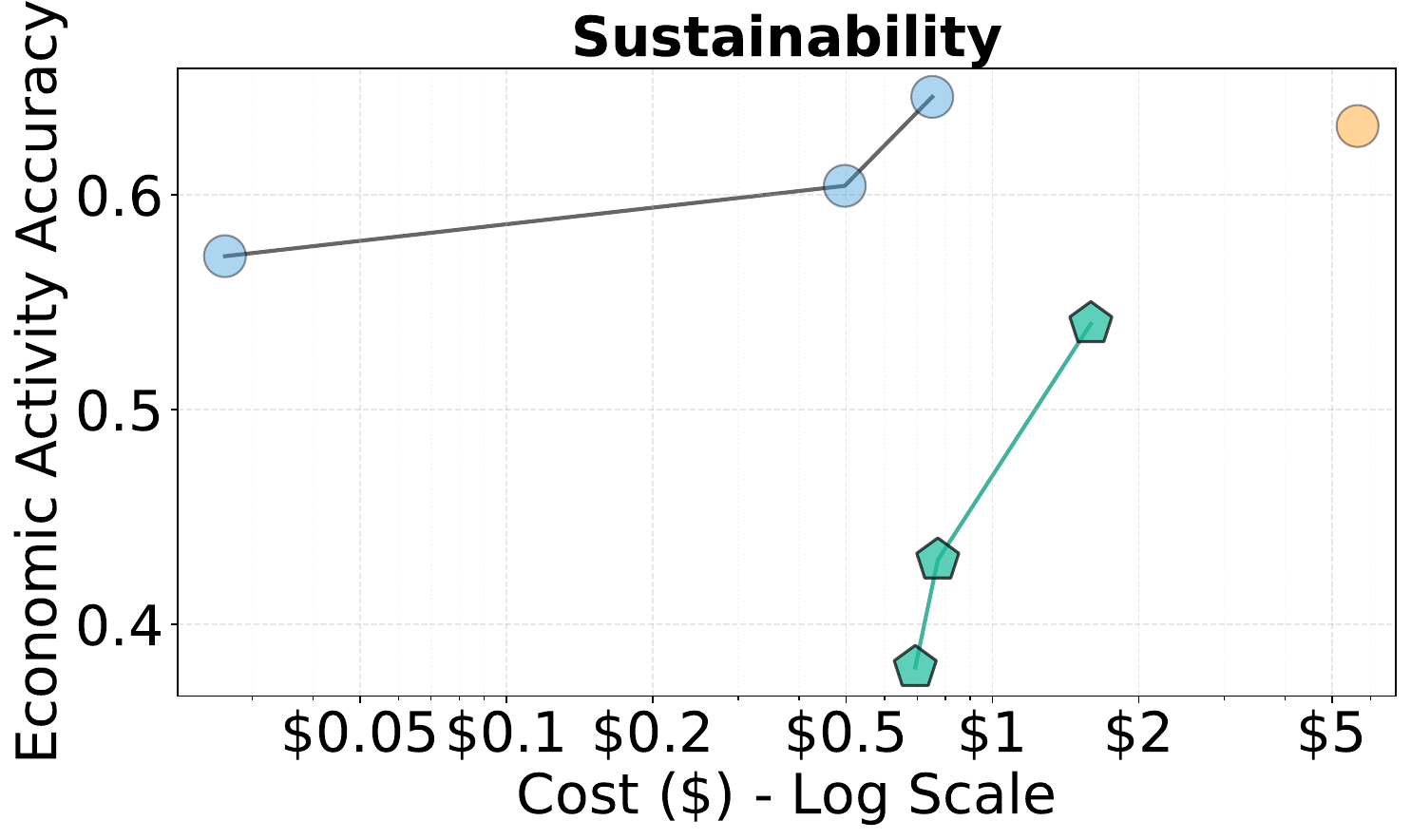}
    \end{subfigure}
    \includegraphics[width=0.6\columnwidth]{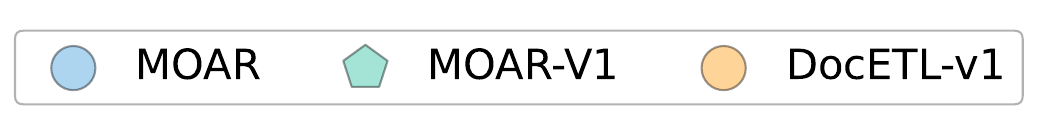}
    \vspace{-3pt}
    \caption{\rfour{Pareto frontiers for MOAR, MOAR-V1 (MOAR's search algorithm restricted to DocETL-V1's directives), and DocETL-V1. On four of six workloads, the expanded directive library produces strictly better frontiers; on two workloads, the new search algorithm alone already improves over DocETL-V1.} }
\label{fig:directive-library}
\vspace{-10pt}
\end{figure}

\topic{New vs.\ Old Directives} \rfour{MOAR introduces both a new search algorithm and an expanded directive library relative to DocETL-V1. To understand how much each component contributes, we run MOAR-V1---MOAR's search algorithm restricted to only DocETL-V1's directives---on all six workloads (\Cref{fig:directive-library}).
On BlackVault and Biodex, even MOAR-V1 already outperforms DocETL-V1, demonstrating that the new search algorithm alone provides gains; adding the new directives further improves the Pareto frontiers on these workloads. On CUAD and Sustainability, the new directives are essential---MOAR's frontier strictly dominates MOAR-V1. On Medec, the two MOAR variants produce comparable frontiers, suggesting that the new search algorithm provides most of the benefit and the existing DocETL-V1 directives are sufficient for this workload. On Game Reviews, MOAR-V1's top-accuracy plan exceeds MOAR's---it applies model substitution and gleaning (both DocETL-V1 directives), but MOAR's search with the full directive library did not happen to explore this configuration.}

\topic{Accuracy Generalization} \rtwo{\Cref{tab:generalization} reports accuracy differences between the optimization sample ($D_o$) and the held-out test set ($D_T$) across all Pareto-optimal pipelines per workload.}

\begin{figure*}
  \centering
  \vspace{-10pt}
  \includegraphics[width=0.95\textwidth]{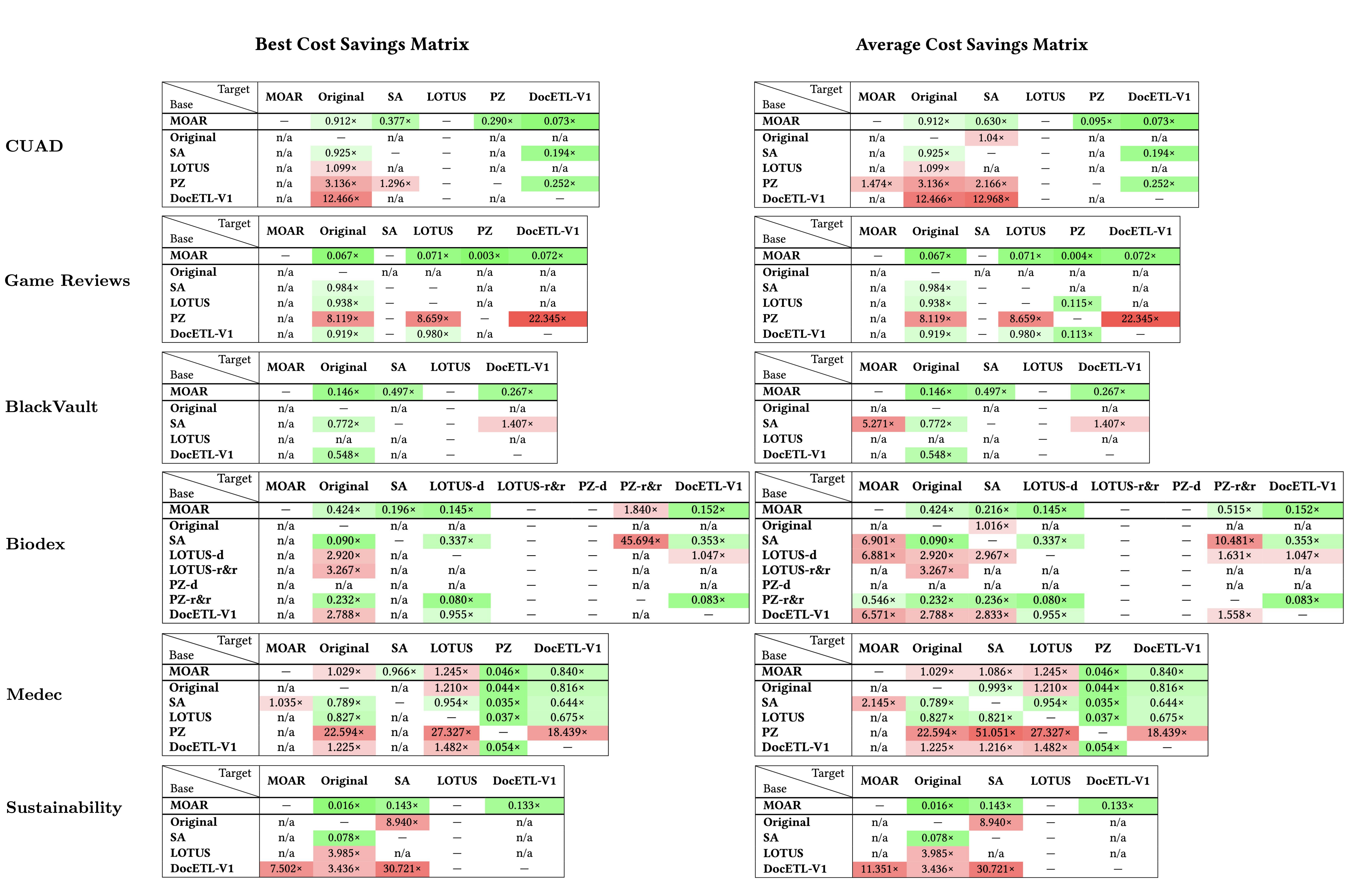}
  \caption{Comparison of cost savings across {\em all pairs} of methods on all workloads. Original represents the initial user-authored pipeline. Rows indicate the base method and columns indicate the target method. In the Best Cost Savings Matrix, each cell shows the monetary cost (in multiples) incurred by the base method to achieve the target method's best accuracy. In the Average Cost Savings Matrix, each cell shows the average monetary cost incurred by the base method to achieve the accuracy of each of the target method's pipelines. ``n/a'' indicates the base method cannot achieve the target method's accuracy; ``--'' indicates the target method does not achieve the original pipeline's accuracy. Diagonal entries are marked with ``--''. }
\label{fig:matrix}
\end{figure*}

\topic{Latency and Cost Comparisons} \Cref{tab:latency} reports test-time latencies for each method's optimized pipelines. \Cref{fig:matrix} presents pairwise cost savings matrices showing how much each method costs to match every other method's accuracy, both for best-accuracy pipelines and averaged across all Pareto-optimal pipelines.

\begin{table}
\centering
\footnotesize
\setlength{\tabcolsep}{4pt}
\caption{\rtwo{LLMs evaluated to select and instantiate rewrite directives on CUAD. gpt-5 is the default. ``---'' indicates the LLM failed to produce any competitive pipeline.}}
\label{tab:agent-models}
\begin{tabular}{lcccc}
\toprule
\textbf{LLM} & \textbf{Reasoning?} & \textbf{Context} & \textbf{Max Acc.} & \textbf{Cost} \\
\midrule
gpt-5 & Yes & 128k & 0.7618 & 8.8865 \\
Kimi K2.5 & Yes & 128k & 0.7637 & 1.9242 \\
Llama-4-Maverick-17B & No & 1M & 0.7630 & 2.0264 \\
Qwen3-8B & No & 32k & --- & --- \\
Llama-3.1-8B & No & 128k & --- & --- \\
\bottomrule
\end{tabular}
\end{table}

\begin{table}
\centering
\footnotesize
\setlength{\tabcolsep}{3pt}
\caption{\rtwo{Representative errors with Llama-3.1-8B to select and instantiate rewrite directives on CUAD.}}
\label{tab:failure-logs}
\begin{tabular}{l l p{3.8cm}}
\toprule
\textbf{Directive} & \textbf{Target} & \textbf{Error} \\
\midrule
(parse failure) & --- & Invalid control character at line 2 col 23 \\
\texttt{reduce\_gleaning} & \texttt{map} & Can only be applied to reduce ops, got map \\
\texttt{swap\_with\_code} & \texttt{map} & Can only be applied to reduce ops, got map \\
\texttt{operator\_fusion} & \texttt{map, map} & List index out of range \\
\bottomrule
\end{tabular}
\end{table}

\begin{table}
\centering
\footnotesize
\setlength{\tabcolsep}{4pt}
\caption{\rtwo{Accuracy generalization from optimization sample ($D_o$) to held-out test set ($D_T$) across all Pareto-optimal pipelines per workload. Mean Diff is the average accuracy difference ($D_T - D_o$); Std Diff is its standard deviation.}}
\vspace{-8pt}
\label{tab:generalization}
\begin{tabular}{l r r r}
\toprule
\textbf{Workload} & \textbf{\#Plans} & \textbf{Mean Diff} & \textbf{Std Diff} \\
\midrule
CUAD & 6 & $+0.006$ & $0.015$ \\
BlackVault & 5 & $-0.002$ & $0.256$ \\
Game Reviews & 3 & $-0.046$ & $0.034$ \\
Sustainability & 4 & $-0.097$ & $0.030$ \\
Biodex & 5 & $-0.049$ & $0.039$ \\
Medec & 11 & $-0.091$ & $0.041$ \\
\midrule
Average & -- & $-0.047$ & -- \\
\bottomrule
\end{tabular}
\end{table}

\begin{table*}[t]
\centering
\footnotesize
\setlength{\tabcolsep}{4pt}
\begin{tabular}{lcccccccccr}
\hline
\textbf{Dataset} & \textbf{MOAR} & \textbf{Original} & \textbf{SA} & \textbf{LOTUS} & \textbf{PZ} & \textbf{PZ-d} & \textbf{PZ-r\&r} & \textbf{LOTUS-d} & \textbf{LOTUS-r\&r} & \textbf{DocETL-V1}\\
\hline
CUAD & 179.24 ± 134.64 & 89.91 & 160.78 ± 126.64 & 92.18 & 60.65 ± 9.82 & -- & -- & -- & -- & 140.00  \\
Game Reviews & 88.04 ± 42.85 & 350.98 & 279.19 ± 8.05 & 1446.50 & 735.57 ± 414.51 & -- & -- & -- & -- & 240.85\\
BlackVault & 92.68 ± 28.51 & 34.89 & 41.70 ± 1.51 & 23.34 & -- & -- & -- & -- & -- & 37.14\\
Biodex & 163.31 ± 123.13 & 150.25 & 245.00 ± 274.87 & -- & -- & 205.49 ± 253.37 & 464.34 ± 20.20 & 307.17 & 12.15 & 402.22 \\
Medec & 51.76 ± 48.67 & 6.95 & 50.10 ± 78.30 & 28.90 & 39.90 ± 11.51 & -- & -- & -- & -- & 6.52\\
Sustainability & 254.72 ± 218.29 & 109.91 & 78.57 ± 7.57 & 2094.90 & -- & -- & -- & -- & -- & 686.52\\
\hline
\end{tabular}
\caption{Test plan latency (seconds) across datasets and methods. ``Original'' refers to the user-specified pipeline prior to optimization. For methods returning multiple pipelines, values show mean ± std across all discovered pipelines; single values indicate one pipeline. ``--'' indicates the method is not evaluated on that dataset.}
\label{tab:latency}
\end{table*}

\end{document}